\documentclass[12pt]{iopart}
\pdfoutput=1
\usepackage{iopams}
\usepackage{graphicx}
\usepackage{dcolumn}
\usepackage[dvipsnames]{xcolor}   
\usepackage[normalem]{ulem} 
\usepackage{enumerate}
\usepackage{yfonts}
\usepackage{longtable}
\usepackage{tabu} 

\expandafter\let\csname equation*\endcsname\relax
\expandafter\let\csname endequation*\endcsname\relax
\usepackage{physics}

\begin{document}

\title{Classical antennae, quantum emitters, and densities of optical states} 

\author{{William~L.~Barnes}}
\address{Department of Physics and Astronomy, University of Exeter, Exeter EX4 4QL, United Kingdom, and}
\address{Complex Photonic Systems (COPS), MESA+ Institute for Nanotechnology, University of Twente, P.O. Box 217, 7500 AE Enschede, Netherlands}

\author{{Simon~A.~R.~Horsley}}
\address{Department of Physics and Astronomy, University of Exeter, Exeter EX4 4QL, United Kingdom}

\author{{Willem~L.~Vos}}
\address{Complex Photonic Systems (COPS), MESA+ Institute for Nanotechnology, University of Twente, P.O. Box 217, 7500 AE Enschede, Netherlands}

\begin{abstract}
We provide a pedagogical introduction to the concept of the local density of optical states (LDOS), illustrating its application to both the classical and quantum theory of radiation.  
We show that the LDOS governs the efficiency of a macroscopic classical antenna, determining how the antenna's emission depends on its environment. 
The LDOS is shown to similarly modify the spontaneous emission rate of a quantum emitter, such as an excited atom, molecule, ion, or quantum dot that is embedded in a nanostructured optical environment. 
The difference between the number density of optical states, the local density of optical states, and the partial local density of optical states is elaborated and examples are provided for each density of states to illustrate where these are required.  
We illustrate the universal effect of the LDOS on emission by comparing systems with emission wavelengths that differ by more than 5 orders of magnitude, and systems whose decay rates differ by more than 5 orders of magnitude. 
To conclude we discuss and resolve an apparent difference between the classical and quantum expressions for the spontaneous emission rate that often seems to be overlooked, and discuss the experimental determination of the LDOS.
\end{abstract}

\date{30th August 2019, in preparation for J. Opt.}

\maketitle

\section{Introduction}\label{sec:intro}
\subsection{Conceptual overview}\label{sec:general}
The local density of optical states (LDOS) measures the availability of electromagnetic (EM) modes at a given point in space.
While this availability of modes is important in many electromagnetic phenomena, we shall here concentrate on the radiation from a small electric dipole antenna.
The antenna radiates in a rather subtle way.
To begin with we force a current to oscillate in the antenna, for example driving it with an external circuit.  As the current varies, an electromagnetic field is produced. Now for the subtlety; this radiated field acts back on the antenna.
This is the so--called \emph{radiation reaction}, where the radiated field does work on the oscillating current, and thereby determines how much power is radiated.
We can imagine the radiation reaction as a pulling of energy out of the oscillating current, putting it into the available modes of the electromagnetic field, whereupon the energy propagates away.
The antenna will emit very differently depending on the local availability of modes into which it may radiate. 
If there are no available modes that can propagate away from the antenna, then no radiation can occur. 
Conversely if a great number of propagating modes have a large intensity where the antenna is placed, then radiation can be emitted very easily. 
From this discussion we see that the amount of radiated power must be governed by the local density of states.  Something analogous to the LDOS can be observed while playing a drum. 
Each mode of the drum skin has nodes---points where the skin does not move---and in hitting the drum you only excite those modes that move the drum skin at the point you hit. 
Strike a bongo drum in the centre and you'll make a bass note, strike it close to the edge and a higher frequency note is produced. 
Strike it right on the edge and no sound comes from the drum skin at all.
%
%
\par
It is clear that the concept of the local density of optical states is a purely classical one, so why do we usually encounter the LDOS concept in conjunction with the spontaneous emission of photons by, for example, quantum dots?  
The answer is both straightforward and subtle. 
The straightforward part is that the exciton in a quantum dot is in essence a very small oscillating current, and the way it radiates is still governed by the availability of EM modes, just as for an antenna broadcasting radio waves. 
But this is not to say that quantum mechanics is completely irrelevant. 
The subtlety is that a quantum emitter -- such as an exciton in a quantum dot -- also experiences the local density of states in a second, non--classical way. 
To appreciate this second contribution we can think of the allowed EM modes in some environment and ask whether quantum mechanics changes them in any way; the answer is both no and yes! 
No, in that in quantum mechanics the allowed modes are the same as those in classical physics. 
Yes, in that in quantum mechanics the allowed modes can never be empty of energy, instead each mode has a zero-point energy or vacuum fluctuation~\cite{Riek_Science_2015_350_420}. 
As a result, in quantum mechanics the electromagnetic field acting on the emitter has two contributions, the radiation reaction is one, and vacuum fluctuations are the other.
For model quantum emitters, these two contributions have an equal magnitude and therefore quantum emitters are twice as bright as their classical equivalent!~\footnote{It is even more subtle: the two contributions are only equal under the assumption of a two level atom (see~\cite{Milonni}, chapter 4), but this is beyond the scope of the present article.} 
Fortunately no disagreement arises when we consider macroscopic antennas, where the contribution from the radiation reaction far exceeds that due to the vacuum field.  Classical and quantum views give the same answer - as demanded by Bohr's correspondence principle.
\par
The local density of states is a concept that can be used to explain a multitude of electromagnetic phenomena, from practical antenna physics~\cite{rudge1982, whitfield2000, agio2013, ARRL2015book}, to fluorescence~\cite{Chew_PRA_1988_38_3410, Snoeks_PRL_1995_74_2459, Sprik_EPL_1996_35_265, Lakowicz1999book, Valeur2012book, demchenko2015}, and even quantum forces due to zero point energy~\cite{Casimir1948PR, Casimir1948Proc, rosa2011, enoch2012}. 
In this article we take a rather unconventional approach by beginning with an analysis of the local density of optical states from a classical perspective. 
We then explore how the quantum picture may be naturally accommodated and finally show the equivalence of the two viewpoints. 
The only really quantum aspect of a quantum emitter, apart from the factor of 2 mentioned above, is in the time domain, where a quantum emitter radiates a probabilistic stream of photons rather than a continuous radiation field.
\par
In the remainder of this first section we introduce the reader to the terminology and the phenomenology of radiation emission in a structured environment. 
We consider the source of radiation to be an oscillating electric dipole, where the source of radiation is small compared to the wavelength, this is by far the commonest type of spontaneous emission source~\cite{Karaveli_PRL_2011_106_193004, Li_PRL_2018_121_227403}. 
The details of the theory necessary to calculate the classical or quantum emission from a small antenna are developed later below. 
We also deal with the distinction between the classical and quantum mechanical emission process, and how to infer the local density of states from a measurement of the emission rate.
If the reader only desires a basic understanding and the essential formulae, then section~\ref{sec:intro} should suffice. Throughout the article we make use of an example system where the density of optical states is easily comprehended, the case of an emitter or antenna in front of a mirror; this system is our leitmotif.

\begin{figure}[ht!]
\centering
\includegraphics[width=1.0\columnwidth]{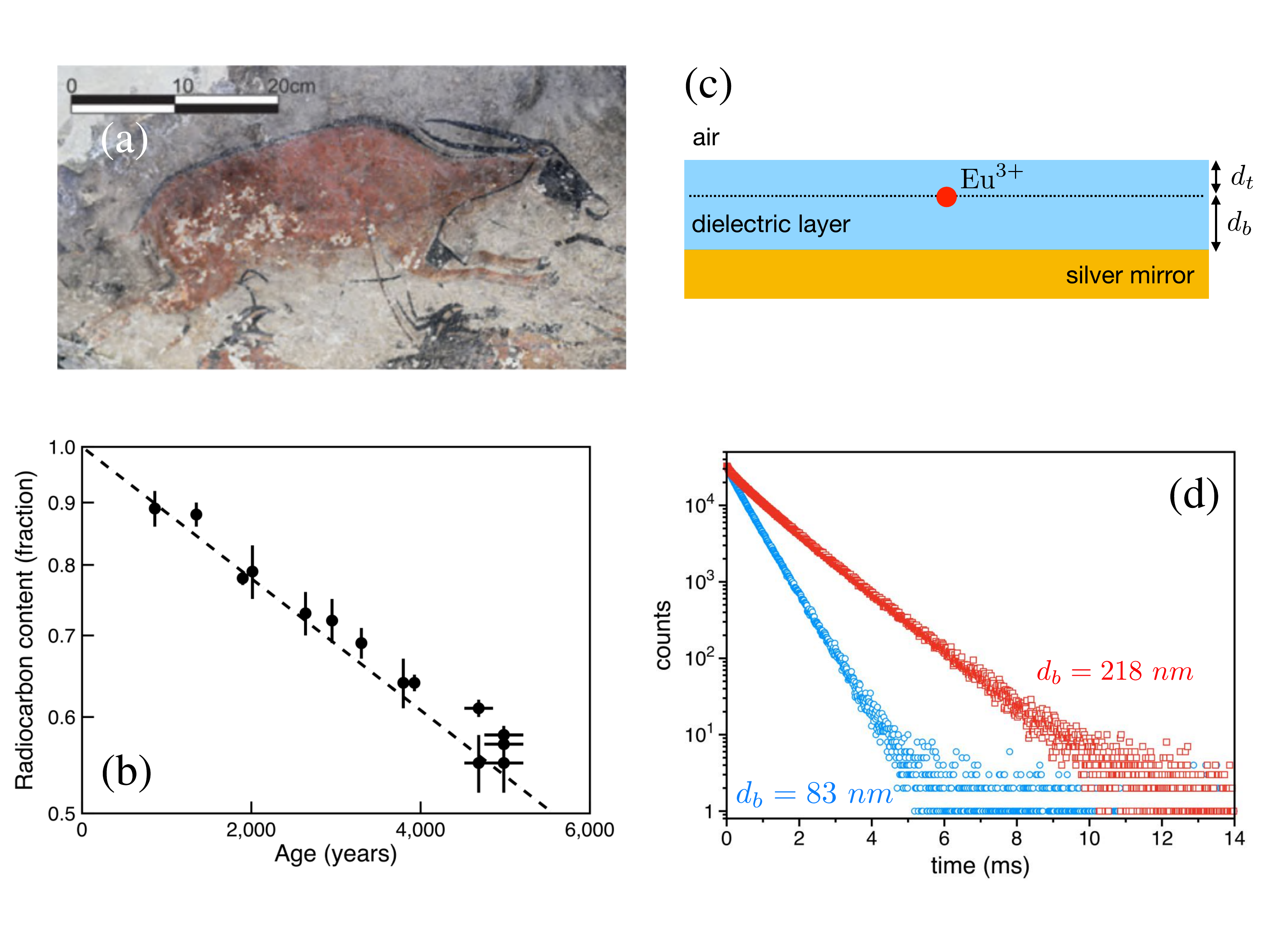}
\caption{
\textbf{Two kinds of spontaneous decay: the decay of the fraction of radioactive \textsuperscript{14}C atoms present in natural (wood) carbon samples (a,b), and the decay of the number of excited $\rm{Eu}^{3+}$ ions after pulsed excitation (c,d)}.
(a) South African rock painting of an age 5723--4420 years, estimated using radiocarbon dating~\cite{Bonneau_Antiquity_2017_91_322}.
(b) An early calibration curve for \textsuperscript{14}C dating adapted from the Nobel address by Libby~\cite{Libby_Nobel}.
The data are the measured fractional \textsuperscript{14}C content of various pieces of wood whose age is already known, with the expected \textsuperscript{14}C decay shown as the dashed line. 
The very good match between expectation and reality would be unlikely if the lifetime of the \textsuperscript{14}C nucleus depended on the local environment. 
(c) Schematic of a $\rm{Eu}^{3+}$ ion placed in close proximity to a planar silver mirror to control the LDOS. 
(d) Number of photons detected as a function of time after a UV excitation pulse for two samples based on the design in (c). 
The samples differ only in the distance of the $\rm{Eu}^{3+}$ ions to the silver mirror: $d_b = 83$ nm for the blue symbols, $d_b = 218$ nm for the red symbols. 
The two data sets show different exponential decays indicating that the excited-state lifetime of the $\rm{Eu}^{3+}$ ions depends on the local optical environment, in this case the distance to a silver mirror. 
The data appear as Fig~6.3 in~\cite{Andrew_PhD} (with kind help of Dr. Andrew). 
Note that for the radiocarbon dating data, (b) it is the number of radioactive particles decaying that is measured; in the fluorescence data (d) it is the number of emitted photons that is measured. }
\label{fig:pigments}
\end{figure}

\subsection{Physical phenomena where densities of optical states play a role}\label{sec:phenomena_with_DOS}
The local density of optical states governs the rate at which an atom in its excited state decays to the ground state, emitting a photon in the process.  Any decay process (such as \textit{e.g.}, radioactive decay) is governed by something like the LDOS, because the products of any decay need to go somewhere, and the LDOS quantifies how easy it is for them to get there.  To highlight the particular importance of the LDOS concept in optics, let us compare the dependence of the following two well--known decay processes on their environment, see Figure~\ref{fig:pigments}: (i) the beta decay of the unstable isotope ${}^{14}{\rm C}$ into the stable ${}^{12}{\rm C}$ isotope through the emission of an electron and an anti--neutrino; and (ii) an excited atom decaying to a lower energy level through the emission of a photon. 
\par
The former process is the basis of radiocarbon dating, an invaluable technique that underpins archaeology and climate studies, relying on the fact that samples with a different age have a different \textsuperscript{14}C content. 
This dating method relies on the highly reproducible half-life of \textsuperscript{14}C, where every \textsuperscript{14}C atom has a half-life of $~5730$ years as shown in Figure~\ref{fig:pigments}(b)~\cite{emery1972}. 
The decay of \textsuperscript{14}C involves nuclear beta decay which is in turn due to the weak force, with a down quark decaying into an up quark and in the process emitting a $W^{-}$ boson (something like a heavy photon). 
This boson then undergoes decay into an electron and an anti--neutrino, see \textit{e.g.} Ref.~\cite{greiner2009}. 
For such decay processes there is very little one can do to change the available electron, neutrino, or $W^{-}$ boson states, and thus no significant position dependence of this half-life, and no possibility to change the local density of states. 
However, in other substances beta decay can occur via the \emph{capture} of an orbital electron by a proton in the nucleus (sometimes called inverse beta decay), emitting a neutrino in the process. 
This kind of decay depends on the value of the orbital electron wavefunction at the position of the nucleus. 
In 1947 Segr\'e noticed~\cite{segre1947} that by placing the decaying atom in a different environment, \textit{e.g.}, in a compound~\cite{segre1949}, or within a C$_{60}$ cage~\cite{ohtsuki2004}, the orbital wavefunctions can be modified, which changes the decay rate. 
Although not commonly discussed in this way, this is an example of the LDOS appearing in nuclear physics. 
It is clear that in nuclear physics the LDOS is only of limited use, simply because it is so difficult to change the parameters that govern the emission rate.
\par
By contrast, one can very much change the availability of electromagnetic modes into which an atom may emit a photon.  For instance if the atom is close to a mirror, the emitted photon is reflected by the mirror, and (depending on the angle of emission, frequency, and precise distance from the mirror) either constructively or destructively interferes at the position of the atom.  
There is no emission into those waves that destructively interfere, and thus the excited state lifetime depends on the distance from the mirror.\footnote{The excited-state lifetime $\tau$ is equivalent to the half-life $t_{1/2}$ that is well-known in nuclear physics: $t_{1/2} = \tau~\mathrm{ln}(2)$.}
This position dependence of the coupling between the atom and the electromagnetic field is quantified by the number of available waves into which emission can occur, which is--roughly speaking--the LDOS.
\par
We will come back to the example of emission in front of a mirror in section~\ref{sec:leightmotif}, but for the moment, imagine what would happen if the half-life of \textsuperscript{14}C were to suddenly depend on the environment of each atom. 
Each atom would have a different half-life, there would be no calibration curve as in Figure~\ref{fig:pigments}(b), and archaeologists and forensic scientists would find themselves in chaos!
Fortunately this does not happen. 
Yet we should not take the dependence of a physical phenomenon on the LDOS as being an obstacle to reliable experiments; indeed quite the opposite is true, it provides a way for us to control the emission of light by structuring the environment, placing the emitters in front of a mirror, putting them in a small box, or in a photonic crystal. 
With the right design, these changes can be radical, we can either bring emission to a halt or accelerate it to produce an avalanche of photons. 
Wishing to sound grand, we might call these spectacular changes a `local engineering of the vacuum'. 
Such engineering gives us exquisite control over the way light is emitted and absorbed, is central to many phenomena in quantum optics, and is important in practical applications such as solid-state lighting~\cite{lozano2013}, collecting photons from single-photon sources~\cite{Barnes_EPJD_2002_18_197,Aharonovich_NatPhot_2016_10_631} and high-speed LEDs~\cite{Tsakmakidis_OE_2016_24_17916}.
\subsection{The single interface as a leitmotif}\label{sec:leightmotif}
\begin{figure}[ht!]
\centering
\includegraphics[width=1.0\columnwidth]{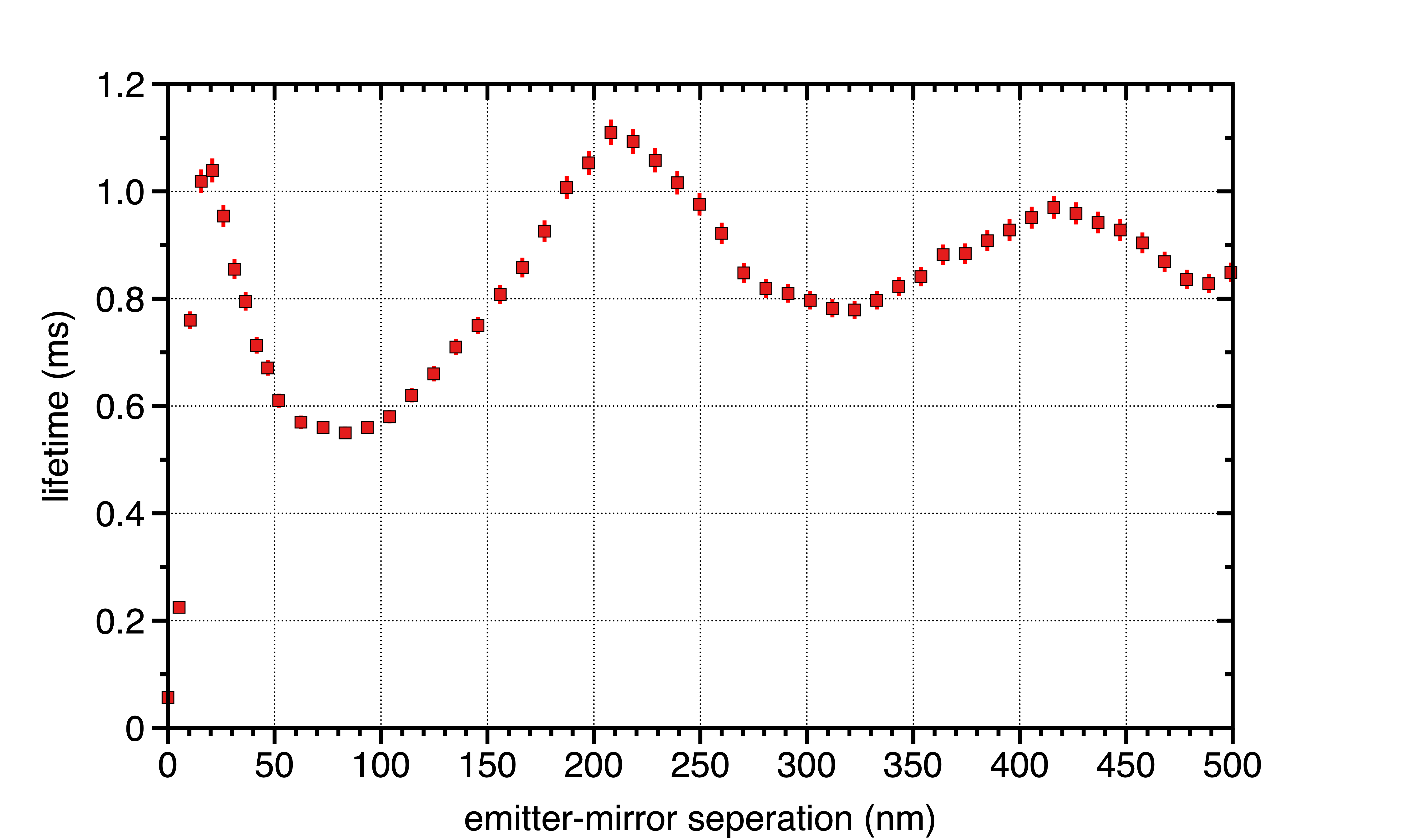}
\caption{
\textbf{Distance dependence of spontaneous emission lifetime in front of a mirror}.
The spontaneous emission lifetime of Eu$^{+3}$ ions following pulsed excitation is shown as a function of the distance between the ions and an adjacent planar silver mirror, for the geometry shown in figure~\ref{fig:pigments} (c). These data are reproduced from~\cite{Amos_PRB_1997_55_7249}.}
\label{fig:leitmotif_1}
\end{figure}

We now return to consider what happens to an emitter above a mirror; this simple reflecting interface provides a convenient way to develop a better understanding about the effect of the local density of states on spontaneous emission, indeed, so powerful is it that we will use it as a leitmotif that runs throughout this article.
In Figure~\ref{fig:pigments}(d) the time evolution of the emission of photons from excited Eu$^{+3}$ ions above a silver mirror were shown, for two different emitter-mirror separations.
The differences in the measured emission lifetimes were ascribed to different local densities of states. 
Data such as those shown in Figure~\ref{fig:pigments}(d) may be acquired as a function of emitter mirror separation. 
Such an experiment was first reported by Drexhage~\cite{Drexhage1968BBPC, Drexhage1970JLumi}, and similar results~\cite{Amos_PRB_1997_55_7249} are shown in Figure~\ref{fig:leitmotif_1}. 

\noindent By examining the key features exhibited by these data many aspects of the role played by the density of states in spontaneous emission may be explored.
The relationship between the spontaneous emission rate $\Gamma$, the spontaneous emission lifetime $\tau$ and the density of states $\rho(\omega)$ at the emission frequency $\omega$ is often written
\begin{equation}
	\Gamma = \frac{1}{\tau} = \rm{const}~\rho(\omega),
\label{eq:SPE__text_book}
\end{equation}
\noindent where, for the moment, we will not concern ourselves about the conditions under which this relationship holds.
Looking at the data in Figure~\ref{fig:leitmotif_1} there are several features we need to explain:

\begin{enumerate}
    \item The oscillation of the lifetime as the distance between the emitters and the surface is increased.
    \item The amplitude of the oscillation, and the way it decays as the distance increases.
    \item The asymptotic value to which the lifetime tends as the separation increases.
    \item The quenching of the lifetime for very small separations.
\end{enumerate}

\noindent 
We discuss these items at length throughout this article, here we want to provide some simple physical arguments to help build intuition, for a fuller discussion see section~\ref{sec:mirror_emission}.
Let us look at each of the features in turn.
\par
\begin{figure}[ht!]
\centering
\includegraphics[width=1.0\columnwidth]{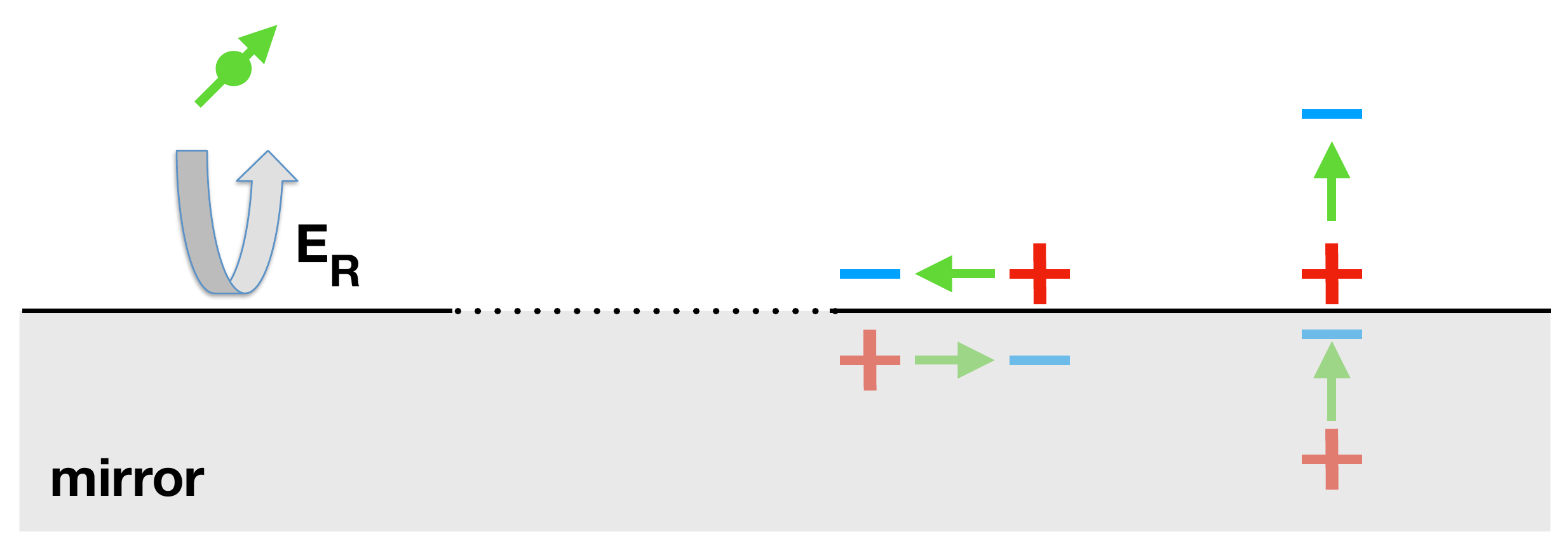}
\caption{
\textbf{Emission next to perfect mirror}.
(Left) A dipole emitter (green) is driven by the field that is reflected from the mirror, this leads to oscillations in the emission lifetime/rate as a function of emitter-mirror separation.
The strength of the reflected field falls with increasing emitter-mirror separation, owing to the point source nature of the emitter.
(Right) Close to the surface, changes to the orientation of the dipole leads to dramatic differences in the lifetime/rate.
For a dipole parallel to the mirror surface the image dipole acts to cancel the emission, for the perpendicular orientation the effective dipole moment is doubled.
}
\label{fig:interface}
\end{figure}

\noindent \textbf{The lifetime oscillates} because emission from the emitters (the Eu$^{+3}$ ions) is reflected by the surface, see Figure~\ref{fig:interface}.
If the reflected field returns in phase with the emission source then the rate will be increased (and the lifetime reduced), and \textit{vice versa}.
The oscillation period will thus be approximately half the emission wavelength in the host material for the ions, in this case approximately 200 nm, in good agreement with what is seen in Figure~\ref{fig:leitmotif_1}.
\par
\noindent \textbf{The amplitude of the oscillation} is dictated by the quantum (radiative) efficiency of the emitter. 
As an extreme case, for very low quantum efficiency the decay is dominated by non-radiative processes so that the influence of the local optical environment is reduced, see section~\ref{sec:quantum_efficiency}.
\textbf{The oscillation amplitude decays} because the emitter is a point source. The strength of the reflected field thus falls as the emitter-mirror separation increases, reducing the contrast of the interference and with it the amplitude of the oscillation.  For a dipole oriented parallel to the interface, the image dipole acts to cancel out the emission process, whilst for a dipole oriented normal to the interface the image dipole acts to double the strength of the emission process, see figure~\ref{fig:interface}. 
\par
\noindent \textbf{The asymptotic value} of the lifetime or rate for large emitter-mirror separations is determined by the effect of the host medium on the emitter and is independent of the mirror, since the emitter is far away.
As we will see in the next section, the density of states appropriate to describe this is the full rather than the local density of states.
\par
\noindent \textbf{Quenching:} 
For a real, e.g. metallic mirror, rather than a perfect one, the situation is more complex, and surface modes also act to quench the emission, see section~\ref{sec:non-perfect-mirror}.
Note that the dipole orientation also has a strong influence on the spatial locations of the maxima and minima in the lifetime oscillation~\cite{Vos2009PRA}, different dipole orientations produce different combinations of polarised electromagnetic fields at the surface that in turn act back on the dipole, see figure~\ref{fig:mirror_emission}.
This dipole-orientational dependence means that we need also to consider a partial local density of states.

\subsection{Number, local and partial local densities of states}\label{sec:different_densities}
With the foundations of the concept introduced we can now be more precise about what we mean by the LDOS.  In fact there is more than one quantity of interest; there are three levels at which we can quantify the number of allowed electromagnetic modes at a frequency $\omega$. 
We call these three increasingly detailed levels of description the \emph{density of optical states} (DOS); the \emph{local density of states} (LDOS); and \emph{the partial local density of states} (PLDOS); all three are number densities. 
Based on our discussion so far, it seems rather confusing that the PLDOS (and not the LDOS) is the most detailed level description.
Nevertheless it seems in practice that the distinction between the LDOS and PLDOS is rarely made, with the term LDOS often being used to denote both.  However, they do represent different things, and in this article we shall distinguish between them. 

(i) The \textit{number density of optical states} $\rho(\omega)$ (DOS) of the system is the number of allowed optical states per unit volume, per unit frequency. 
This density of states does not account for the spatial distribution of the modes, and would not be able to tell us, for example, how the distance from a mirror affects the radiation from an antenna. 
Typically, the DOS is invoked for infinite free space. 
An example of where this is the appropriate quantity governing the physics of emission is when calculating the blackbody spectrum~\cite{Loudon_TQTL}, or for calculating the free-space decay rate of emitters that are distributed within a certain volume.

(ii) The \textit{local density of states} (LDOS) $\rho_{l}(\boldsymbol{r},\omega)$, includes a contribution from each mode in proportion to its electric field intensity at the point of interest $\boldsymbol{r}$, and thus---to some degree---accounts for the local availability of modes into which emission may take place. 
The LDOS is the appropriate density when the electric field intensity varies in space and, for example, the emitter(s) sample all directions in space randomly and on a time scale that is fast compared with the emission lifetime. 
This is the case for example with the $Eu^{3+}$ ions mentioned above and employed by Drexhage~\cite{Drexhage_ProgOpt_1974_12_163}, see also Figures~\ref{fig:microcavity_figure}, and \ref{fig:Drexhage} below. 

(iii) The \textit{partial} (or \textit{projected}) \textit{local density of states} (PLDOS) is relevant when the emitter is sensitive to both the polarization and the intensity of the available modes. 
For example, a dipole aligned along $\boldsymbol{e}_{d}$ will interact most strongly with those modes where the electric field is polarized in the same direction, and least strongly with those modes with an electric field polarized perpendicularly. 
At this level of detail the PLDOS $\rho_{\rm p}(\boldsymbol{e}_{d},\boldsymbol{r},\omega)$ includes each mode in proportion to the strength of the electric field projected along the dipole direction $\boldsymbol{e}_{d}$, an example is shown in Figure~\ref{fig:mirror_emission} below. 

These three levels of description form a hierarchy of detail. 
If we average the PLDOS over all orientations of the emitter, we obtain the LDOS. 
And if we further average the LDOS over all positions, we obtain the DOS. 

\subsection{What the LDOS does and does not tell us about emission}
\par
It might be tempting to think that the PLDOS tells us everything about radiation from a small antenna (or quantum emitter), but this is not true. 
Let us first discuss what the PLDOS does describe before turning to what it does not describe.
Consider again our leitmotif where the emitter is placed in front of a mirror. 
Despite its simplicity, this system is actually the archetype of one where the coupling between the emitter and the field depends on position, it was also used in Drexhage's pioneering measurements of the spontaneous emission lifetimes of excited rare-earth ions~\cite{Drexhage_ProgOpt_1974_12_163}. 
To be concrete, the formula for the rate of spontaneous emission from a dipolar quantum emitter $\Gamma^{\rm Q}$ in terms of the PLDOS $\rho_{\rm p}$ is
\begin{equation}
	\Gamma^{\rm Q}=\frac{\pi\omega|\mathcal{P}|^{2}}{\hbar\epsilon_0\epsilon}\rho_{\rm p}(\boldsymbol{e}_{d},\boldsymbol{r}_0,\omega),\label{eq:SPE_rate_PLDOS}
\end{equation}
indicating that the rate of decay of a single emitter is determined by the PLDOS (which depends on the emitter position $\boldsymbol{r}_0$, the orientation of the dipole moment $\boldsymbol{e}_{d}$ and the emission frequency $\omega$), and is \emph{proportional} to it, with a proportionality constant that depends on the physical constants $\hbar$ and $\epsilon_0$ as well as the permittivity of the surrounding medium $\epsilon$, the frequency of emission $\omega$, and the emitter's transition dipole moment $\mathcal{P}$. 
The derivation of this formula is given below in Section~\ref{sec:quantum emitter}.

\begin{figure}[ht!]
\includegraphics[width=15.5cm]{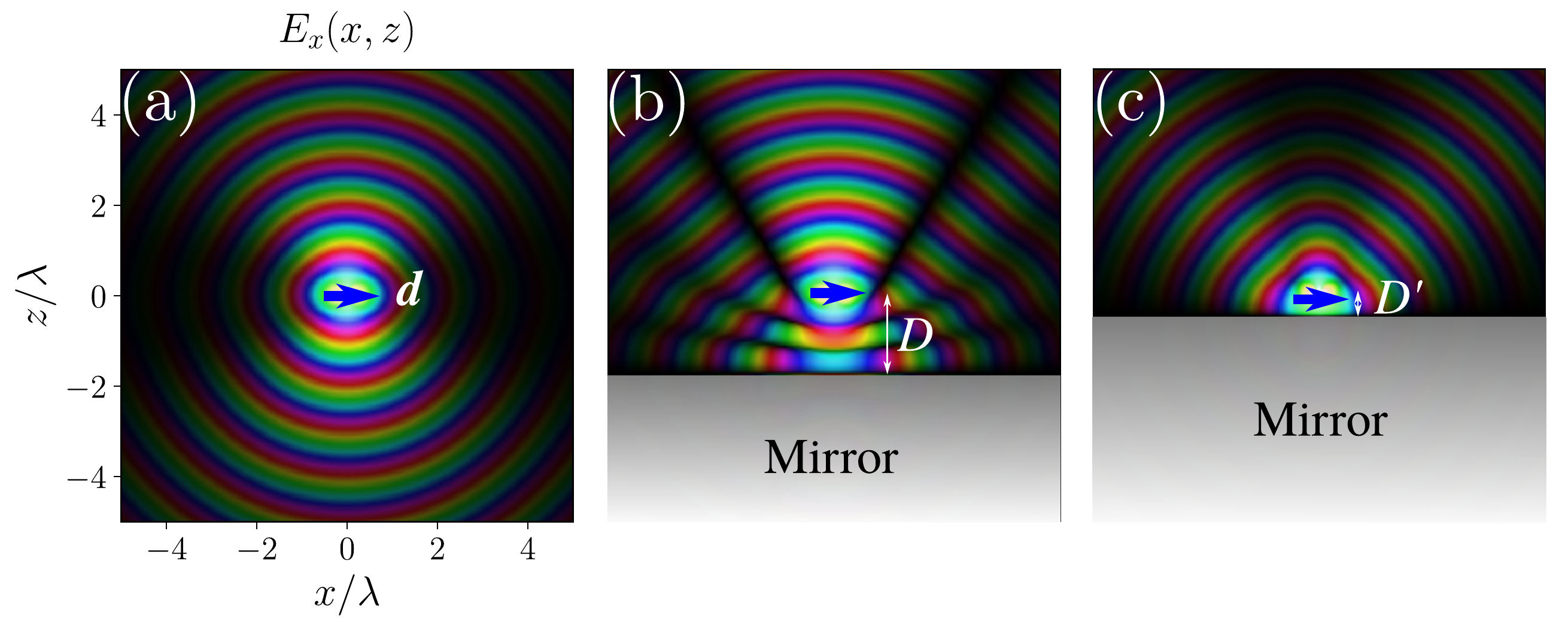}
\caption{{\textbf{Radiation from an antenna depends on its environment}}.
Panels (a--c) show the $x$--component of the radiation pattern from an antenna with an oscillating dipole moment $\boldsymbol{d}(t)$ radiating at fixed frequency $f=c/\lambda$, the dipole moment points in the $x$ direction.
Brightness represents amplitude and colour represents phase, advancing as red, green, blue. 
(a) Radiation pattern of an antenna in free space, and (b--c) close to a mirror, with the dipole-mirror separation $D=1.75\lambda$ and $D'=0.5\lambda$ respectively.
As an antenna of this orientation approaches the mirror, the net radiated power reduces to zero. 
Details of the calculation are given in Sec~\ref{sec:classical-dipole}. 
\label{fig:dipole_free-space_near-mirror}
}
\end{figure}

What the PLDOS does not tell us is into which modes the radiation goes.
Figure~\ref{fig:dipole_free-space_near-mirror} shows the result of calculating one component of the field produced by a classical dipole emitter, in three different scenarios. 
In Fig.~\ref{fig:dipole_free-space_near-mirror}(a) the dipole sits in free space, with the blue arrow indicating the instantaneous orientation of the dipole moment, \textit{i.e.}, the direction in which the current flows through the antenna.
Figures~\ref{fig:dipole_free-space_near-mirror}(b,c) show how this pattern changes when the dipole is placed near a reflecting surface at two different distances.
%
Changing the distance from the mirror results in two important changes to the emission.
Firstly the spatial pattern of the radiating field is altered, as is evident in Figure~\ref{fig:dipole_free-space_near-mirror}(b), which shows enhanced radiation normal to the surface (compared to free space), and almost no radiation normal to the surface in Figure~\ref{fig:dipole_free-space_near-mirror}(c).
However, this change in distribution should \emph{not} be taken as an indication that either the PLDOS, the LDOS or the DOS have changed, \textit{i.e.}, we should not--- on this basis alone---conclude that the rate of emission given by Eq.~(\ref{eq:SPE_rate_PLDOS}) has changed either. 
The two different distributions may have changed, but if they carry the same net power away from a dipole of fixed orientation then the numerical value of the PLDOS will be the same. 
The second change between the two right hand panels of Figure~\ref{fig:dipole_free-space_near-mirror} is in the overall brightness of the plot.
This brightness indicates the overall intensity of the emitted radiation, and thus the power leaving the dipole.
The overall intensity \emph{is} governed by the PLDOS, and concerning figure ~\ref{fig:dipole_free-space_near-mirror} we can say that the PLDOS at distance $D'$ from the mirror must be lower than at distance $D$.
Yet, having considered only a single dipole orientation, we cannot yet say anything about the LDOS (in Sec~\ref{sec:classical-dipole} we shall see that it has the same dependence).  It might be initially rather surprising that the density of states (DOS) is identical in all three panels of Figure~\ref{fig:dipole_free-space_near-mirror}: however, this is because exactly the same continuum of mode frequencies are available.

At the time of writing, the convergence of improved nanofabrication techniques and improved understanding of how nanostructures modify the vacuum has enabled much more extensive control over the spontaneous emission of photons by quantum emitters than can be achieved with a simple mirror. 

%
\subsection{The essential formulae}\label{sec:essentials}
So far the discussion has been rather qualitative.  In this section we shall provide the necessary formulae to calculate our trio of densities of states, as well as formulae for the classical and quantum emission processes they govern.

Suppose we have an optical system with a finite volume $V$ that supports a set of normalized electromagnetic modes with frequencies $\omega_n$ and electric fields $\boldsymbol{\mathcal{E}}_{n}$.
For example, these could be the solutions to Maxwell's equation inside a metal box~\cite{Hasan2018PRL}. 
Provided the system contains no dissipative elements, the electric field modes are orthogonal, and we choose to normalize them such that
\begin{equation}
\int_{V}\boldsymbol{\mathcal{E}}_{n}\cdot\boldsymbol{\mathcal{E}}_{m}^{\star}d^{3}\boldsymbol{r}=\delta_{n,m}.
\label{eq:orthonormality}
\end{equation}
i.e. we normalize them in a way that is convenient for counting the number of modes. By defining the modes in this way we can write the PLDOS as
\begin{equation}
{\rm PLDOS:}\hspace{1cm}\rho_{\rm p}(\boldsymbol{e}_{d},\boldsymbol{r},\omega) d\omega = \sum_{n}\delta(\omega-\omega_n)\left|\boldsymbol{e}_{d}\cdot\boldsymbol{\mathcal{E}}_{n}(\boldsymbol{r})\right|^{2} d\omega.
\label{eq:pldos_defn}
\end{equation}
The PLDOS governs the coupling of the electromagnetic field to a sub-wavelength scale current at position $\boldsymbol{r}$ and directed along $\boldsymbol{e}_{d}$.
We can account for this expression fairly easily.  The delta function simply records the frequencies of the available states, under an integral sign it contributes $1$ at the allowed frequencies, and $0$ elsewhere. 
These delta functions are weighted by the normalized electric field intensity in the direction of the dipole moment $|\boldsymbol{e}_{d}\cdot\boldsymbol{\mathcal{E}}_{n}(\boldsymbol{r})|^{2}$, which records how strongly an emitter at position $\boldsymbol{r}$ and with orientation $\boldsymbol{e}_{d}$ interacts with the mode of frequency $\omega_{n}$.
This is the mathematical expression of our verbal definition of the PLDOS in section~\ref{sec:different_densities}.

The other two densities of states discussed in section~\ref{sec:different_densities} can be obtained by appropriate averaging of the PLDOS. 
For instance, the LDOS is three times the orientational average of the PLDOS (relevant when the orientation of the emitter's dipole moment samples different directions in space on a time scale that is faster than the emission lifetime),
\begin{equation}
{\rm LDOS:}\hspace{1cm}\rho_{l}(\boldsymbol{r},\omega) = \sum_{\boldsymbol{e}_{d}}\rho_{\rm p}(\boldsymbol{e}_{d},\boldsymbol{r},\omega)=\sum_{n}\delta(\omega-\omega_n)\left|\boldsymbol{\mathcal{E}}_{n}(\boldsymbol{r})\right|^{2}\label{eq:ldos_defn}
\end{equation}
and thus weights each delta function with the electric field \emph{intensity} of each mode.  Note that by convention the pre--factor of $1/3$ that one might expect from an orientational average is omitted. 
Historically, early papers that pointed out the position dependence of the LDOS include Sprik \textit{et al.}~\cite{Sprik_EPL_1996_35_265}, and Snoeks \textit{et al.}~\cite{Snoeks_PRL_1995_74_2459}, whilst Chew pointed out the position-dependent emission rate without invoking the LDOS~\cite{Chew_PRA_1988_38_3410}. 

Finally the DOS is simply the volume average of the LDOS,
\begin{equation}
{\rm DOS:}\hspace{1cm}\rho(\omega) = \frac{1}{V}\int_{V} d^{3}\boldsymbol{r}\rho_{l}(\boldsymbol{r},\omega)=\frac{1}{V}\sum_{n}\delta(\omega-\omega_n),\label{eq:dos_defn}
\end{equation}
which, when integrated over a given frequency range, gives us the total number of available electromagnetic modes per unit volume in that frequency range. 

The factor of $V$ appearing in Eq.~(\ref{eq:dos_defn}) is the volume over which the electromagnetic modes are normalized, as shown in Eq.~(\ref{eq:orthonormality}). 
We note that in practice determining what this volume should be can require some thought.  In an idealized cavity with impenetrable walls, it is simply the volume of the cavity.
For an infinite medium, we should use an infinite volume in Eq.~(\ref{eq:dos_defn}), which ensures that the density of states is well defined when the spectrum $\omega_n$ is continuous rather than discrete. 
This latter situation occurs, for example, in the experimentally relevant case of a finite material embedded in an infinite environment~\cite{Hasan2018PRL}.  Note that there are some important subtleties in dealing with mode volumes in the presence of loss - see Ref~\cite{Yan_PRB_2018_97_205422,Chen_PRApp_2019_11_044018}.

While the above definitions are quite simple and intuitive, they have the drawback that they cannot be applied when dissipation is significant (for example if the mirror in figure~\ref{fig:dipole_free-space_near-mirror} absorbs some of the incident light).  This is because in such systems there are no real valued eigenfrequencies (due to dissipation, the mode amplitudes must decrease over time, corresponding to a complex value of $\omega_n$), and thus no obvious meaning for the delta functions in equations (\ref{eq:pldos_defn}--\ref{eq:dos_defn}).  In these cases it is easier to use an alternative and more powerful definition for our densities of states in Eqs.~~(\ref{eq:pldos_defn}--\ref{eq:dos_defn}), 
\begingroup
\addtolength{\jot}{10pt}
\begin{align}
{\rm PLDOS}:&\;\;
\rho_{\rm p}(\boldsymbol{e}_{d},\boldsymbol{r},\omega) \equiv \frac{2\omega{\rm n}^{2}}{\pi c^{2}}\boldsymbol{e}_{d}\cdot{\rm Im} \left[\overleftrightarrow{\boldsymbol{G}}(\boldsymbol{r},\boldsymbol{r},\omega)\right]\cdot\boldsymbol{e}_{d},
\label{eq:defn_pldos_green}\\
{\rm LDOS}:&\;\;
\rho_{l}(\boldsymbol{r},\omega) \equiv 
\frac{2\omega{\rm n}^{2}}{\pi c^{2}}{\rm Tr}\,{\rm Im} \left[\overleftrightarrow{\boldsymbol{G}}(\boldsymbol{r},\boldsymbol{r},\omega)\right],
\label{eq:defn_ldos_green}\\
{\rm DOS}:&\;\;
\rho(\omega) \equiv \frac{2\omega{\rm n}^{2}}{\pi c^{2}}\int_{V} d^{3}\boldsymbol{r}{\rm Tr}\,{\rm Im}\left[\overleftrightarrow{\boldsymbol{G}}(\boldsymbol{r},\boldsymbol{r},\omega)\right],
\label{eq:defn_dos_green}
\end{align}
\endgroup
where ${\rm n} = \sqrt{\epsilon\mu}$ is the refractive index of the background material, \textit{e.g.}, the material filling a cavity. 
The electromagnetic Green function $\overleftrightarrow{\boldsymbol{G}}$ is a dyadic quantity (see, \textit{e.g.}, Refs.~\cite{Economou2006book,novotny2006book}), which is a mathematical object with two indices, \textit{i.e.}, $\overleftrightarrow{\boldsymbol{G}}\equiv G_{i j}$. 
The physical interpretation of the Green function $\overleftrightarrow{\boldsymbol{G}}$ is that it gives the electric field at position $\boldsymbol{r}$ due to a point dipole (described by a delta function) at position $\boldsymbol{r}_0$, where the first column of $\overleftrightarrow{\boldsymbol{G}}$ is the electric field vector for emission from an $x$-oriented dipole, the second column for $y$ oriented dipole, and the third column for $z$ oriented one.  The Green function obeys the same vector Helmholtz equation that the electric field obeys in a medium of index $\textrm n$ (see section~\ref{sec:classical-dipole} for a derivation of this formula from Maxwell's equations), 
\begin{equation}
\boldsymbol{\nabla}\times\boldsymbol{\nabla}\times\overleftrightarrow{\boldsymbol{G}}(\boldsymbol{r},\boldsymbol{r}_0,\omega)-{\rm n}^{2}k_0^{2}\overleftrightarrow{\boldsymbol{G}}(\boldsymbol{r},\boldsymbol{r}_0,\omega) = \boldsymbol{1}_{3}\delta^{(3)}(\boldsymbol{r}-\boldsymbol{r}_0).\label{eq:green_function_definition}
\end{equation}
but with the identity matrix and a delta function on the right hand side, rather than zero. 
The utility of the Green function is that from it we can calculate the electric field due to \emph{any} current distribution $\boldsymbol{j}(\boldsymbol{r})$.  In a non--magnetic material ($\mu=1$) the formula for the electric field is
\begin{equation}
    \boldsymbol{E}(\boldsymbol{r})={\rm i}\mu_0\omega\int\overleftrightarrow{\boldsymbol{G}}(\boldsymbol{r},\boldsymbol{r}',\omega)\cdot\boldsymbol{j}(\boldsymbol{r}')\,d^{2}\boldsymbol{r}'\label{eq:field_from_G}
\end{equation}
where the prefactor of ${\rm i}\omega$ (\textit{i.e.}, a $90^{\circ}$ phase shift) is present because radiation arises from the time derivative of the current, and $\mu_0$ is the constant determining the strength of interaction between the current and the field, we derive this expression in section~\ref{sec:pldos_Gf}.
\begin{figure}[ht!]
\centering
\includegraphics[width=0.5\columnwidth]{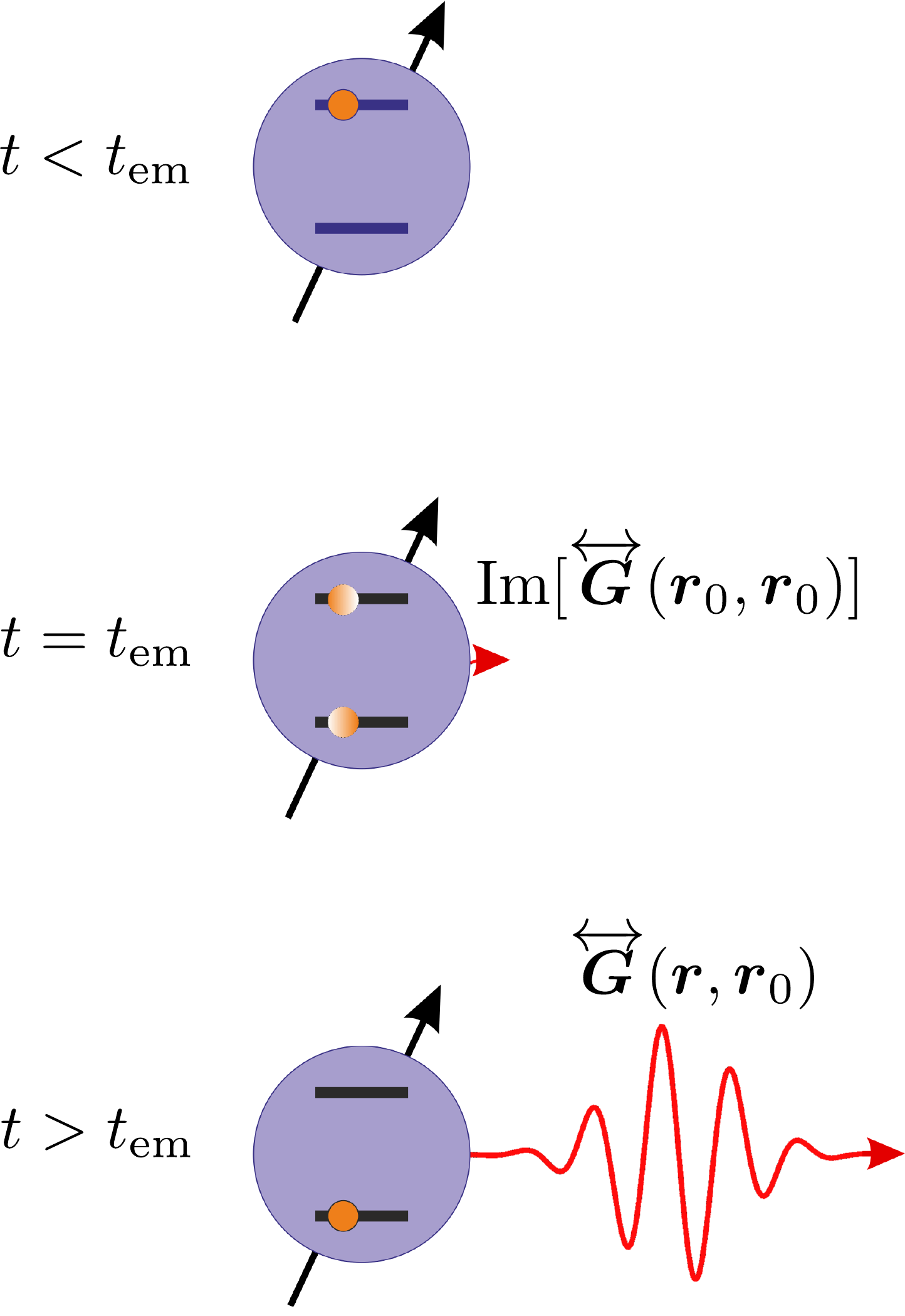}
\caption{\label{fig:cartoon-emission}
Cartoon of a dipolar two-level quantum emitter at different moments in time while spontaneously decaying through the emission of a photon.
Top: just before emission $(t < t_{em})$ the emitter is in the excited state as shown by the orange electron in the upper energy state. 
Middle: at the moment of emission $(t = t_{em})$, the emitter at location $\boldsymbol{r}_0$ goes from the upper to the lower state and the proverbial "head" of the photon appears at the emitter location $\boldsymbol{r}_0$; the cartoon shows that the creation of light is described by the imaginary part of the Green function, with $\boldsymbol{r}_0$ appearing twice: ${\rm Im}[\overleftrightarrow{\boldsymbol{G}}(\boldsymbol{r}_0,\boldsymbol{r}_0)]$. 
Bottom: after emission $(t > t_{em})$, the emitter's electron is in the lower energy state and the propagation of the emitted electric field to a different point $\boldsymbol{r}$ is described by a Green function where the different emitter position and the target position as arguments of the Green function: $\overleftrightarrow{\boldsymbol{G}}(\boldsymbol{r},\boldsymbol{r}_0)$. 
}
\end{figure}

The densities of states in Eqs.~(\ref{eq:defn_pldos_green}--\ref{eq:defn_dos_green}) do \textbf{\textit{not}} depend on the complete Green function, with $\boldsymbol{r}_{0}$ appearing in both spatial arguments. 
To try to give an intuitive feeling for why only the imaginary part appears, let us consider the cartoon sequence of spontaneous emission in Figure~\ref{fig:cartoon-emission}.
At some time before the emission process begins $(t < t_{em})$ the emitter is prepared in the excited state. 
At the moment of emission $(t = t_{em})$, the emitter at location $\boldsymbol{r}_0$ goes from the upper to the lower state, and we might say that the proverbial "head" of the photon appears at the emitter location $\boldsymbol{r}_0$. 
If a movie of the appearance of a new photon is played backwards (time-reversed) then it looks like the absorption of a photon. 
In the case of absorption of electromagnetic energy into, for example, a dielectric material the absorbed energy is proportional to the imaginary part of the electric susceptibility ${\rm Im}[\chi(\omega)]$. 
Similarly, the energy lost from an oscillating current into the electromagnetic field is proportional to ${\rm Im}[\overleftrightarrow{\boldsymbol{G}}(\boldsymbol{r}_0,\boldsymbol{r}_0,\omega)]$. 
This is indeed what we have in Eqs~ (\ref{eq:defn_pldos_green}--\ref{eq:defn_dos_green}).
Therefore the cartoon shows that the creation of a photon is described by the imaginary part of the Green function. 
The reason for the two spatial arguments of the Green function being equal here $(\boldsymbol{r} = \boldsymbol{r}_{0})$ is that the start of the process (no photon) and the end of the process (1 photon just appears) both occur at the point of emission $\boldsymbol{r}_{0}$. 
Thus, the emission process is described by ${\rm Im}[\overleftrightarrow{\boldsymbol{G}}(\boldsymbol{r}_0,\boldsymbol{r}_0)]$. 
After its appearance $(t > t_{em})$, the photon propagates somewhere else (\textit{e.g.}, to an observer's detector) and this propagation is described by the Green function with different initial and final positions $(\boldsymbol{r} \neq \boldsymbol{r}_{0})$: $\overleftrightarrow{\boldsymbol{G}}(\boldsymbol{r},\boldsymbol{r}_0)$. 
Although there are only a few systems where the Green function - and hence the PLDOS - can be calculated exactly, namely a flat mirror, a slab, and a sphere, there are numerical tools that serve to calculate this quantity in other situations. 
One such freely available tool is the FDTD code MEEP from the MIT group~\cite{oskooi2010}, see also the table below. 

As the field coming from any source of radiation can be written in terms of the Green function (\ref{eq:field_from_G}), ${\rm Im}[\overleftrightarrow{\boldsymbol{G}}]$ turns up in a variety of physical phenomena. 
As a result the LDOS implicitly controls many more physical processes than just the emission from a point source. 
An interesting example is provided by the scattering of electromagnetic waves. 
The amplitude of scattering from an object can be specified by a complex angle--dependent scattering amplitude $f(\theta,\phi)$. 
The so--called optical theorem states that the imaginary part of the forward scattering amplitude is proportional to the total cross section of the scatter $\sigma_{s}$: in simple terms it states that what is removed from the field in front of the object is proportional to ${\rm Im}[f]$ evaluated in front of the object. 
What is removed is equal to the sum of all the scattered and absorbed radiation. 
See Ref.~\cite{newton1976} for an interesting discussion of the history and derivation of this theorem. 

We can already imagine that the LDOS plays some role in scattering, because it determines the strength of interaction between the scattering object and the field. 
To sketch the importance of the LDOS to the optical theorem, consider the scattering of an EM wave from a small object at $\boldsymbol{r}_{0}$, which does not absorb EM energy. 
The incident field polarizes the object, causing radiation to be emitted. 
Some of this radiation acts back on the object and changes this polarization, with this backaction described through the Green function $\overleftrightarrow{\boldsymbol{G}}(\boldsymbol{r}_{0},\boldsymbol{r}_{0})$. 
Despite being lossless, the total polarizability $\alpha$ of the object must therefore be a complex quantity that depends on the Green function for equal spatial arguments, $\overleftrightarrow{\boldsymbol{G}}(\boldsymbol{r}_{0},\boldsymbol{r}_{0})$\footnote{Note that the real part of $\overleftrightarrow{\boldsymbol{G}}(\boldsymbol{r}_{0},\boldsymbol{r}_{0})$ is infinite. 
This reflects the difficulty of assuming a point--like object. 
For a small but finite size object the real part of the Green function must be averaged over the object, which produces a finite result.}. 
To prevent misunderstanding we note that this complex polarizability is the object's response to the \emph{incident} field and \emph{not} the total field.
The imaginary part of $\alpha$ is thus determined by ${\rm Im}[\overleftrightarrow{\boldsymbol{G}}(\boldsymbol{r}_{0},\boldsymbol{r}_{0})]$, \textit{i.e.}, the LDOS. 
The forward scattering amplitude $f$ equals $\alpha$ times a real constant, and thus ${\rm Im}[f]$ (and hence the scattering cross section) is also determined by the LDOS. 
This reveals the beautiful self--consistency of Maxwell's equations: however one chooses to calculate the energy radiated by the object, one always reaches the conclusion that it is governed by the LDOS. 

When a system does not exhibit any significant dissipation of electromagnetic energy - a perfectly conducting cavity, for instance - our two sets of definitions (\ref{eq:pldos_defn}--\ref{eq:dos_defn}) and (\ref{eq:defn_pldos_green}--\ref{eq:defn_dos_green}) are entirely equivalent, as we will show in section~\ref{sec:pldos_Gf}. 
This is because in a lossless system the Green function can be expanded in terms of the system's orthonormal eigenmodes (for the details of such expansions see~\cite{barton1989}), such an expansion makes (\ref{eq:defn_pldos_green}--\ref{eq:defn_dos_green}) identically equal to (\ref{eq:pldos_defn}--\ref{eq:dos_defn}). 
However, when there is significant dissipation (for example, a cavity with absorbing walls) the eigenmodes are no longer orthonormal, preventing any straightforward modal expansion of the PLDOS. 
In this case we should restrict ourselves to formulae (\ref{eq:defn_pldos_green}--\ref{eq:defn_dos_green}), and we lose our simple understanding of the local density of states as the number of locally available modes. 
We shall treat systems with loss, and explain some of the potential pitfalls of including dissipation in section~\ref{sec:non-perfect} where the densities of states (\ref{eq:defn_pldos_green}--\ref{eq:defn_dos_green}) can even become infinite!

Let us now turn to the simplest physical scenarios where the densities of states (\ref{eq:pldos_defn}--\ref{eq:defn_dos_green}) play a role.  Edward Purcell was the first to suggest that the emission rate of an emitter could be modified from its free space value (\ref{eq:SPE_rate_free}) by placing it inside a resonant cavity~\cite{Purcell_PhysRev_1946_69_681}, and although he did not say so directly, the cavity's role is to modify the local density of states. 
Controlling spontaneous emission in this way has developed into a field known as cavity quantum electrodynamics, an overview of which can be found in an extensive book chapter by Haroche~\cite{Haroche_1992_CQED}. 
Nice recent examples include the use of a 3D photonic crystal to inhibit the spontaneous emission from quantum dots embedded in the crystal~\cite{Leistikow_PRL_2011_107_193903}, an effect predicted in 1987~\cite{Yablonovitch_PRL_1987_58_2059}, and control over the exciton radiative ilfetime in a Van der Waals heterostructure~\cite{Fang_PRL_2019_123_067401}. 
More recently the influence of the local density of states has been explored in other optical processes, for example energy transfer (ET) between molecules, when such transfer takes place via a dipole-dipole interaction~\cite{Blum_2012_PRL_109_203601}.  The role of the LDOS (and the PLDOS) in this and other processes continues to be an area of vigorous research~\cite{Wubs_NJP_2016_18_053037}.  It is worth mentioning that the concept of the local density of states is not specific to the emission of electromagnetic waves, it has also been applied in acoustics~\cite{Langguth_PRL_2016_116_224301} and metamaterial research~\cite{Hoi_NatPhys_2015_11_1045}. 

Consider next the average power $\langle P^{\rm C}\rangle$ leaving a classical dipole antenna (again, for more detail see section~\ref{sec:classical-dipole}).  The instantaneous rate of energy leaving the antenna, $P^{\rm C}$, is given by the negative rate of work (force $\times$ velocity) done by the electric field on the charge carriers in the antenna,
\begin{equation}
	P^{\rm C}=-\int d^{3}\boldsymbol{r}\,\boldsymbol{v}\cdot\sigma \boldsymbol{E}=-\int d^{3}\boldsymbol{r}\,\boldsymbol{j}\cdot\boldsymbol{E},\label{eq:emitted_power}
\end{equation}
\textit{i.e.}, minus the energy entering the antenna from the field.  In the above equation  $\boldsymbol{v}$ is the local velocity of the charge density $\sigma$, and $\boldsymbol{j}=\sigma\boldsymbol{v}$ is the electrical current.  As we shall show in Sec~\ref{sec:classical-dipole}, the time average of this quantity for a point dipole at $\boldsymbol{r}_{0}$, oriented along $\boldsymbol{e}_{d}$ and oscillating at frequency $\omega$ is given by,
\begin{align}
	\langle P^{\rm C}\rangle=\frac{1}{T}\int_0^{T} P^{\rm C}\,dt&=\frac{\pi\omega^{2}|\tilde{d}|^{2}}{4\epsilon_0\epsilon}\rho_{\rm p}(\boldsymbol{e}_{d},\boldsymbol{r}_0,\omega),\label{eq:intro_classical_power}
\end{align}
where the integration is over many optical cycles ($T \gg 2\pi/\omega^{-1}$), and $\tilde{d}$ is the complex amplitude of the dipole moment of the antenna.

\begin{figure}[ht!]
	\includegraphics[width=15.5cm]{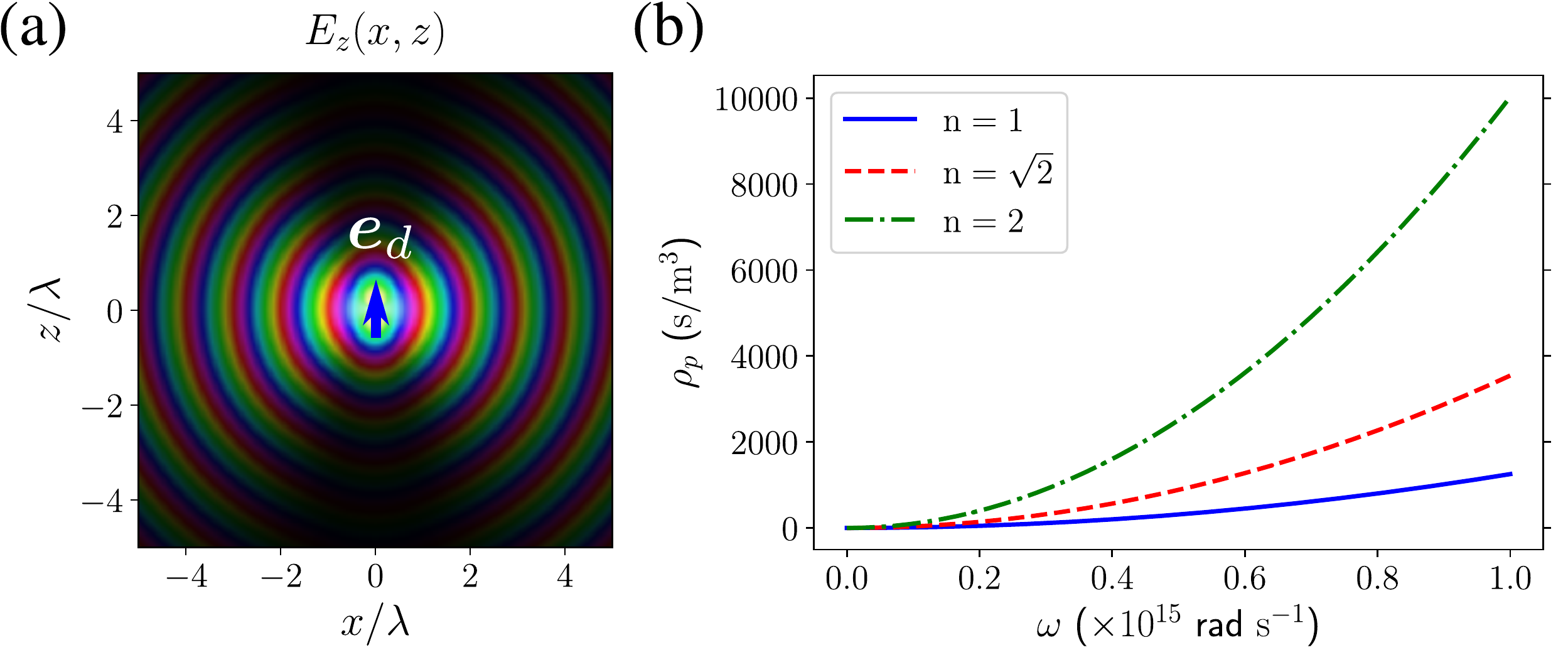}
    \caption{In a homogeneous environment the partial local density of states $\rho_{p}$ (\ref{eq:pldos_defn}), the local density of states $\rho_l$ (\ref{eq:ldos_defn}), and the number density $\rho$ (\ref{eq:dos_defn}) all contain the same information.  (a) The electric field component $E_z$ for an antenna oriented along the $z$--axis, in free space (colours and brightness defined as in figure~\ref{fig:dipole_free-space_near-mirror}). (b) the partial local density of states $\rho_p$ in a homogeneous environment (\ref{eq:free_space_continuum_ldos}) as a function of frequency, for three different values of the refractive index.  Note that although the partial local density of states scales as the refractive index cubed, the emitted power (\ref{eq:free_space_radiation}) scales with the relative impedance times the index squared.\label{fig:classical_free_space}}
\end{figure}

Just as for the rate of emission from an atom (\ref{eq:SPE_rate_PLDOS}), the power radiated by a classical antenna (\ref{eq:intro_classical_power}) is proportional to the PLDOS, with the proportionality constant given by the square of the magnitude of the current $\omega^{2}|\tilde{d}|^{2}$, times the constant $\pi/4\epsilon_0\epsilon$.
Figures~\ref{fig:classical_free_space},~\ref{fig:mirror_emission} and~\ref{fig:cavity_emission} in the sections below show calculations of $\rho_{\rm p}$ for three simple cases, illustrating that by just adding mirrors one can enhance or suppress the emission from the antenna, relative to emission into free space.  To understand how this works, suppose we want to stop an antenna emitting at frequency $\omega$.  From the definition of the PLDOS (\ref{eq:pldos_defn}) we see that all the modes $\boldsymbol{\mathcal{E}}_{n}$ of the system where $\omega=\omega_n$ must have zero field component along the antenna axis.  We imagine doing this for example, by putting our emitter between closely spaced parallel mirrors.  Provided the mirrors are close enough (relative to the wavelength of emission), there is only one available mode, where the electric field is normal to the surface of the mirrors, like that of a capacitor.  Thus a dipole antenna oriented \emph{in the plane} of these closely spaced mirrors cannot emit.  Similarly, to enhance the emission we need to enhance the electric field at the position of the emitter.  This is realized, for example, by placing an object close to the antenna that supports a localized electromagnetic mode, where the normalized field amplitude $\boldsymbol{\mathcal{E}}_{n}$ will be large.  The fluorescence from molecules can be increased in this manner through, for example, placing them close to metallic nanoparticles that support plasmonic modes~\cite{tam2007}.  The rate of energy emitted from a small quantum system is ruled by almost exactly the same physics as the macroscopic (classical) antenna just discussed.  The main difference is that the state of the atom is governed by quantum mechanics and is thus represented by a wave--function $|\psi\rangle$.  As we shall demonstrate (using first-order perturbation theory) in Sec~\ref{sec:quantum emitter}, while the atom makes a transition  from excited state $|1\rangle_{\rm at}$ to the lower energy one $|0\rangle_{\rm at}$, the average emitted power $\langle P^{\rm Q}\rangle$ is given by an expression that is almost identical to (\ref{eq:intro_classical_power})
\begin{equation}
	\label{eq:intro_quantum_power}
	\langle P^{\rm Q}\rangle = \frac {\pi\omega^{2}|\mathcal{P}|^2}{\epsilon_0\epsilon}\rho_{\rm p}(\boldsymbol{e}_{d},\boldsymbol{r}_0,\omega),
\end{equation}
where,
\begin{equation}
	\mathcal{P}=\langle 0|_{\rm at}\hat{d}|1\rangle_{\rm at},
    \label{eq:trans_dip_matr_elem}
\end{equation}
is the dipole matrix element associated with transitions between the atomic excited state $|1\rangle_{\rm at}$, and ground state $|0\rangle_{\rm at}$.
The difference of a factor of 4 will be discussed in section~\ref{sec:the_difference}
We consider the transition to be associated with a dipole moment along $\boldsymbol{e}_{d}$, and $\hat{d}$ is the operator that gauges the dipole amplitude.  Experiments typically measure the rate of emitted photon clicks, and thus one is more interested in the number of photons per second $\Gamma^{\rm Q}=\langle P^{\rm Q}\rangle/\hbar\omega$.  Dividing Eq.~(\ref{eq:intro_quantum_power}) by $\hbar\omega$, one obtains the rate of spontaneous emission given earlier, see Eq.~(\ref{eq:SPE_rate_PLDOS}).
%
%
We note that for the case where the orientation of the emitter is effectively isotropic, \textit{e.g.}, the orientation  of the dipole moment tumbles freely in space on a time scale faster than the emission rate, the spontaneous emission rate becomes the orientational average of (\ref{eq:SPE_rate_PLDOS}), and thus depends on the LDOS,
\begin{equation}
	\langle\Gamma^{\rm Q}\rangle = \frac{1}{3}\sum_{\boldsymbol{e}_{q}}\Gamma^{\rm Q}=\frac {\pi\omega|\mathcal{P}|^2}{3\epsilon_0\epsilon\hbar}\rho_{l}(\boldsymbol{r}_0,\omega).\label{eq:SPE_rate_LDOS}
\end{equation}
If in addition, our emitters are randomly distributed in position and orientation then the overall emission rate is an average over both position and orientation,
\begin{equation}
	\langle\Gamma^{\rm Q}\rangle = \frac{1}{3}\sum_{\boldsymbol{e}_{q}}\int_{V} d^{3}\boldsymbol{r}_{0}\Gamma^{\rm Q}=\frac {\pi\omega|\mathcal{P}|^2}{3\epsilon_0\epsilon\hbar}\rho(\omega),\label{eq:SPE_rate_DOS}
\end{equation}
\textit{i.e.}, the rate depends on the DOS.  If the emitter position is not completely unknown then the above spatial integral is over only a portion of the space occupied by the field modes, and thus not equal to the DOS. 

Free space is a special case, where all three densities of states are proportional to each other,
\begin{equation}
\rho = \rho_{l} = 3\rho_{\rm p},
\label{eq:densities}
\end{equation}
with the factor of 3 arising because of the isotropy of free space.  In this special case, the partial density of states appearing can be replaced by $1/3$ of the free space number density of states, $\rho(\omega)=\rho_{0}(\omega) = \left({\omega^2}/{\pi^2 c^3}\right) $ (see section~\ref{sec:DOS}).  This replacement in Eqs.~(\ref{eq:intro_classical_power}) and (\ref{eq:intro_quantum_power} gives the following expressions for the emitted power,
\begin{equation}
	\langle P^{\rm Q}\rangle=\frac{\omega^{4}|\mathcal{P}|^{2}}{3\pi\epsilon_0 c^{3}},\qquad \langle P^{\rm C}\rangle=\frac{\omega^{4}|\tilde{d}|^{2}}{12\pi\epsilon_0 c^{3}},
\end{equation}
and the free-space decay rate equals,
\begin{equation} \label{eq:SPE_rate_free}
\Gamma_0 =\frac{\langle P^{Q}\rangle}{\hbar\omega}=\frac {\omega^{3}|\mathcal{P}|^2}{3\pi\epsilon_0 c^3\hbar},
\end{equation}
where the detailed derivation is provided in Sec~\ref{sec:quantum emitter}.

\subsection{Orders of magnitude}\label{sec:pract-examples}
\begin{figure}[ht!]
\begin{center}
\includegraphics[width=11.5cm]{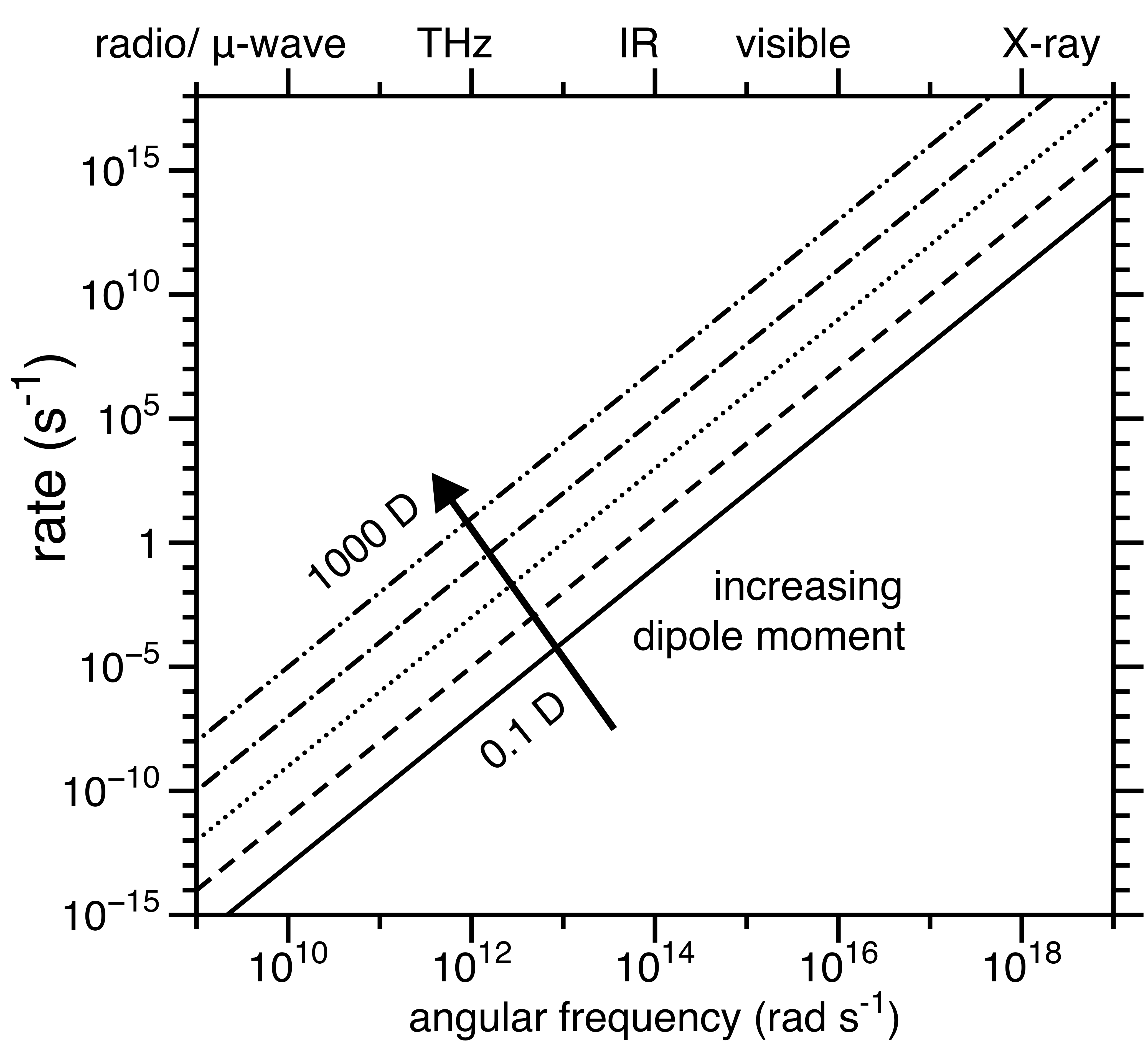}
\end{center}
\caption{The calculated spontaneous emission rate in free space as a function of emission frequency. 
Data were calculated using Equation~(\ref{eq:SPE_rate_free}) for increasing transition dipole moments $\mathcal{P} = 0.1$~D, 1~D, 10~D, 100~D and 1000~D.}
\label{fig:rates_graph}
\end{figure}

It is instructive to inspect the magnitude of the emission rate from equation~\eqref{eq:SPE_rate_free}. 
In Fig.~\ref{fig:rates_graph}, Eq.~\eqref{eq:SPE_rate_free} has been used to calculate the decay rate in free space as a function of the angular frequency of the radiation. 
Five data sets are shown, each line corresponding to a different value of the dipole moment. 
Spontaneous emission associated with Rydberg states, typically at microwave frequencies, involves large dipole moments, of order $1000$~D\footnote{Dipole moments are usually given in Debye (D), a unit defined as $1.10^{−18}$ statcoulomb.centimetre, and equivalent to $0.393 e.a_0$, with $e$ the electronic charge and $a_0$ the Bohr radius $a_0 = 5.3\times 10^{-11}\rm m$.}, owing to the large radii of Rydberg atoms; their associated spontaneous emission rate corresponds to a photon being emitted once every day. 
Dye molecules such as the laser dye R6G emit in the visible and have dipole moments of order 10D~\cite{Baranov_ACSPhot_2017}; for such molecules the emission rate corresponds to a photon being emitted every 10 nanoseconds or so.

%
%
\subsection{Different viewpoints on emission: from microscopic to macroscopic sources}\label{sec:3ways}

\begin{figure}[ht!]
\centering
\includegraphics[width=0.9\columnwidth]{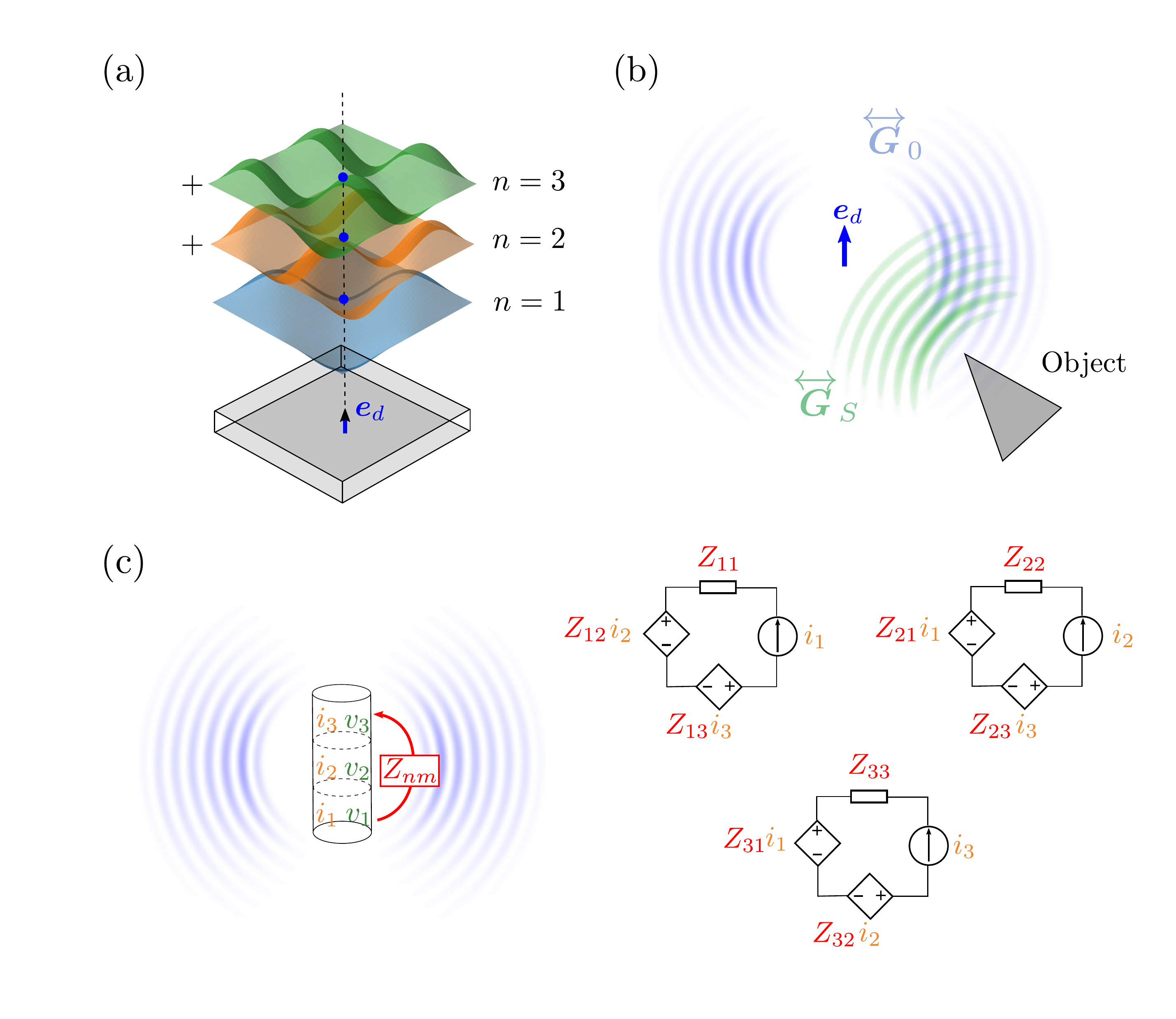}
\caption{
\textbf{Three ways of viewing the effect of the local environment on the emission of radiation by an antenna.}
(a) In the `modal viewpoint' (commonly encountered in quantum optics), the local environment -- in this case a planar cavity -- determines the allowed EM modes to which the emitter may couple.
(b) In the `scattering viewpoint' (commonly encountered in nanophotonics) the local environment scatters some of the emitted radiation back towards the antenna. 
The scattered fields (green) provide an extra driving force on the current in the antenna, thereby changing the power radiated.
(c) In the \emph{impedance viewpoint} (commonly encountered in electrical engineering) the current and voltage throughout a finite size antenna is discretized (left), and then understood as a multi--port network (right). 
In this picture the real part of the impedance matrix $Z_{n m}$--that is, the loss from the circuit--plays a role analogous to the local density of states, becoming identical to it in the limit of a vanishingly small antenna. For each segment shown in the left hand side of the figure a corresponding part of the impedence network is shown on the right.}
\label{fig:emission_pictures}
\end{figure}
\par
As already established, the LDOS concept concerns a modification to the rate of emission, due to a change in the environment surrounding the emitter. 
In the literature one commonly encounters at least three different pictures for describing this modification, which are each sketched in figure~\ref{fig:emission_pictures}. 
All of these viewpoints can be understood in terms of the imaginary part of the Green function ${\rm Im}[\overleftrightarrow{\boldsymbol{G}}]$.

The first is the \emph{modal viewpoint}, used in our introduction, and is shown in Fig~\ref{fig:emission_pictures}.
From this viewpoint modifications to the local density of states are visualized by considering the sum over the eigenmodes of the system, evaluated at the emitter position $\boldsymbol{r}_0$. 
This picture corresponds to the modal expansion of the Green function (and thus equations (\ref{eq:pldos_defn}--\ref{eq:dos_defn})), and is best suited to any lossless closed system, such as an ideal cavity.

The second description is the \emph{scattering viewpoint}, and is sketched in Fig~\ref{fig:emission_pictures}b where changes to the local density of states are instead understood in terms of the field scattered from the environment. This corresponds to writing the Green function $\overleftrightarrow{\boldsymbol{G}}$ as a sum of two parts,
\begin{equation}
\overleftrightarrow{\boldsymbol{G}}(\boldsymbol{r},\boldsymbol{r}_0,\omega)=\overleftrightarrow{\boldsymbol{G}}_0(\boldsymbol{r},\boldsymbol{r}_0,\omega)+\overleftrightarrow{\boldsymbol{G}}_S(\boldsymbol{r},\boldsymbol{r}_0,\omega),\label{eq:G0Gs}
\end{equation}
where $\overleftrightarrow{\boldsymbol{G}}_{0}$ is the Green function for a homogeneous environment with the same refractive index as that in which the emitter is embedded, and  $\overleftrightarrow{\boldsymbol{G}}_{S}$ is the contribution due to scattering by the inhomogeneities in the environment surrounding the emitter (we shall see examples of this decomposition in Sec~\ref{sec:examples}). 
This picture is not well suited to lossless closed systems such as an ideal cavity, because at resonance, the field scatters with equal strength an infinite number of times from cavity walls and it is thus unnatural to separate $\overleftrightarrow{\boldsymbol{G}}_0$ from $\overleftrightarrow{\boldsymbol{G}}_{S}$. 
However, it is well suited to understanding modifications to the emission due to the presence of isolated metallic or dielectric bodies, where the modal approach becomes more difficult to use. 
From Eqs. (\ref{eq:defn_pldos_green}--\ref{eq:defn_dos_green}) we see that the imaginary part of the scattered Green function ${\rm Im}[\overleftrightarrow{\boldsymbol{G}}_{S}]$ represents all such environmental modifications to \textit{e.g.}, spontaneous emission rates. 
This quantity also appears when calculating \textit{e.g.}, the Casimir force between dielectric bodies~\cite{landau2002}.
Although not explicitly discussed, both of these first two equally valid viewpoints can be found side--by--side in Chapter 8 of Novotny and Hecht's textbook~\cite{novotny2006book}.

The third viewpoint, the \emph{impedance viewpoint}, shown in Fig~\ref{fig:emission_pictures}c, is the one found in the electrical engineering literature, where the `input impedance' of the antenna replaces the concept of the local density of states (see, \textit{e.g.}, section 1.8 of~\cite{rudge1982}). 
This connection can be understood directly by examining a calculation of the antenna input impedance using the \emph{method of moments}~\cite{pozar1982,gibson2014}. 
To see how this viewpoint works, consider a thin wire of length $\ell$ aligned along the $z$ axis and driven with a voltage oscillating at a frequency $\omega$. 
Using formula~(\ref{eq:field_from_G}), we can use the Green function to relate the electric field $\tilde{\boldsymbol{E}}$ and the current $\tilde{\boldsymbol{j}}=\boldsymbol{e}_{z}\delta(x)\delta(y)I(z)$ along the wire,
\begin{equation}
    \tilde{\boldsymbol{E}}(\boldsymbol{r})={\rm i}\omega\mu_0\int_{-\ell/2}^{\ell/2} d z'\overleftrightarrow{\boldsymbol{G}}(\boldsymbol{r},z'\boldsymbol{e}_{z},\omega)\cdot\boldsymbol{e}_{z}I(z')\label{eq:wire_field}.
\end{equation}
We expand the current in the wire in terms of a set of basis functions $f_n(z')$ that are set equal to one over some small patch of the wire $z\in[z_n,z_n+1]$ and zero elsewhere,
\begin{equation}
    I(z)=\sum_n i_n f_n(z).
\end{equation}
Multiplying (\ref{eq:wire_field}) by $\boldsymbol{e}_{z}f_m(z)$ and integrating over the length of the wire gives the voltage $v_m$ across segment $m$ as 
\begin{equation}
    v_m=-\int_{z_{m}}^{z_{m+1}}E_z(z)\,dz=-\int_{-\ell/2}^{\ell/2}\boldsymbol{e}_{z}\cdot\tilde{\boldsymbol{E}}(z)f_m(z)dz.
\end{equation}
Using our expression for the electric field in terms of the current (\ref{eq:wire_field}), this expression for the voltage can be written in matrix form as (see fig \ref{fig:emission_pictures}c, right-hand panel),
\begin{equation}
    v_m=\sum_n Z_{m n}i_n,
\end{equation}
where the impedance matrix $Z_{mn}$ relates the small difference in voltage across each segment to every current element in the wire,
\begin{equation}
    Z_{mn}=-{\rm i}\omega\mu_0\int_{-\ell/2}^{\ell/2} dz\int_{-\ell/2}^{\ell/2} dz'\boldsymbol{e}_{z}\cdot\overleftrightarrow{\boldsymbol{G}}(z\boldsymbol{e}_{z},z'\boldsymbol{e}_{z},\omega)\cdot\boldsymbol{e}_{z}f_m(z)f_n(z').\label{eq:mom_impedance}
\end{equation}
The impedance (\ref{eq:mom_impedance}) plays a fundamental role in (macroscopic) electrical engineering, allowing the continuous field problem to be replaced with a collection of coupled circuits (as shown on the right of figure~\ref{fig:emission_pictures}c). 
From the matrix $Z_{m n}$ one can calculate the input impedance, which relates, e.g., the induced current to a given applied voltage (see~\cite{pozar1982}, and for some simple examples see~\cite{newman1988}). 
The real part of this discretized impedance tells us about the energy lost from the electrical current into radiation, 
\begin{equation}
    {\rm Re}[Z_{m n}]=\omega\mu_0\int_{-\ell/2}^{\ell/2}dz\int_{-\ell/2}^{\ell/2}dz'\boldsymbol{e}_{z}\cdot{\rm Im}\left[\overleftrightarrow{\boldsymbol{G}}(z\boldsymbol{e}_{z},z'\boldsymbol{e}_{z},\omega)\right]\cdot\boldsymbol{e}_{z}f_m(z)f_n(z')
\end{equation}
and as expected this is proportional to ${\rm Im}[\overleftrightarrow{\boldsymbol{G}}]$, although not with equal spatial arguments. 
The non--equal spatial arguments arise because the theory encountered in electrical engineering is more general than the local density of states, with the ability to treat a source of any size. 
Only in the limit of a small source $(\ell \to 0)$ does the real part of the impedance matrix (\ref{eq:mom_impedance}) become a single number (because there is no need to discretize the source into more than one piece) proportional to ${\rm Im}[\overleftrightarrow{\boldsymbol{G}}(\boldsymbol{0},\boldsymbol{0},\omega)]$ and hence become proportional to the PLDOS. 
While this description has greater generality, it also lumps together the effect of the environment and antenna geometry. 
The concept of the local density of states is valid for an infinitesimal antenna, and thus isolates the dependence of the emission on the environment. 

%
%
\subsection{A clarification of terms}\label{sec:terms}
It is perhaps useful at this point to remind ourselves of some of the terms we have used in this section. 
First, we have applied two different words to refer to the same thing; 'modes' and 'states'. 
In electromagnetism we usually speak of the allowed modes of a system, \textit{e.g.}, optical cavity modes, waveguide modes, and so forth. 
In solid state physics we usually speak of allowed states, \textit{e.g.}, for electrons in a crystal potential, or phonons on a lattice. 
While the two terms are synonymous, we have adopted the term density of states, while often referring to the 'modes' of the system.
\par
Second, and rather more importantly, we have already talked about the lifetime of an excited state, and will also employ the terms emission (or decay) probability, and emission (or decay) rate. 
In the case of an atom, the process we are interested in is that of the emission of a photon that may accompany the transition from an excited atomic state into a lower energy state. 
The probability $p$ that any one atom emits a photon over the time $T$ is quantified by the spontaneous emission rate $\Gamma=p/T$. 
For $N^e$ identical emitters, the average number of emitted photons per second is then given by $N^e\Gamma$. 
If we consider a fixed time interval then the emission probability $p$ is proportional to the emission rate $\Gamma$. 
With this understanding, the terms probability and rate can be used interchangeably. 
\par
The lifetime of the emitter is the quantity that is usually measured in experiments (see section~\ref{sec:rate}). 
The spontaneous emission lifetime $\tau$ is related to the spontaneous emission rate by,
\begin{equation}
\label{eq:rate-lifetime}
\Gamma = 1/\tau.
\end{equation}
For example, the lifetime $\tau$ given in Fig~\ref{fig:pigments}d was obtained by detecting the arrival time of photons following repeated pulsed excitation of the Eu$^3+$ ions.
In this case the distribution of arrival times matches a single exponential, with decay constant $\Gamma$. 
%
%
\subsection{Guide for the reader\label{sec:guide}}
\par
The remainder of this paper explains the origins and subtleties of the formulae given in this present section, along with an elaboration of several examples. 
In Sec~\ref{sec:DOS} we show how to calculate the DOS (Eq.~(\ref{eq:dos_defn})) for free space. 
In Sec~\ref{sec:classical-dipole} we treat the physics of the classical dipole antenna, showing that the PLDOS governs the rate at which it radiates energy. 
Section~\ref{sec:examples} treats several examples where the PLDOS, LDOS and DOS can be found exactly. 
In Sec~\ref{sec:quantum emitter} we derive expressions for the rate of emission from a quantum emitter, showing that it is also governed by the PLDOS, and in Sec~\ref{failings} we  discuss some of the approximations underlying our calculations. 
In Sec~\ref{sec:the_difference} we resolve the origin of the factor of $4$ between the expressions for the rate of emission from the classical (\ref{eq:intro_classical_power}) and quantum (\ref{eq:intro_quantum_power}) antenna, finding that its origin is tied to the existence of the vacuum field.  
In Sec~\ref{sec:rate} we illustrate how the LDOS can be extracted from an experimental measurement of the rate of decay.
In Sec~\ref{sec:conclusions} we offer some conclusions.  A glossary of terms is given at the end of this document.

\section{The density of states (DOS)\label{sec:DOS}}
\par
As established in Sec~\ref{sec:different_densities}, the coarsest level of description of coupling an emitter to the electromagnetic field is given by the density of states (DOS).
Here we show how to calculate this quantity, since it provides a useful introduction to both classical and quantum analyses.
We will see later that this quantity emerges from a suitable averaging over the rate of emission from a dipole antenna.
%
%
\begin{figure}[ht!]
\begin{center}
\includegraphics[width=10.0cm]{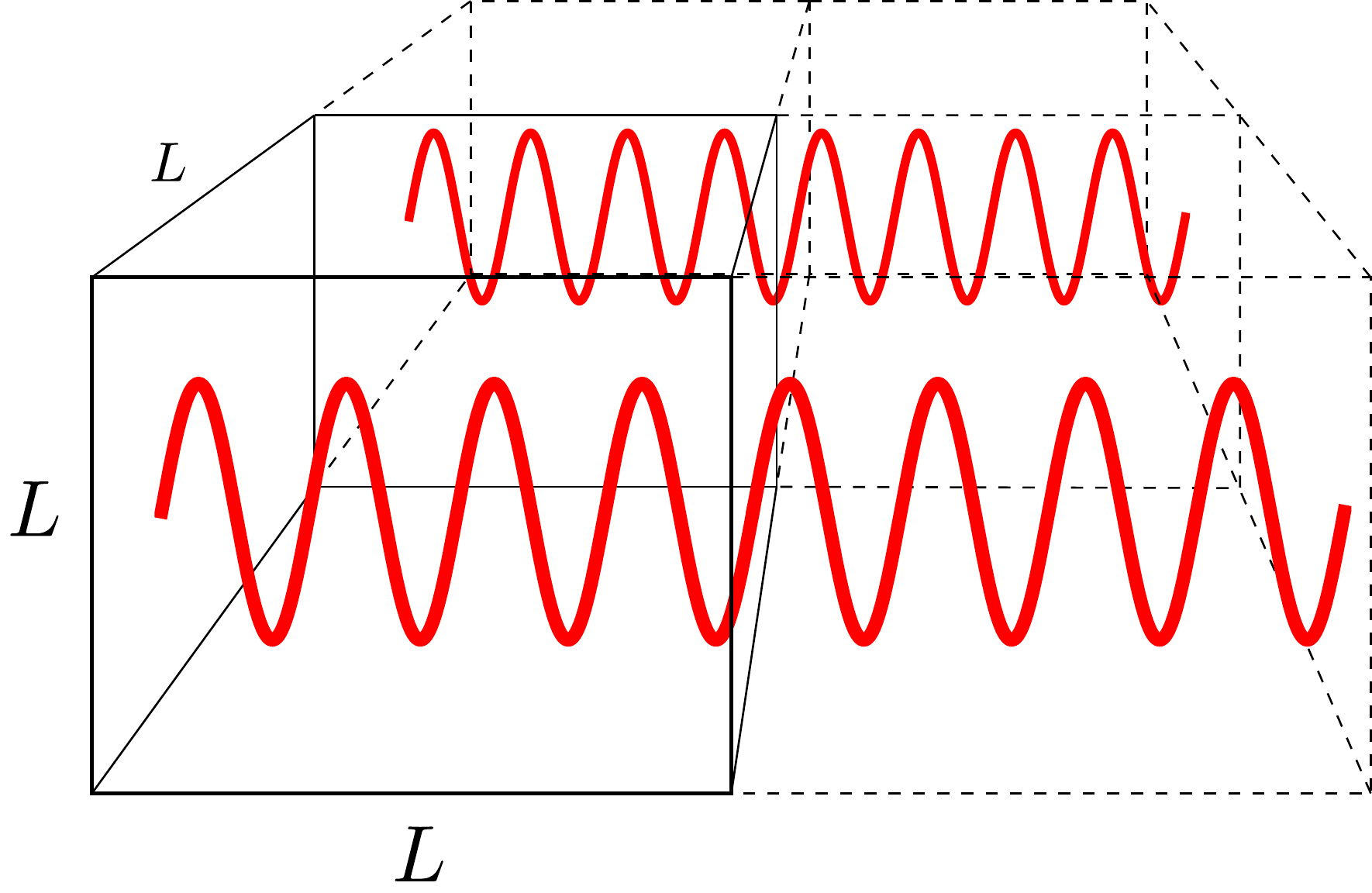}
\end{center}
\caption{Schematic of our treatment of infinite free space, which we consider as an infinite lattice of $L\times L\times L$ cubes with periodic boundary conditions on their edges.  Taking $L$ as arbitrarily large we obtain an LDOS that is independent of $L$.\label{fig:universe_vac_fluct}
}
\end{figure}
\par
To calculate $\rho(\omega)$ in free space (which for reference we denote as $\rho_0(\omega)$), we can use (\ref{eq:dos_defn}), but we need to make use of a limiting procedure to avoid problems with the infinite size of free space. 
To avoid having to face this problem directly we treat the infinite volume as being made up of many smaller $L\times L\times L$ cubes, taking the field in each box to be the identical.  This assumption is equivalent to the use of periodic boundary conditions on the edges of each cube, and entails a quantization of the allowed values of the wavevector $\boldsymbol{k}$,
\begin{equation}
	\boldsymbol{k}(n_x,n_y,n_z)=\frac{2\pi}{L}\left(n_x\boldsymbol{e}_{x}+n_y\boldsymbol{e}_{y}+n_z\boldsymbol{e}_{z}\right).\label{eq:periodic_k}
\end{equation}
where $n_x$, $n_y$ and $n_z$ are integers.
Later we will take the length $L$ to be infinity, and we will find that the result tends to something that is independent of $L$.
It is evident that such a choice of wavevector leads to waves that are identical in each of our imagined boxes, e.g.,
\begin{equation}
	{\rm e}^{{\rm i}\boldsymbol{k}(n_x,n_y,n_z)\cdot(\boldsymbol{x}+L\boldsymbol{e}_{x})}={\rm e}^{{\rm i}\boldsymbol{k}(n_x,n_y,n_z)\cdot\boldsymbol{x}}{\rm e}^{2\pi{\rm i}n_x}={\rm e}^{{\rm i}\boldsymbol{k}(n_x,n_y,n_z)\cdot\boldsymbol{x}}.
\end{equation}
In free space the dispersion relation that connects the wavevector $\boldsymbol{k}$ and the frequency $\omega$ is $\boldsymbol{k}^{2}=\omega^{2}/c^{2}$.  Therefore our discrete set of allowed wavevectors leads to a discrete set of allowed frequencies,
\begin{equation}
	\omega(n_x,n_y,n_z)=c|\boldsymbol{k}(n_x,n_y,n_z)|=\frac{2\pi c}{L}\sqrt{n_x^{2}+n_y^{2}+n_z^{2}}.\label{eq:allowed_frequencies}
\end{equation}
We can now use this expression for the allowed frequencies (\ref{eq:allowed_frequencies}) in our formula for the DOS (\ref{eq:dos_defn}). 
We then find that the number of modes per box in an infinitesimal bandwidth $d\omega$ is given by
\begin{equation}
	\rho_0(\omega)d\omega=\frac{2}{L^{3}}\sum_{n_x,n_y,n_z}\delta(\omega-\omega(n_x,n_y,n_z))d\omega,
\end{equation}
where the factor of two arises from the two independent polarizations of EM waves. 
This expression is zero except at the allowed frequencies (\ref{eq:allowed_frequencies}). 
If we now take the limit $L\to\infty$ then this comb of discrete frequencies becomes a continuous distribution, \textit{i.e.}, it is non--zero at every frequency. 
Using the fact that the difference in two neighboring values of $k_x$ is $\Delta k_x=2\pi/L$, the above expression can be written as an integral,
\begin{equation}
	\rho_0(\omega)d\omega=2\sum_{n_x,n_y,n_z}\frac{\Delta k_x\Delta k_y\Delta k_z}{(2\pi)^{3}}\delta(\omega-\omega(n_x,n_y,n_z))d\omega\to 2\int\frac{d^{3}\boldsymbol{k}}{(2\pi)^{3}}\delta(\omega-c|\boldsymbol{k}|)d\omega.
\end{equation}
The integral is easily evaluated in spherical coordinates $d^{3}\boldsymbol{k} = \sin(\theta_{k}) k^2 dk\,d\theta_{k}\,d\phi_{k}$, yielding 
\begin{align}
	\rho_0(\omega)d\omega&= \frac{2}{c}\int_{0}^{2\pi}\frac{d\phi_{k}}{2\pi} \int_{0}^{\pi}\frac{d\theta_{k}}{2\pi} \int_0^{\infty}\frac{k^2 dk}{2\pi}\sin(\theta_{k}) \delta(k-\omega/c)d\omega,\nonumber\\[5pt]
    &= \frac{\omega^{2}}{\pi^{2}c^{3}} d\omega.
    \label{eq:dos_free_space}
\end{align}
We thus see that whilst our definition of all the densities of states (\ref{eq:pldos_defn}--\ref{eq:dos_defn}) was given in terms of a discrete set of modes, they extend to a continuum, provided care is taken with the limiting process.
\par
It is important to note that the density of states, whether it be the number, local or partial density of states, in essence counts a number of modes for a certain physical situation. In contrast, the excitation level of these modes, \textit{i.e.}, the number of photons populating the mode, is irrelevant. 
Hence, illuminating the structure with a laser, or exciting surface plasmon polaritons, and so forth, does \emph{not} modify a density of states in any way; the density of states is about the \textit{existence of modes, not their population.} 
%
\begin{figure}[ht!]
\begin{center}
\includegraphics[width=11.5cm]{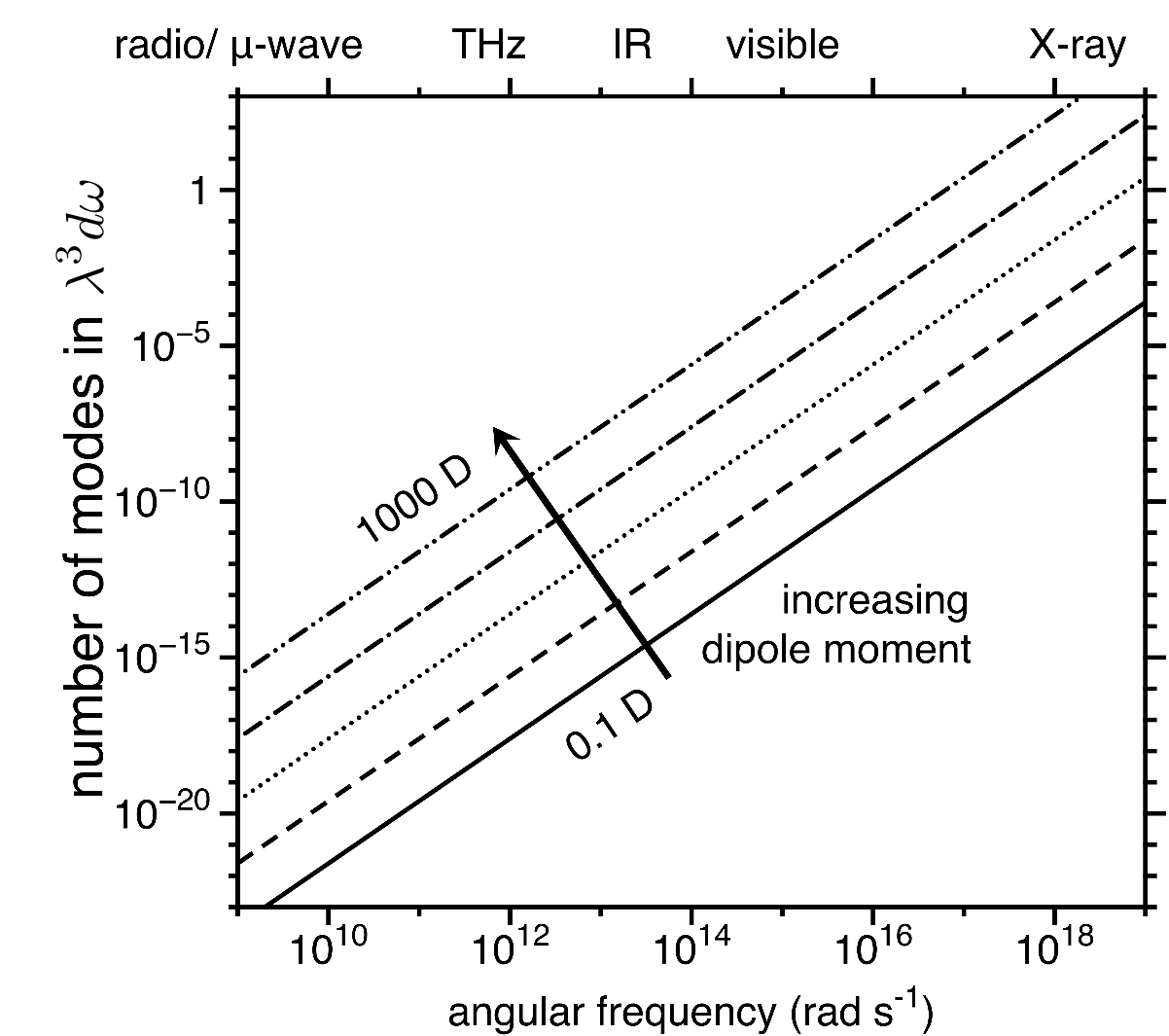}
\end{center}
\caption{The calculated number of modes in a volume $\lambda^3$ and a bandwidth $d\omega = \Gamma_0$. Data were calculated using equations \eqref{eq:num_modes} for five different dipole moments, 0.1D, 1D, 10D, 100D and 1000D.\footnote{{\color{blue}A mild ontological inconvenience of this otherwise natural choice is that a property of the emitter (namely the transition dipole moment) will appear in a property of the EM field, namely the number of modes, through our choice of bandwidth.}} 
Transition dipole moments of $0.1$ to $10$D are typical for emitters in the visible range (atoms, molecules, quantum dots), $100$ to $1000$D is feasible for emitters in the microwave range, whereas in the X-ray range $0.1$ to $1$D are typical. 
}
\label{fig:num_modes}
\end{figure}
\par
At this point we consider the numbers associated with the density of states. 
How many modes does an emitter interact with? 
We are now in a position to calculate this number. 
To go from a number density of modes---as given by equation (\ref{eq:dos_free_space})---to a number of modes, we need to multiply by a chosen volume $V$, and by a chosen frequency interval $d\omega$. 
We thus write the number of modes at frequency $\omega$ as
\begin{equation}
	\label{eq:num_modes_1}
N = \rho_0(\omega) V \Delta\omega.
\end{equation}
Since the wavelength is the natural length scale for wave phenomena, we pick for the volume $V=\lambda^3$.\footnote{Our choice of mode volume is of course arbitrary; for the cavities used in atom optics the volume is typically $\gg\lambda_0^3$, whilst in nanophotonics the use of plasmonic cavities enables cavity volumes to be very significantly smaller than $\lambda_0^3$, where $\lambda_0$ is the free-space wavelength~\cite{Chikkaraddy_Nature_2016_535_127}. } 
For the frequency bandwidth $\Delta\omega$ it seems natural to pick the natural linewidth $\Gamma$, this is the linewidth that is related by the `uncertainty principle' to the excited state lifetime $\tau$: $\Delta\omega \approx \Gamma = 1/\tau$.
With these choices the number of modes (\ref{eq:num_modes_1}) is 
\begin{equation}
	\label{eq:num_modes}
	N = \frac{8\omega^2|\mathcal{P}|^2}{3\varepsilon_0 c^3\hbar}.
\end{equation}
The number of modes $N$ is plotted in figure \ref{fig:num_modes} as a function of angular frequency, and for 5 different values of the dipole moment. 
We see that the number of states is quite small and typically $\ll1$. 
In the microwave range, the number of modes lies between about $N = 10^{-22}$ and $10^{-14}$, in the visible range between about $N = 10^{-12}$ and $10^{-4}$. 
Only in the X-ray range, where $N$ is between $10^{-6}$ and $10^{2}$, does the number of modes approach $N \approx 1$. 
Do these results match what you expected? 
Since most unwary colleagues we quizzed expected $N \approx 1$, the answer is probably not. 
We think that the answer to this non-existent puzzle is that there is nothing 'magic' about equation~\ref{eq:dos_free_space}. 

\section{The classical dipole antenna and the PLDOS\label{sec:classical-dipole}:}
\par
In this section we will discuss emission by a classical dipole and show that the PLDOS governs the power emitted from a small classical antenna, thereby justifying the formulae given in Sec~\ref{sec:essentials}. 
The way the antenna radiates will depend on the surrounding environment, as is illustrated in Figure~\ref{fig:dipole_free-space_near-mirror}. 

\subsection{Essential electrodynamics}
\par
Our first task is to find a general expression for the electromagnetic field produced by a small antenna. 
We consider the simple case where the antenna sits in a region of free space, as shown in Fig~\ref{fig:dipole_free-space_near-mirror}. 
The time-variation of the electrical current $\boldsymbol{j}$ flowing through the antenna produces electromagnetic radiation.  The equation governing this radiation process is the Maxwell equation containing the electric current,
\begin{align}
	\boldsymbol{\nabla}\times\boldsymbol{B}=\mu_0\boldsymbol{j}+\frac{1}{c^{2}}\frac{\partial\boldsymbol{E}}{\partial t}.\label{eq:maxwell4}
\end{align}
To solve this equation we work in terms of two auxiliary quantities, the vector potential $\boldsymbol{A}$ and the scalar potential $\phi$, that are related to the field strengths by 
\begin{align}
	\boldsymbol{B}&=\boldsymbol{\nabla}\times\boldsymbol{A},\nonumber\\
    \boldsymbol{E}&=-\boldsymbol{\nabla}\phi-\frac{\partial\boldsymbol{A}}{\partial t}\label{eq:potentials}.
\end{align}
Substituting these expressions for the fields in terms of the potentials into Eq.~(\ref{eq:maxwell4}) we find the following version of the vector wave equation
\begin{align}
	\boldsymbol{\nabla}\times\boldsymbol{\nabla}\times\boldsymbol{A}+\frac{1}{c^{2}}\frac{\partial^{2}\boldsymbol{A}}{\partial t^{2}}+\frac{1}{c^{2}}\boldsymbol{\nabla}\frac{\partial\phi}{\partial t}&=\mu_0\boldsymbol{j},\nonumber\\
    &=\mu_0\frac{\partial\boldsymbol{d}}{\partial t}\delta^{(3)}(\boldsymbol{r}-\boldsymbol{r}_0).\label{eq:wave_equation}
\end{align}
In the second line we have written the electrical current in terms of its time-dependent dipole moment $\boldsymbol{d} (t)$ that quantifies the separation of positive and negative charges across the antenna, and we have assumed that the current is only flowing within a very small region of space close to the point $\boldsymbol{r}_{0}$, so that the current is given by,
\begin{equation}
	\boldsymbol{j}=\frac{d\boldsymbol{d}}{dt}\delta^{(3)}(\boldsymbol{r}-\boldsymbol{r}_0).\label{eq:defn_current}
\end{equation}
We have also made the assumption that the current in the antenna is a fixed function of time, \textit{i.e.}, we have ignored how it is modified by the radiation it produces. 
This is a reasonable approximation when the electro-motive force (EMF) driving the current is large in comparison to the EMF due to the radiated field. 
When we come to treat things quantum mechanically we shall make the same assumption, which in that context is known as the \emph{weak coupling} regime or the Markovian regime, which is where first-order perturbation theory holds.
\par
Our aim is now to solve equation (\ref{eq:wave_equation}) for the vector potential $\boldsymbol{A}$, by writing it as a sum over the allowed electromagnetic waves (modes) in our system. 
In the absence of a driving current, the electric field can be one of a set of possible eigenmodes $\boldsymbol{\mathcal{E}}_{n}$ oscillating at a fixed frequency $\omega=\omega_{n}>0$.  Combining Maxwell's two curl equations in the absence of a source, these eigenmodes satisfy  
\begin{equation}
\boldsymbol{\nabla}\times\boldsymbol{\nabla}\times\boldsymbol{\mathcal{E}}_{n}-k_{n}^2\boldsymbol{\mathcal{E}}_{n}=0,\label{eq:modes}
\end{equation}
with $k_n$ the modulus of the wave vector $k_n = \omega_n/c$. 
The subscript $n$ labels all these possible modes of the system. 
For example, if the electromagnetic field was confined within a conducting box then $n$ would stand for the three indices $n_x$, $n_y$, $n_z$, labeling the quantization of the modes across the three spatial directions, and for which we could label the two polarizations $\zeta=1,2$. 
These modes are orthogonal, and we choose to normalize them as shown in equation (\ref{eq:orthonormality}) of Sec~\ref{sec:essentials}. 
Taking the divergence of both sides of (\ref{eq:modes}) we see that these modes all have zero divergence; they are purely transverse. 
Note that the remaining longitudinal part of the field (\textit{i.e.}, near-field components), which is due to any charge on the antenna, cannot be described in terms of the propagating field $\boldsymbol{\mathcal{E}}_{n}$. 
We now expand the vector potential in equation (\ref{eq:wave_equation}) as a sum over these modes, with expansion coefficients $c_n(t)$,
\begin{equation}
	\boldsymbol{A}=\sum_{n}c_n(t)\boldsymbol{\mathcal{E}}_{n}(\boldsymbol{r}),\label{eq:field_expansion}
\end{equation}
where we must take the real part if the $\boldsymbol{\mathcal{E}}_{n}$ are complex valued.  Substituting (\ref{eq:field_expansion}) into (\ref{eq:wave_equation}) we find the following equation for the expansion coefficients $c_n(t)$
\begin{align}
	\sum_{n}\left(\ddot{c}_n+\omega_n^{2}c_n\right)\boldsymbol{\mathcal{E}}_{n}+\boldsymbol{\nabla}\frac{\partial\phi}{\partial t}=\frac{1}{\epsilon_{0}}\boldsymbol{e}_{d}\dot{d}\delta^{(3)}(\boldsymbol{r}-\boldsymbol{r}_0),\label{eq:wave_equation_components}
\end{align}
where we have assumed that the dipole has a fixed orientation $\boldsymbol{d}(t)=d(t)\boldsymbol{e}_{d}$. 
Multiplying both sides of (\ref{eq:wave_equation_components}) by $\boldsymbol{\mathcal{E}}_{m}^{\star}$, integrating, and applying (\ref{eq:orthonormality}) we find the equation governing the time evolution of the coefficients $c_n$ to be
\begin{equation}
	\ddot{c}_{n}+\omega_n^{2}c_n=\frac{1}{\epsilon_{0}}\dot{d}(t)\boldsymbol{e}_{d}\cdot\boldsymbol{\mathcal{E}}_{n}^{\star}(\boldsymbol{r}_{0}),\label{eq:driven_oscillator}
\end{equation}
which shows that the amplitudes of the electromagnetic modes each behave as simple harmonic oscillators, driven by a `force' which is proportional to the time derivative of the dipole moment.
Assuming that the antenna is switched on at $t=0$, and that the $c_n$ were zero before this, the solution of (\ref{eq:driven_oscillator}) is,
\begin{equation}
	c_n(t)=\frac{1}{\epsilon_0\omega_n}\boldsymbol{e}_{d}\cdot\boldsymbol{\mathcal{E}}^{\star}_n(\boldsymbol{r}_0)\begin{cases}
    0&(t<0)\\[10pt]
    \int_{0}^{t}dt'\sin(\omega_{n}(t-t'))\dot{d}(t')\label{eq:solution_part1}& (t>0).
    \end{cases}
\end{equation}
Similarly, taking the divergence of both sides of (\ref{eq:wave_equation_components}) we find the equation governing the behaviour of the scalar potential $\phi$,
\begin{equation}
	\boldsymbol{\nabla}^{2}\phi_s(\boldsymbol{r},t)= \frac{1}{\epsilon_{0}}\boldsymbol{e}_{d}\cdot\boldsymbol{\nabla}\delta^{(3)}(\boldsymbol{r}-\boldsymbol{r}_0),\label{eq:poisson}
\end{equation}
where we have written $\phi=d(t)\phi_s$.
Evidently (\ref{eq:poisson}) is Poisson's equation for the electrostatic potential due to a charge density $\sigma=-\boldsymbol{e}_{d}\cdot\boldsymbol{\nabla}\delta^{(3)}(\boldsymbol{r}-\boldsymbol{r}_0)$.  The electric field (\ref{eq:potentials}) is thus made up of; (\ref{eq:field_expansion}), corresponding to the radiation in the system, with the amplitude of each mode behaving as a driven simple harmonic oscillator (\ref{eq:driven_oscillator}); plus $-\boldsymbol{\nabla}\phi$, which is the non--propagating part of the field due to the instantaneous separation of electric charge across the antenna (\ref{eq:poisson}).
\par
\subsection{The PLDOS in terms of fields:}
As described in Sec~\ref{sec:essentials}, we quantify the radiation leaving the antenna through the rate at which the radiated electric field $\boldsymbol{E}$ does work on the current $\boldsymbol{j}$, 
\begin{align}
	P^{\rm C}(t)&=-\int d^{3}\boldsymbol{r} \boldsymbol{j}\cdot\boldsymbol{E},\nonumber\\[10pt]
    &=\dot{d}(t)d(t)\boldsymbol{e}_{d}\cdot\boldsymbol{\nabla}\phi_s(\boldsymbol{r}_0,t)+\frac{1}{\epsilon_0}\sum_{n}\int_{0}^{t}dt'\frac{\sin(\omega_{n}(t-t'))}{\omega_{n}}\ddot{d}(t')\dot{d}(t)\left|\boldsymbol{e}_{d}\cdot\boldsymbol{\mathcal{E}}_n(\boldsymbol{r}_0)\right|^{2},\label{eq:antenna_power}
\end{align}
where in the second line we used our expression for the field (\ref{eq:potentials}) as well as the expansion (\ref{eq:field_expansion}). 
Assuming that the dipole oscillates at a fixed frequency $\omega>0$, then
\begin{equation}
	d(t)={\rm Re}\left[\tilde{d}{\rm e}^{-{\rm i}\omega t}\right],\label{eq:monochromatic_dipole}
\end{equation}
so that by averaging equation (\ref{eq:antenna_power}) over a long time $T\gg1/\omega$ we find that the average emitted power can be expressed as
\begin{align}
	\langle P^{\rm C}\rangle&=\frac{1}{T}\int_0^{T}P^{\rm C}(t)dt,\nonumber\\[10pt]
    &=\frac{\omega^{3}|\tilde{d}|^{2}}{4\epsilon_0}\sum_{n}\frac{1}{2\omega_{n}}\frac{\sin^{2}[(\omega-\omega_{n})T/2]}{[(\omega-\omega_n)/2]^{2}T}\left|\boldsymbol{e}_{d}\cdot\boldsymbol{\mathcal{E}}_n(\boldsymbol{r}_0)\right|^{2}+\omega_{n}\to-\omega_n,\label{eq:average_emitted_power}
\end{align}
where the $+\omega_{n}\to-\omega_n$ indicates that a second term should be added to the equation, the same as the first, but with $+\omega_{n}$ replaced by $-\omega_{n}$. Notice that the final term on the right of (\ref{eq:antenna_power}) averaged to zero over time: in the cases we consider here, this `electrostatic' part of the field $\boldsymbol{\nabla}\phi$ carries no energy away from the antenna and is thus not of interest~\footnote{There are cases (notably when the antenna is embedded in an absorbing material) where this part of the field does contribute to the emitted power, causing divergences that can make it very difficult to compute the spontaneous emission rate from atoms~\cite{Barnett_JPhysB_1996_29_3763,Scheel_1999_PRA_60_4094}.}.
As the averaging time $T$ is made longer and longer, the sinc-squared function within the summation of (\ref{eq:average_emitted_power}) becomes more and more sharply peaked around the points where $\omega=\omega_n$, and in the limit becomes a delta function~\cite{arfken2013},
\begin{equation}
	\lim_{T\to\infty}\frac{\sin^{2}[(\omega-\omega_{n})T/2]}{[(\omega-\omega_n)/2]^{2}T}=2\pi\delta(\omega-\omega_n)\label{eq:delta_formula}.
\end{equation}
Given that both $\omega$ and $\omega_n$ are assumed positive, the long-time average of the power emitted from the antenna is thus given by,
\begin{align}
	\langle P^{\rm C}\rangle&=\frac{\pi\omega^{2} |\tilde{d}|^{2}}{4\epsilon_{0}}\sum_{n}\left|\boldsymbol{e}_{d}\cdot\boldsymbol{\mathcal{E}}_{n}(\boldsymbol{r}_0)\right|^{2}\delta(\omega-\omega_{n}),\\
    &=\frac{\pi\omega^{2} |\tilde{d}|^{2}}{4\epsilon_{0}}\rho_{\rm p}(\boldsymbol{e}_{d},\boldsymbol{r}_0,\omega),\label{eq:classical_power}
\end{align}
and varies in proportion to the PLDOS defined in Sec~\ref{sec:essentials}, in agreement with similar expressions in the literature e.g.~\cite{Vats_PRA_2002_65_043808}.  If the emitter is not embedded in free space, but instead sits in an homogeneous medium, one can obtain the equivalent of (\ref{eq:classical_power}) through the substitutions $\epsilon_0\to\epsilon\epsilon_0$ and $\mu_0\to\mu\mu_0$, leaving,
\begin{equation}
	\langle P^{\rm C}\rangle=\frac{\pi\omega^{2} |\tilde{d}|^{2}}{4\epsilon\epsilon_{0}}\rho_{\rm p}(\boldsymbol{e}_{d},\boldsymbol{r}_0,\omega).\label{eq:classical_power_medium}
\end{equation}
In section~\ref{sec:examples} we evaluate $\rho_{\rm p}$ in some simple cases.
%
%
\subsection{The PLDOS in terms of the Green function:\label{sec:pldos_Gf}}
\par
As mentioned in section~\ref{sec:essentials}, the above method for calculating the rate of emission from the antenna (and hence the PLDOS) is no longer applicable when there is significant dissipation in the system.  There is a more general expression for the PLDOS in terms of the electromagnetic Green function $\overleftrightarrow{\boldsymbol{G}}$, defined in section~\ref{sec:essentials}.  In this article we have so far avoided using the Green function, since the sum over modes is likely to give those less familiar with the problem greater insight into the physics. However, the power of the Green function approach is such that now makes a good time to introduce it since, among other things, it allows us to incorporate dissipation in a natural way.
\par
To represent our results in terms of a Green function, we combine Maxwell's equations to eliminate the magnetic field.  We also assume the current is oscillating at a fixed frequency $\omega$, writing quantities as \textit{e.g.}, $\boldsymbol{j}={\rm Re}[\tilde{\boldsymbol{j}}{\rm e}^{-{\rm i}\omega t}]$.  The electric field is then governed by the vector wave equation 
\begin{equation}
\boldsymbol{\nabla}\times\boldsymbol{\nabla}\times\tilde{\boldsymbol{E}}-\epsilon\mu k_0^{2}\tilde{\boldsymbol{E}}={\rm i}\mu\mu_0\omega\tilde{\boldsymbol{j}}\label{eq:vector_helmholtz}
\end{equation}
where we have included the permittivity $\epsilon$ and permeability $\mu$ of the medium.  The average emitted power (\ref{eq:emitted_power}) is given by
\begin{equation}
	\langle P^{\rm C}\rangle=-\frac{1}{T}\int d^{3}\boldsymbol{r}\int_0^T {\rm Re}[\tilde{\boldsymbol{j}}{\rm e}^{-{\rm i}\omega t}]\cdot{\rm Re}[\tilde{\boldsymbol{E}}{\rm e}^{-{\rm i}\omega t}]dt=-\frac{1}{2}\int d^{3}\boldsymbol{r}{\rm Re}\left[\tilde{\boldsymbol{j}}^{\star}\cdot\tilde{\boldsymbol{E}}\right].\label{eq:emitted_power_gf}
\end{equation}
Using the definition of the Green function as the solution to (\ref{eq:green_function_definition}), the electric field in (\ref{eq:vector_helmholtz}) can be written as an integral,
\begin{equation}
	\tilde{\boldsymbol{E}}={\rm i}\mu\mu_0 \omega\int d^{3}\boldsymbol{r}'\overleftrightarrow{\boldsymbol{G}}(\boldsymbol{r},\boldsymbol{r}',\omega)\cdot\tilde{\boldsymbol{j}}(\boldsymbol{r}'),\label{eq:E_field_Gf}
\end{equation}
and therefore the average emitted power (\ref{eq:emitted_power_gf}) is given by,
\begin{align}
	\langle P^{\rm C}\rangle&=\frac{\mu\mu_0 \omega}{2}{\rm Im}\left[\int d^{3}\boldsymbol{r}\int d^{3}\boldsymbol{r}'\tilde{\boldsymbol{j}}^{\star}(\boldsymbol{r})\cdot\overleftrightarrow{\boldsymbol{G}}(\boldsymbol{r},\boldsymbol{r}',\omega)\cdot\tilde{\boldsymbol{j}}(\boldsymbol{r}')\right]\nonumber,\\
    &=\frac{\mu\mu_0 \omega^{3} |\tilde{d}|^{2}}{2}{\rm Im}\left[\boldsymbol{e}_{d}\cdot\overleftrightarrow{\boldsymbol{G}}(\boldsymbol{r}_{0},\boldsymbol{r}_{0},\omega)\cdot\boldsymbol{e}_{d}\right],\label{eq:average_power_green_function}
\end{align}
where we used the expression for the current in terms of the dipole moment (\ref{eq:defn_current}).  Comparing (\ref{eq:average_power_green_function}) with our earlier expression (\ref{eq:classical_power_medium}) we see now that the partial local density of states can also be written in terms of the imaginary part of the Green function,
\begin{equation}
	\rho_{\rm p}(\boldsymbol{e}_{d},\boldsymbol{r}_0,\omega)=\frac{2\omega{\rm n}^{2}}{\pi c^{2}}\boldsymbol{e}_{d}\cdot{\rm Im}\left[\overleftrightarrow{\boldsymbol{G}}(\boldsymbol{r}_0,\boldsymbol{r}_0,\omega)\right]\cdot\boldsymbol{e}_{d},\label{eq:pldos_green_function}
\end{equation}
where the refractive index of the host medium is given by ${\rm n}=\sqrt{\epsilon\mu}$.  Expression (\ref{eq:pldos_green_function}) is a form often used in the literature on spontaneous emission (e.g. \cite{Tomas_PRA_1997_56_4197,NandH}).  Given the relationship between the PLDOS, LDOS and DOS described in section~\ref{sec:different_densities}, one can---through suitable averaging---obtain the three expressions (\ref{eq:defn_pldos_green}) given in section~\ref{sec:essentials}.
Although the Green function is neat and compact, for the uninitiated it is not always clear how it relates to the spatial distribution of the allowed modes in the system.
Here this relationship can be seen by comparing the two expressions for the partial local density of states (\ref{eq:pldos_defn}) and (\ref{eq:pldos_green_function}). (For the mathematics of Green functions, see Barton~\cite{barton1989}, and for something more specific to the dyadic Green functions used here see Tai~\cite{tai1993}.)
\par
There are several cases where analytic expressions for the Green function are known, \textit{e.g.}, systems such as conducting spheres and cylinders~\cite{milton2001}, coaxial waveguides~\cite{Tai_IEEETransAntProp_1983_AP31_355}, and graphene~\cite{hanson2008}), and these can be applied to obtain expressions for the PLDOS. 
The simplest case is probably an infinite homogeneous medium of permittivity $\epsilon$ and permeability $\mu$, which we give for reference (and shall use later). 
In this case the Green function $\overleftrightarrow{\boldsymbol{G}}$ is given by
\begin{equation}
	\overleftrightarrow{\boldsymbol{G}}_0=\left[\frac{1}{k_0^{2}{\rm n}^{2}}\boldsymbol{\nabla}\otimes\boldsymbol{\nabla}+\boldsymbol{1}\right]\frac{{\rm e}^{{\rm i}k_0{\rm n}|\boldsymbol{r}-\boldsymbol{r}_{0}|}}{4\pi|\boldsymbol{r}-\boldsymbol{r}_{0}|},\label{eq:free_space_green_function}
\end{equation}
where '$\otimes$' indicates a tensor product, \textit{e.g.},
\begin{equation}
\boldsymbol{\nabla}\otimes\boldsymbol{\nabla} = \left(\begin{matrix}\frac{\partial^{2}}{\partial x^{2}}&\frac{\partial^{2}}{\partial x\partial y}&\frac{\partial^{2}}{\partial x\partial z}\\
\frac{\partial^{2}}{\partial y\partial x}&\frac{\partial^{2}}{\partial y^{2}}&\frac{\partial^{2}}{\partial y\partial z}\\
\frac{\partial^{2}}{\partial z\partial x}&\frac{\partial^{2}}{\partial y\partial z}&\frac{\partial^{2}}{\partial z^{2}}\end{matrix}\right).
\end{equation}
The imaginary part of (\ref{eq:free_space_green_function}) determines the PLDOS (see Fig.~\ref{fig:cartoon-emission}), and after expanding $\sin(k_0{\rm n}|\boldsymbol{r} - \boldsymbol{r}_0|)/(4\pi|\boldsymbol{r} - \boldsymbol{r}_0|)$ to the first two leading terms in $r = |\boldsymbol{r} - \boldsymbol{r}_0|$ (assuming that $\boldsymbol{e}_{d}$ points along the z--axis) one obtains
\begin{equation}
	\boldsymbol{e}_{d}\cdot{\rm Im}\left[\overleftrightarrow{\boldsymbol{G}}_0(\boldsymbol{r}_{0},\boldsymbol{r}_{0},\omega)\right]\cdot\boldsymbol{e}_{d}=\frac{k_0{\rm n}}{4\pi}\left[1-{\rm Re}\left[\frac{1}{{\rm n}^{2}}\right]\frac{{\rm n}^{2}}{3}\right]+\frac{1}{k_0^{2}}{\rm Im}\left[\frac{1}{{\rm n}^{2}}\right]\lim_{r\to 0}\frac{\partial^{2}}{\partial z^{2}}\frac{\cos\left(k_0{\rm n}r\right)}{4\pi r},\label{eq:homogeneous_gf}
\end{equation}
which for lossless media ($\epsilon$ and $\mu$ real) reduces to
\begin{equation}
		\boldsymbol{e}_{d}\cdot{\rm Im}\left[\overleftrightarrow{\boldsymbol{G}}_0(\boldsymbol{r}_{0},\boldsymbol{r}_{0},\omega)\right]\cdot\boldsymbol{e}_{d}=\frac{k_0{\rm n}}{6\pi},\label{eq:ImG0}
\end{equation}
which, when substituted into the expression for the PLDOS (\ref{eq:pldos_green_function}) gives,
\begin{equation}
	\rho_{\rm p}(\boldsymbol{e}_{d},\boldsymbol{r}_0,\omega)=\frac{{\rm n}^{3}\omega^{2}}{3\pi^{2} c^{3}}=\frac{{\rm n}^{3}}{3}\rho_0(\omega).\label{eq:pldos_gf_fs}
\end{equation}
As expected, when $\epsilon\mu=1$, as in free space, this is $1/3$ the value of the free space DOS (\ref{eq:dos_free_space}), as the emitter only probes one of the three spatial dimensions, whereas the DOS counts all the modes irrespective of dimension (or dipole orientation).  
The factor of ${\rm n}^{3}$ is due to the effect of the surrounding dielectric medium, where the wavelength in each direction is compressed by a factor of  $1/{\rm n}$.
%
%
\section{Examples\label{sec:examples}}
\par
In this section we illustrate how the emission from an antenna depends on the environment by evaluating the partial local density of states (\ref{eq:pldos_defn}) in some simple cases.  We treat atomic systems in Sec.~\ref{sec:quantum emitter}.
%
%
\subsection{Emission in a homogeneous environment\label{sec:free_space_emission}}
\par
The very simplest case to begin with is the PLDOS in a homogeneous isotropic environment, \textit{i.e.}, for an emitter embedded in an infinite medium of constant scalar permittivity and permeability.  We have just evaluated this using the Green function (\ref{eq:pldos_gf_fs}), but to show the equivalence of the modal approach (\textit{i.e.}, the equivalence between the two descriptions sketched in figure~\ref{fig:emission_pictures}(a,b)), we shall also now evaluate it using a summation over modes.

Note that due to the isotropy and homogeneity of the system, the PLDOS, LDOS and DOS all contain the same information.
We imagine dividing space up into $L\times L\times L$ cubes, then taking the limit $L\to\infty$.  This restricts the wavevector to the form (\ref{eq:periodic_k}) and the frequency to
\begin{equation}
	\omega(n_x,n_y,n_z)=\frac{2\pi c}{L\sqrt{\epsilon\mu}}\sqrt{n_x^{2}+n_y^{2}+n_z^{2}}.
\end{equation}
The two polarizations of the electric field are represented by unit vectors that are orthogonal to the wavevector (\ref{eq:periodic_k}), and to each other.  We choose these polarization unit vectors to be
\begin{align}
	\boldsymbol{e}_{1}(n_{x},n_{y},n_{z})& = \frac{n_{x}\boldsymbol{e}_{y}-n_{y}\boldsymbol{e}_{x}}{\sqrt{n_{x}^{2}+n_{y}^{2}}},\nonumber\\[10pt]
    \boldsymbol{e}_{2}(n_{x},n_{y},n_{z})& = \frac{1}{\sqrt{n_{x}^{2}+n_{y}^{2}+n_{z}^{2}}}\left[\sqrt{n_{x}^{2}+n_{y}^{2}}\boldsymbol{e}_{z}-n_{z}\frac{n_{x}\boldsymbol{e}_{x}+n_{y}\boldsymbol{e}_{y}}{\sqrt{n_{x}^{2}+n_{y}^{2}}}\right].\label{eq:two_polarizations}
\end{align}
Because free space is isotropic we can choose the unit vector $\boldsymbol{e}_{d}$ in the definition of the PLDOS (\ref{eq:pldos_defn}) to point in the $\boldsymbol{e}_{z}$ direction, without affecting the answer. 
This eliminates the polarization $\boldsymbol{e}_{1}$ from the formulae. 
The relevant modes of our system are then plane waves with an electric field given by
\begin{equation}
	\boldsymbol{\mathcal{E}}_{n_{x},n_{y},n_{z}} = \frac{1}{L^{3/2}}\boldsymbol{e}_{2}(n_x,n_y,n_z)\exp\left({\rm i}\boldsymbol{k}(n_x,n_y,n_z)\cdot\boldsymbol{r}\right),\label{eq:free_space_mode}
\end{equation}
where the prefactor of $1/L^{3/2}$ is chosen so that the modes are orthonormal (\ref{eq:orthonormality}) when integrated over the $L\times L\times L$ cubic volume.  Substituting (\ref{eq:free_space_mode}) into the PLDOS (\ref{eq:pldos_defn}) then gives us,
\begin{equation}
	\rho_{\rm p}(\boldsymbol{e}_{d},\boldsymbol{r}_0,\omega)=\frac{1}{L^{3}}\sum_{n_{x},n_{y},n_{z}}\frac{n_{x}^{2}+n_{y}^{2}}{n_{x}^{2}+n_{y}^{2}+n_{z}^{2}}\delta\left(\omega-\frac{2\pi c}{L \rm n}\sqrt{n_x^{2}+n_y^{2}+n_z^{2}}\right).\label{eq:free_space_ldos_sum}
\end{equation}
We now take the same limit as in section~\ref{sec:DOS}, letting $L\to\infty$ thereby turning the summation into an integral. 
Evaluating this integral in spherical coordinates $k,\theta_k,\phi_k$ leaves us with
\begin{align}
	\rho_{\rm p}(\boldsymbol{e}_{d},\boldsymbol{r}_0,\omega)&=\int_0^{2\pi}\frac{d\phi_k}{2\pi}\int_0^{\pi}\sin^{3}(\theta_k)\frac{d\theta_k}{2\pi}\int_0^{\infty}\frac{k^{2}dk}{2\pi}\delta\left(\omega-\frac{c}{\rm n}k\right),\nonumber\\
    &=\frac{{\rm n}^{3}\omega^{2}}{4\pi^{2} c^{3}}\int_0^{\pi}\sin^{3}(\theta_k)d\theta_k,\nonumber\\
    &=\frac{{\rm n}^{3}\omega^{2}}{3\pi^{2}c^{3}}\label{eq:free_space_continuum_ldos},
\end{align}
in agreement with (\ref{eq:pldos_gf_fs}).
Given that in this system the PLDOS is independent of both the emitter orientation $\boldsymbol{e}_{d}$ and the emitter position $\boldsymbol{r}_0$, the LDOS $\rho_{l}$ (\ref{eq:ldos_defn}) is simply three times the partial local density of states, and is in turn equal to the number density of states, $\rho(\omega)$ (\ref{eq:dos_defn}), \textit{i.e.},
\begin{equation}
3\rho_p(\boldsymbol{e}_{d},\boldsymbol{r}_0,\omega)=\rho_{l}(\boldsymbol{r}_0,\omega)=\rho(\omega)=\frac{{\rm n}^{3}\omega^{2}}{\pi^{2}c^{3}}={\rm n}^{3}\rho_0(\omega).
\label{eq:ndos_ldos_pdos}
\end{equation}
\noindent In a homogeneous environment all three measures of the density of states contain exactly the same information, the only difference being a factor of $1/3$ which arises because the partial local density of states is only concerned with one out of the three possible antenna orientations.
\par
Applying our formula for the partial local density of states (\ref{eq:free_space_continuum_ldos}) to our expression for the power radiated from an antenna (\ref{eq:classical_power_medium}) we find that the average power emitted from a dipole antenna in a homogeneous medium is equal to,
\begin{equation}
	\langle P^{\rm C}\rangle=\frac{|\tilde{d}|^{2}\omega^{4}}{12\pi\epsilon_{0}}\frac{{\rm n}^{3}}{\epsilon}=\frac{{\rm n}}{\epsilon} {\rm n}^{2}\left(\frac{|\tilde{d}|^{2}\omega^{4}}{12\pi\epsilon_{0}}\right)=Z {\rm n}^{2}\langle P_0^{\rm C}\rangle,\label{eq:free_space_radiation}
\end{equation}
where $\langle P_0^{\rm C}\rangle=|\tilde{d}|^{2}\omega^{4}/12\pi\epsilon_0$ is the time averaged power emitted from a classical dipole antenna in free space, and $Z=\sqrt{\mu/\epsilon}$ is the relative impedance of the medium. 
Note that the emitted power scales as $\omega^{4}$, which can be understood as the combination of one factor of $\omega^{2}$ coming from the time variation of the dipole moment, and another $\omega^{2}$ coming from the local density of states. 
The rate of energy lost from an antenna embedded in a homogeneous medium, relative to the rate of emission in free space also increases as the relative impedance of the medium multiplied by the refractive index squared. 
This is the combined effect of both an increase of the partial local density of states, due to the factor of $n^3$ in (\ref{eq:free_space_continuum_ldos}), and the modified coupling between the antenna and the field, arising from the factor of $1/\varepsilon$ in (\ref{eq:classical_power_medium}); the improved polarizability of the medium (relative to vacuum) screens the dipole moment of the antenna, reducing it by a factor of $1/\epsilon$.
%
%
%
\subsection{Emission close to a perfect mirror\label{sec:mirror_emission}}
\par
The simplest inhomogeneous case returns us to our leitmotif, the PLDOS close to a perfectly reflecting mirror (in this section we assume $\epsilon=\mu=1$ outside the mirror). 
While the allowed frequencies remain continuous in this system, all positions and directions are no longer equivalent. 
An antenna will emit differently depending on its distance from the mirror, and its orientation relative to the plane of the mirror. 
In this system we will begin to see the distinction between the three measures of the density of states (\ref{eq:pldos_defn}), (\ref{eq:ldos_defn}) and (\ref{eq:dos_defn}).
\begin{figure}[ht!]
\begin{center}
\includegraphics[width=7.5cm]{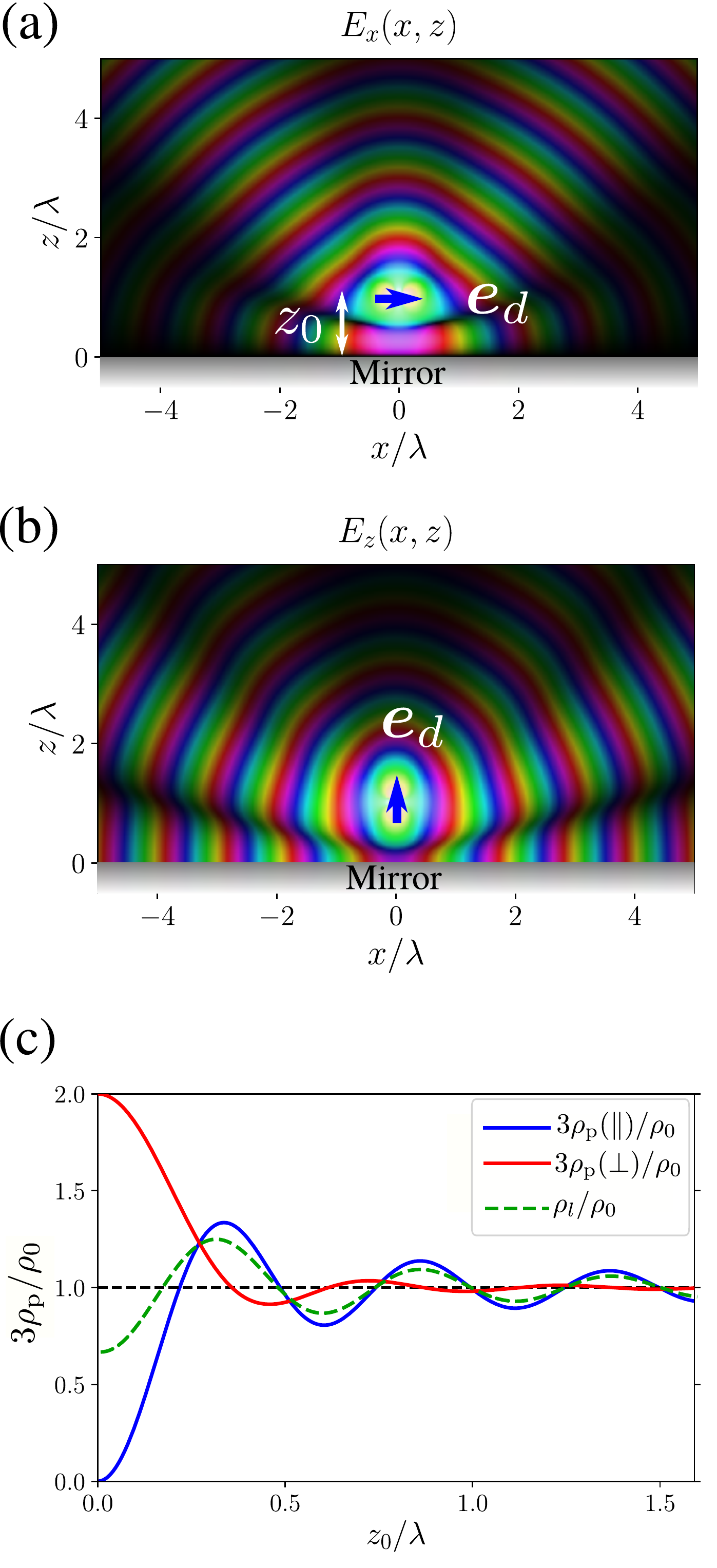}
\end{center}
\caption{Local density of states in the vicinity of a mirror. 
(a) The electric field component $E_x$ ($\parallel$ orientation) from an antenna pointing along the $x$--axis at a position $z_0=\lambda$ above a mirror (colour and brightness similar as in figure~\ref{fig:dipole_free-space_near-mirror}).
(b) Electric field component $E_z$ ($\perp$ orientation) for an antenna pointing along the $z$--axis at position $z_0=\lambda$.
(c) The PLDOS (Eqs.~\ref{eq:mirror_pldos} and~\ref{eq:out_of_plane_integral}) in units of the free space number density of states $\rho_0$ (\ref{eq:dos_free_space}) as a function of antenna position $z_0$ for the two orientations. 
The dashed green curve shows the LDOS (\ref{eq:mirror_ldos}) as the sum $\rho_{l}=2\rho_{p}(\parallel)+\rho_{p}(\perp)$.
\label{fig:mirror_emission}}
\end{figure}
A good mirror is typically a good conductor.  In the presence of an electric field the charges within the conductor move, canceling the in--plane electric field at the surface.  A mirror can therefore be mimicked using the boundary condition that the in-plane electric field is zero.  We take the surface of the mirror to be at $z=0$, and sum together pairs of the allowed modes in free space (\ref{eq:free_space_mode}) such that the in--plane electric field is zero in this plane.  For the two types of polarization defined in (\ref{eq:two_polarizations}) we have either (for polarization $\boldsymbol{e}_{2}$),
\begin{multline}
	\boldsymbol{\mathcal{E}}_{n_{x},n_{y},n_{z}}^{(2)}=\frac{N_{n_{z}}}{\sqrt{L^{3}}}\bigg[\boldsymbol{e}_{2}(n_{x},n_{y},n_{z})\exp\left({\rm i}\boldsymbol{k}(n_{x},n_{y},n_{z})\cdot\boldsymbol{r}\right)\\
    +\boldsymbol{e}_{2}(n_{x},n_{y},-n_{z})\exp\left({\rm i}\boldsymbol{k}(n_{x},n_{y},-n_{z})\cdot\boldsymbol{r}\right)\bigg],\label{eq:mirror_mode1}
\end{multline}
(where the normalization constant is given by $N_{n_{z}>0}=1$ and $N_{n_{z}=0}=1/\sqrt{2}$) or for polarization $\boldsymbol{e}_{1}$,
\begin{multline}
	\boldsymbol{\mathcal{E}}_{n_{x},n_{y},n_{z}}^{(1)}=\frac{1}{\sqrt{L^{3}}}\bigg[\boldsymbol{e}_{1}(n_{x},n_{y},n_{z})\exp\left({\rm i}\boldsymbol{k}(n_{x},n_{y},n_{z})\cdot\boldsymbol{r}\right)\\
    -\boldsymbol{e}_{1}(n_{x},n_{y},-n_{z})\exp\left({\rm i}\boldsymbol{k}(n_{x},n_{y},-n_{z})\cdot\boldsymbol{r}\right)\bigg],\label{eq:mirror_mode2}
\end{multline}
where again we initially assume that free space is periodic with spatial period $L$, taking the limit of large $L$ at the end of the calculation.  Given that these modes vanish on a series of $z={\rm const.}$ planes, separated by a distance $L/2$, they are normalized over the $L\times L\times L/2$ volume above the mirror at $z=0$.  Note that we need only consider positive values for the integer $n_z$; the negative values do not correspond to different modes.
\par
For a general orientation $\boldsymbol{e}_{d}$, an antenna will couple to some combination of both modes (\ref{eq:mirror_mode1}) and (\ref{eq:mirror_mode2}).  To simplify the discussion we consider two cases, where the antenna either points entirely in the plane of the mirror ($\parallel$), or entirely perpendicular to this plane ($\perp$).
%
%
\subsubsection{Antenna pointing in the plane of the mirror ($\parallel$)\\[10pt]}
All directions in the plane of the mirror are equivalent, and we can therefore choose the antenna to point along the $x$--axis: $\boldsymbol{e}_{d}=\boldsymbol{e}_{x}$ (see Figure~\ref{fig:mirror_emission}a). 
Substituting the above expressions for the two polarizations of the electric field (\ref{eq:mirror_mode1}--\ref{eq:mirror_mode2}) into the definition of the partial local density of states (\ref{eq:pldos_defn}), after a few manipulations it can be reduced as follows
\begin{align}
\rho_p(\parallel,\boldsymbol{r}_0,\omega)&=\sum_{n_{x},n_{y},n_{z}\geq0}\sum_{\zeta=1,2}\left|\boldsymbol{e}_{x}\cdot\boldsymbol{\mathcal{E}}_{n_{x},n_{y},n_{z}}^{(\zeta)}(\boldsymbol{r}_0)\right|^{2}\delta(\omega-\omega(n_{x},n_{y},n_{z}))\nonumber\\[10pt]
    &\approx\frac{4}{L^{3}}\sum_{n_{x},n_{y}}\sum_{n_{z}>0}\frac{n_{y}^{2}+n_{z}^{2}}{n_{x}^{2}+n_{y}^{2}+n_{z}^{2}}\sin^{2}\left(\frac{2\pi n_{z} z_{0}}{L}\right)\delta(\omega-\omega(n_{x},n_{y},n_{z})),\label{eq:pldos_in_plane}
\end{align}
where in the second line we neglected the $n_{z}=0$ term in the sum, the contribution of which is zero in the large $L$ limit. 
Taking this limit as we did previously in equation (\ref{eq:free_space_continuum_ldos}), the sum becomes an integral, which can again be evaluated in spherical polar coordinates leaving the expression
\begin{equation}
	\rho_p(\parallel,\boldsymbol{r}_0,\omega)=\frac{\omega^2}{\pi^{2}c^{3}}\int_{0}^{\pi/2}d\theta_k\left[\sin(\theta_k)-\frac{1}{2}\sin^{3}(\theta_k)\right]\sin^{2}\left(\frac{\omega}{c}\cos(\theta_k) z_{0}\right).\label{eq:mirror_pldos}
\end{equation}
the $z_0$ dependence of which is plotted in figure~\ref{fig:mirror_emission}b.  When the antenna is close to the mirror ($\omega z_0/c\ll1$) then the sine-squared dependence within the integrand can be approximated by the quadratic $\omega^{2} z_{0}^{2}\cos(\theta_{k})/c^{2}$, allowing us to evaluate the integral analytically,
\begin{align}
	\text{Close to mirror:}\;\;\rho_p(\parallel,\boldsymbol{r}_0,\omega)&\approx\frac{\omega^{4}z_{0}^{2}}{\pi^{2}c^{5}}\int_{0}^{\pi/2}d\theta_k\left[\sin(\theta_k)\cos^{2}(\theta_k)-\frac{1}{2}\sin^{3}(\theta_k)\cos^{2}(\theta_k)\right],\nonumber\\
    &=\frac{4\omega^{2}z_{0}^{2}}{15c^{2}}\rho_{0},
    \label{eq:close_mirror}
\end{align}
where $\rho_0$ is the free space DOS (\ref{eq:free_space_continuum_ldos}). 
Therefore close to the mirror, the emitted power (\ref{eq:classical_power}) will be very low compared to free space.  We can understand this in terms of the image dipole induced in the mirror, shown in Fig.~\ref{fig:interface}.
Given that the dipole lies in the plane, as we bring the dipole emitter towards the mirror the image dipole ultimately reduces the net dipole moment to zero, extinguishing the radiation.
\par
Meanwhile when the antenna is far away from the mirror ($\omega z_0/c\gg1$), the sine-squared oscillates between $0$ and $1$ so rapidly that its effect is just to scale the rest of the integrand.  To see this, we can examine the general form of the integral appearing in (\ref{eq:mirror_pldos}) at a point where $\omega z_{0}/c=\pi N$ ($N$ integer), with $N\gg1$,
\begin{align}
	\int_{0}^{\pi/2}d\theta_k \sin(\theta_{k})f(\cos(\theta_{k}))\sin^{2}\left(\frac{\omega z_{0}}{c}\cos(\theta_{k})\right)&=\int_{0}^{1}d\eta f(\eta)\sin^{2}\left(\pi N\eta\right),\nonumber\\
    &\approx\sum_{n=0}^{N-1}f(n/N)\int_{0}^{\frac{1}{N}}d\eta\sin^{2}\left(\pi\eta N\right),\nonumber\\
    &\approx\frac{1}{2}\int_{0}^{1}f(\eta)d\eta,\label{eq:large_z_result}
\end{align}
where $f(\cos(\theta_k))$ is an arbitrary function that is independent of $z$, and the approximations tend to exact results in the limit of infinite $N$.  Using this result (\ref{eq:large_z_result}) we can evaluate (\ref{eq:mirror_pldos}) in the limit where the antenna is very far from the mirror
\begin{equation}
	\text{Far from the mirror:}\qquad\rho_p(\parallel,\boldsymbol{r}_0,\omega)\sim\frac{\omega^2}{3\pi^{2}c^{3}}=\frac{\rho_0}{3}.
\end{equation}
As the antenna is moved ever further away from the mirror, the partial local density of states and therefore the rate of emission (\ref{eq:classical_power}), approaches that of free space.  The onset of this behaviour is evident on the far right of figure~\ref{fig:mirror_emission}b.
%
%
\subsubsection{Antenna pointing out of the mirror plane ($\perp$)\\[10pt]}
For the opposite case $\boldsymbol{e}_{d}=\boldsymbol{e}_{z}$ (figure~\ref{fig:mirror_emission}c), the antenna only couples to one of the two polarizations defined in (\ref{eq:mirror_mode2}).  This results in the following expression for the partial local density of states,
\begin{align}
	\rho_p(\perp,\boldsymbol{r}_0,\omega)&=\sum_{n_{x},n_{y},n_{z}\geq0}\left|\boldsymbol{e}_{z}\cdot\boldsymbol{\mathcal{E}}_{n_{x},n_{y},n_{z}}^{(2)}(\boldsymbol{r}_0)\right|^{2}\delta(\omega-\omega(n_{x},n_{y},n_{z})),\nonumber\\
    &\approx\frac{4}{L^{3}}\sum_{n_{x},n_{y},n_{z}>0}\frac{n_{x}^{2}+n_{y}^{2}}{n_{x}^{2}+n_{y}^{2}+n_{z}^{2}}\cos^{2}\left(\frac{2\pi n_{z} z_{0}}{L}\right)\delta(\omega-\omega(n_{x},n_{y},n_{z})).
    \label{eq:pldos_out_of_plane}.
\end{align}
Again taking the large $L$ limit, converting (\ref{eq:pldos_out_of_plane}) to an integral, and using spherical polar coordinates to evaluate this integral we can reduce (\ref{eq:pldos_out_of_plane}) to,
\begin{equation}
	\rho_p(\perp,\boldsymbol{r}_0,\omega)=\frac{\omega^{2}}{\pi^{2}c^{3}}\int_{0}^{\pi/2}d\theta_{k}\sin^{3}(\theta_{k})\cos^{2}\left(\frac{\omega}{c}\cos(\theta_{k})z_0\right),\label{eq:out_of_plane_integral}
\end{equation}
the $z_0$ dependence of which is plotted in figure~\ref{fig:mirror_emission}d.
From equation (\ref{eq:out_of_plane_integral}) we see that an antenna pointing into or out of a mirror emits rather differently to one pointing in the mirror plane, especially very close to the mirror; in particular whilst the integrand of (\ref{eq:mirror_pldos}) tends to zero as the dipole approaches the mirror, the integrand of (\ref{eq:out_of_plane_integral}) tends to a maximum! 
Explicitly, the expression for $\rho_p$ tends to \text{double} the free space density of states
\begin{equation}
	\text{Close to mirror:}\qquad\rho_p(\perp,\boldsymbol{r}_0,\omega)\approx\frac{\omega^{2}}{\pi^{2}c^{3}}\int_{0}^{\pi/2}d\theta_{k}\sin^{3}(\theta_{k})=\frac{2\rho_{0}}{3}.
    \label{eq:out_of_plane_integral_close}
\end{equation}
We can again understand the implied doubling of the radiated power (\ref{eq:classical_power}) in terms of the image dipole induced within the mirror, shown in Fig.~\ref{fig:interface}. 
This image doubles the apparent length of the antenna. 
One would expect such a stretched antenna to radiate four times more power but since we only calculate the power radiated into the upper half space, this leads to the factor of two in Eq.~(\ref{eq:out_of_plane_integral_close}).
\par
Again, just as for the parallel dipole orientation, when the distance from the mirror is large ($\omega z_0/c\gg1$) then the partial local density of states reduces to the free space value (\ref{eq:free_space_continuum_ldos}) by the same argument as given in (\ref{eq:large_z_result})
\begin{equation}
	\text{Far from mirror:}\qquad\rho_p(\perp,\boldsymbol{r}_0,\omega)\approx\frac{\rho_{0}}{3}.
    \label{eq:out_of_plane_integral_far}
\end{equation}
\par
Having calculated the partial local density of states for the two antenna orientations (\ref{eq:mirror_pldos}) and (\ref{eq:out_of_plane_integral}) as a function of position and frequency, we can calculate the LDOS (\ref{eq:ldos_defn}) and the DOS (\ref{eq:dos_defn}). 
The LDOS is the sum of $\rho_p$ over the three possible antenna orientations
\begin{align}
	\rho_{l}(\boldsymbol{r}_{0},\omega)&=2\rho_{p}(\parallel,\boldsymbol{r}_{0},\omega)+\rho_{p}(\perp,\boldsymbol{r}_{0},\omega),\nonumber\\[10pt]
    &=\frac{\omega^{2}}{\pi^{2}c^{3}}\int_{0}^{\pi/2}d\theta_{k}\left[2\sin(\theta_{k})\sin^{2}\left(\frac{\omega}{c}\cos(\theta_{k})z_{0}\right)+\sin^{3}(\theta_{k})\cos\left(\frac{2\omega}{c}\cos(\theta_{k})z_{0}\right)\right]
    \label{eq:mirror_ldos}
\end{align}
as plotted in Figure~\ref{fig:mirror_emission}d. 
This plot shows that as one approaches the mirror the local density of states reduces to $\rho_{l}/3\rho_0 = 2/3$, meaning that were we to randomly choose our antenna orientation then we would find an average reduction in emission close to the mirror. 
This, however, misses the fact--evident in the more detailed quantity $\rho_p$--that emission for the two in--plane antenna orientations is zero at the mirror, whilst the out-of-plane emission is doubled.
\par
Finally, at the crudest level, we can compute the DOS as the volume average of (\ref{eq:mirror_ldos}),
\begin{align}
	\rho(\omega)&=\frac{2\omega^{2}}{\pi^{2}c^{3}}\int_{0}^{\pi/2}d\theta_{k}\sin(\theta_{k})\lim_{L\to\infty}\frac{1}{L}\int_{0}^{L}dz_0\sin^{2}\left(\frac{\omega}{c}\cos(\theta_{k})z_{0}\right),\nonumber\\
    &=\frac{\omega^{2}}{\pi^{2}c^{3}},
\end{align}
which is unchanged from the free space value (\ref{eq:dos_free_space}). This is simply because as a function of frequency we have the same continuum of modes that we had in free space.  The number density of states clearly misses the presence of the mirror altogether!  Were we to try to use $\rho(\omega)$ to predict the behaviour of an antenna of any orientation close to a mirror we would be completely wrong.  However, what this does show, is that were we to distribute a large collection of randomly oriented antennas over all space, the average emitted power would be the same as in free space.  This means that the oscillations of $\rho_l$ evident in figure~\ref{fig:mirror_emission}d are, on average, evenly spread around the free space value. 
%
%
%
\subsection{Emission inside a planar microcavity with perfect mirrors, Purcell, and Casimir}\label{sec:cavity_emission}
\par
The next simplest case is that of an antenna within a planar cavity, \textit{i.e.}, sandwiched between two parallel mirrors at $z=0$ and $z=h$. 
It is reasonably straightforward to generalize the above results to this case. 
Assuming a spatial period $L$ in the $x$--$y$ plane, and a quantization of the wavevector along the $z$ axis, the set of allowed wavevectors are
\begin{equation}
	\boldsymbol{k}(n_{x},n_{y},n_{z})=\frac{2\pi}{L}\left(n_{x}\boldsymbol{e}_{x}+n_{y}\boldsymbol{e}_{y}\right)+\frac{\pi n_{z}}{h}\boldsymbol{e}_{z},\label{eq:wave_vector_cavity}
\end{equation}
the corresponding polarization vectors that are orthogonal to the wavevector and to each other are
\begin{align}
	\boldsymbol{e}_{1}(n_{x},n_{y},n_{z})&=\frac{n_{x}\boldsymbol{e}_{y}-n_{y}\boldsymbol{e}_{x}}{\sqrt{n_{x}^{2}+n_{y}^{2}}},\nonumber\\
    \boldsymbol{e}_{2}(n_{x},n_{y},n_{z})&=\frac{1}{\sqrt{n_{x}^{2}+n_{y}^{2}+\left(\frac{L n_{z}}{2h}\right)^{2}}}\left[\sqrt{n_{x}^{2}+n_{y}^{2}}\boldsymbol{e}_{z}- \frac{L n_{z}}{2h}\frac{n_{x}\boldsymbol{e}_{x}+n_{y}\boldsymbol{e}_{y}}{\sqrt{n_{x}^{2}+n_{y}^{2}}}\right],\label{eq:polarization_cavity}
\end{align}
Imposing the boundary condition that the in--plane electric field vanishes on the surfaces of both mirrors, the allowed modes take the same form as for the case of a single mirror (\ref{eq:mirror_mode1}--\ref{eq:mirror_mode2}), but with modifications to the polarization and wavevector given by (\ref{eq:wave_vector_cavity}--\ref{eq:polarization_cavity}), as well as a change to the normalization prefactor, where one of the factors of $L$ must be replaced with a factor of $2h$ (which can be understood by comparing equations (\ref{eq:wave_vector_cavity}) and (\ref{eq:periodic_k})). 
The electric field is then given by,
\begin{align}
	\boldsymbol{\mathcal{E}}^{(1)}_{n_{x},n_{y},n_{z}}&=\frac{1}{\sqrt{2h L^{2}}}\left[\boldsymbol{e}_{1}(n_{x},n_{y},n_{z}){\rm e}^{{\rm i}\boldsymbol{k}(n_{x},n_{y},n_{z})\cdot\boldsymbol{r}}-\boldsymbol{e}_{1}(n_{x},n_{y},-n_{z}){\rm e}^{{\rm i}\boldsymbol{k}(n_{x},n_{y},-n_{z})\cdot\boldsymbol{r}}\right],\nonumber\\[10pt]
    \boldsymbol{\mathcal{E}}^{(2)}_{n_{x},n_{y},n_{z}}&=\frac{N_{n_{z}}}{\sqrt{2h L^{2}}}\left[\boldsymbol{e}_{2}(n_{x},n_{y},n_{z}){\rm e}^{{\rm i}\boldsymbol{k}(n_{x},n_{y},n_{z})\cdot\boldsymbol{r}}+\boldsymbol{e}_{2}(n_{x},n_{y},-n_{z}){\rm e}^{{\rm i}\boldsymbol{k}(n_{x},n_{y},-n_{z})\cdot\boldsymbol{r}}\right],
\end{align}
where $N_{n_{z}}$ is defined as given after equation (\ref{eq:mirror_mode1}).
\begin{figure}[ht!]
	\includegraphics[width=15.5cm]{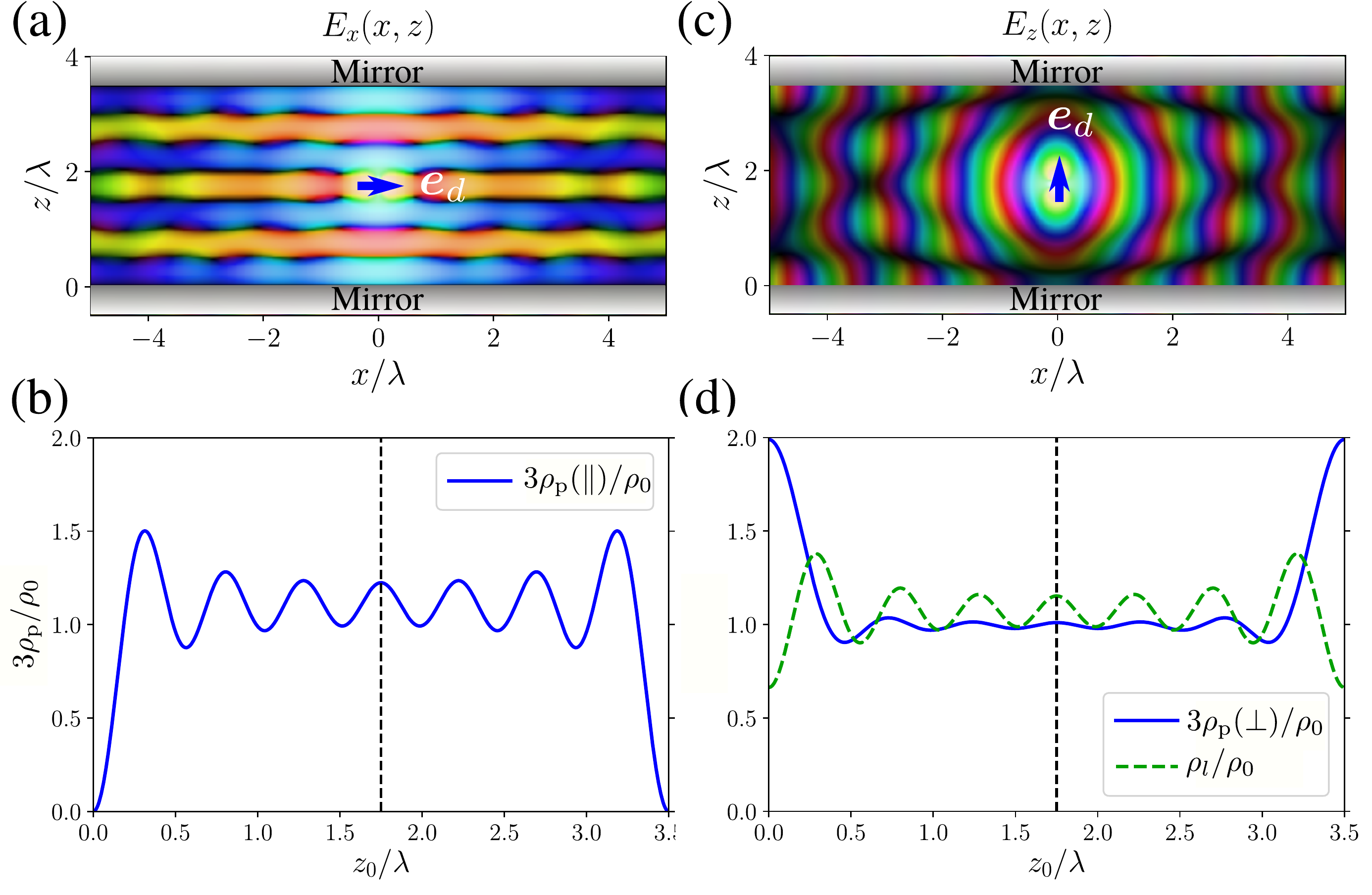}
	\caption{The partial local density of states in a planar cavity.  (a) The electric field component $E_x$ for an antenna positioned at $z_0=1.75\lambda$ in a cavity of width $h=3.5\lambda$ (colours and brightness as in figure~\ref{fig:dipole_free-space_near-mirror}). 
	(c) As in (a) but for an $\boldsymbol{e}_{z}$ oriented antenna. 
	(b) The partial local density of states (Eq.~(\ref{eq:rho_par_cavity})) for the same cavity width, for an antenna oriented in the plane of the mirrors. 
	(d) Same parameters, but for the perpendicular antenna orientation, with the green dashed curve giving the local density of states (\ref{eq:cavity_ldos}). 
	Data in panels (b) and (d) are normalized to the free-space number density of modes $\rho_0$ (Eq.~(\ref{eq:dos_free_space})).
	\label{fig:cavity_emission}}
\end{figure}
\par
Considering the same two antenna orientations as in section~\ref{sec:mirror_emission}, and performing the same $L\to\infty$ limiting procedure, after a few algebraic manipulations we obtain the partial local density of states for the parallel emitter orientation to be,
\begin{equation}
	\rho_{p}(\parallel,\boldsymbol{r}_{0},\omega)=\frac{\omega}{2\pi h c^{2}}\sum_{0<n_{z}<\frac{h\omega}{\pi c}}\left[1+\left(\frac{c \pi n_{z}}{\omega h}\right)^{2}\right]\sin^{2}\left(\frac{\pi n_{z} z_0}{h}\right)\label{eq:rho_par_cavity}
\end{equation}
and for the perpendicular one we find,
\begin{equation}
	\rho_{p}(\perp,\boldsymbol{r}_{0},\omega)=\frac{\omega}{\pi h c^{2}}\sum_{0<n_{z}<\frac{h\omega}{\pi c}}\left[1-\left(\frac{c\pi n_{z}}{h\omega}\right)^{2}\right]\cos^{2}\left(\frac{\pi n_{z} z_0}{h}\right)+\frac{\omega}{2\pi h c^{2}}.\label{eq:rho_perp_cavity}
\end{equation}
The two expressions (\ref{eq:rho_par_cavity}) and (\ref{eq:rho_perp_cavity}) are plotted for a particular cavity size $h=3.5\lambda$ in figure~\ref{fig:cavity_emission}b and~\ref{fig:cavity_emission}d. 
Panels a and c of this figure illustrate the emission pattern for a parallel and perpendicularly oriented antenna respectively, placed at the centre of the cavity.  
The dashed vertical line in panels b and d indicates the position at the centre of the cavity.
\par
A striking difference between $\rho_{p}$ for the two antenna orientations is evident when the frequency of the antenna is reduced below $\pi c/h$.  In this case the summations over $n_z$ in both (\ref{eq:rho_par_cavity}) and (\ref{eq:rho_perp_cavity}) reduce to zero. 
This means that while the PLDOS for the parallel orientation is zero, it becomes uniform in space and equal to $\omega/2\pi h c^{2}$ for the perpendicular orientation. 
This is because, however close two mirrors get there is always one mode, polarized along $\boldsymbol{e}_{z}$ and propagating in the $x$--$y$ plane, that can propagate in the gap. 
This capacitor-like mode is effectively two dimensional, being completely insensitive to the $z$ position between the mirrors~\cite{Gramotnev_NatPhot_2013_8_13}.
\par
From the PLDOS (\ref{eq:rho_par_cavity}) and (\ref{eq:rho_perp_cavity}) we can again derive an expression for the LDOS 
\begin{equation}
	\rho_l(\boldsymbol{r}_{0},\omega)=\frac{\omega}{\pi h c^{2}}\sum_{0<n_{z}<\frac{h\omega}{\pi c}}\left[1-\left(\frac{c\pi n_{z}}{h\omega}\right)^{2}\cos\left(\frac{2\pi n_{z} z_0}{h}\right)\right]+\frac{\omega}{2\pi h c^{2}},\label{eq:cavity_ldos}
\end{equation}
which is plotted as the red dashed line in figure~\ref{fig:cavity_emission}d, for the same fixed cavity width.  Again this quantity tells us that close to the cavity mirrors emission is suppressed, missing the fact that it is actually zero for two of the antenna orientations. 
Also, it tells us that for low enough frequencies a randomly oriented antenna will emit in a way that is independent of the position in the cavity.
\par
Performing a volume average of the LDOS (\ref{eq:cavity_ldos}) we obtain the number density of states 
\begin{align}
	\rho(\omega)&=\frac{\omega}{2\pi h c^{2}}+\frac{\omega}{\pi h c^{2}}\sum_{0<n_{z}<\frac{h\omega}{\pi c}}1,\nonumber\\
    &=\frac{\omega}{\pi h c^{2}}\left(\left\lfloor{\frac{h\omega}{\pi c}}\right\rfloor+\frac{1}{2}\right),\label{eq:cavity_ndos}
\end{align}
where `$\lfloor x \rfloor$' indicates the largest integer less than or equal to $x$. 
Unlike the case of a single mirror, this number density of states \textit{is} modified from its value in free space (\ref{eq:free_space_continuum_ldos}). 
In a cavity sandwiched between perfect mirrors, a collection of randomly oriented, randomly positioned antennas will on average emit differently compared to free space. 
However, as soon as the mirrors become partially transmitting (but still lossless), we have $\rho=\rho_0$ because, again,  we have the same continuum of modes as free space.

\par
It is useful at this point to make a connection between the density of states and another common way of characterizing the effect of changing the optical environment on spontaneous emission, the Purcell factor. 
Originally introduced by Purcell in the context of radio-frequency work~\cite{Purcell_PhysRev_1946_69_681}, the Purcell factor $F_P$, as it is now known, provides one measure of the effect of a cavity on the spontaneous emission process.
It is the ratio of the radiative emission rate with the cavity present~$\Gamma_{R(in~cavity)}$, and the radiative emission rate in the absence of the cavity~ $\Gamma_{R(no~cavity)}$. 
(The distinction between radiative and non-radiative rates is discussed in section~\ref{sec:rate}). 
To connect the Purcell factor with the density of states, we re-write equation~\eqref{eq:dos_defn} as
\begin{equation}
V \rho(\omega) = \sum_{n}\delta(\omega-\omega_n).\label{eq:dos_integ}
\end{equation}
Next we consider a cavity that has only one cavity resonance at frequency $\omega_c$. 
Further, we introduce loss/damping into our considerations, \textit{e.g.}, through the use of imperfect mirrors, something we have not done yet -- but will look at in detail in section~\ref{sec:non-perfect} below -- and recognize that our single resonance will have a spectral width $\Delta\omega_c$. 
Finally, if we assume the spectral width of our mode to be Lorentzian in form, and that the emission frequency is matched to the cavity resonance, \textit{i.e.}, $\omega_0=\omega_c$ then the density of states at the emission frequency can be written as~\cite{Fox_QO}
\begin{equation}
\rho(\omega_0) = \frac{2}{\pi \Delta\omega_c V}.
\label{eq:Purcell_a}
\end{equation}
Next we introduce the quality factor $Q$ of the cavity resonance through $Q=\omega/\Delta\omega$ to re-write~\eqref{eq:Purcell_a} as
\begin{equation}
\rho(\omega_0) = \frac{6Q}{\pi \omega_0 V}.
\label{eq:Purcell_b}
\end{equation}
We have multiplied this last equation by a factor of 3, as we assume here that the dipole moment of the emitter is aligned with the field so that there is no orientational averaging involved. 
With the Purcell factor defined as 
\begin{equation}
F_P \equiv \frac{\Gamma_{R(in~cavity)}}{\Gamma_{R(no~cavity)}},\label{eq:Purcell_c}
\end{equation}
we now take $\Gamma_{R(no~cavity)}$ to be the free space value $\Gamma_0$ given by~(\ref{eq:SPE_rate_DOS}), and use~(\ref{eq:SPE_rate_free}) for $\Gamma_{R(in~cavity)}$ to find 
\begin{equation}
F_P = \frac{\Gamma_Q}{\Gamma_0}= \frac{\pi \omega_c|\mathcal{P}|^2\rho(\omega_c)}{3\epsilon_0\epsilon\hbar}.\frac{3\pi \epsilon_0c^3\hbar}{\omega^3_c|\mathcal{P}|^2}= \frac{\pi^2 c^3}{\omega^2}\rho(\omega_c) = \frac{\rho(\omega_c)}{\rho_o(\omega_c)}.\label{eq:Purcell_bb}
\end{equation}
Combining (\ref{eq:Purcell_bb}) with (\ref{eq:Purcell_b}) we have
\begin{equation}
F_P = 6\pi Q c^3/V\omega^3.\label{eq:Purcell_bbb}
\end{equation}
If we now note that $(c/\omega)^3 = \lambda^3/8\pi^3n^3$, where $n$ in the refractive index of the medium inside the cavity, then we find the well-known expression for the Purcell factor~\cite{Fox_QO}, \textit{i.e.},
\begin{equation}
F_P = \frac{3}{4n^3}\frac{Q}{V}\left(\frac{\lambda}{n}\right)^3.
\label{eq:Purcell_d}
\end{equation}
Although often useful, and frequently used, the Purcell factor can be difficult to interpret - care is needed, for example in considering whether the single mode analysis of Purcell is appropriate~\cite{Koenderink_OL_2010_35_4208}, or in dealing with the concept of a mode volume in the presence of dissipation~\cite{Lalanne_LPR_2018_12_1700113}.

The discussion above on the density of states between perfect planar mirrors also has consequences for a different application to that of spontaneous emission control, namely the Casmir effect and the Casimir-Polder force~\cite{Casimir1948PR, Casimir1948Proc}. 
By introducing two parallel mirrors, we only allow optical states with momenta that 'fit' the distance between the mirrors and conversely, we exclude the many states from the continuum that do not fit. 
Here it is relevant to remind ourselves that the lowest excitation of any field mode has zero photons but still has a zero-point energy, and therefore a non-zero energy $\frac{1}{2} \hbar \omega$ that is attributed to vacuum fluctuations~\cite{Milonni}. 
Therefore, the exclusion of modes between the mirrors represents a change in total energy as compared to the vacuum outside, and thus the vacuum is effectively pushing on the mirrors. 
In a similar spirit, vacuum fluctuations also induce forces on two atoms in close proximity, or on an atom near a mirror, also known as Casimir-Polder forces. 
Another way to view these phenomena is to consider van der Waals dispersion forces, taking into account retardation effects. 

Now that we have introduced the idea of a cavity with loss we can investigate in more detail the consequences of changing the perfect mirrors we have looked at here for more realistic imperfect mirrors, that we discuss in section~\ref{sec:non-perfect}.

\subsection{Imperfect mirrors, imperfect cavities\label{sec:non-perfect}}
\subsubsection{Imperfect mirrors\label{sec:non-perfect-mirror}}
So far the discussion has been rather idealized, considering modifications to the PLDOS due to mirrors that reflect all incident waves at all frequencies. 
A real mirror is characterized by a pair of complex reflection coefficients $r_{1,2}(\omega,k_\parallel)$ for the two polarizations, TE and TM. 
These reflection coefficients are functions of both frequency $\omega$ and in-plane wavevector $k_{\parallel}$. 
In general a real mirror is also lossy, meaning that the inclusion of losses in the calculation of the PLDOS is easier in the Green function formalism than it is in the summation over modes, we therefore adopt the Green function approach here.
\begin{figure}[ht!]
\centering
\includegraphics[width=0.9\columnwidth]{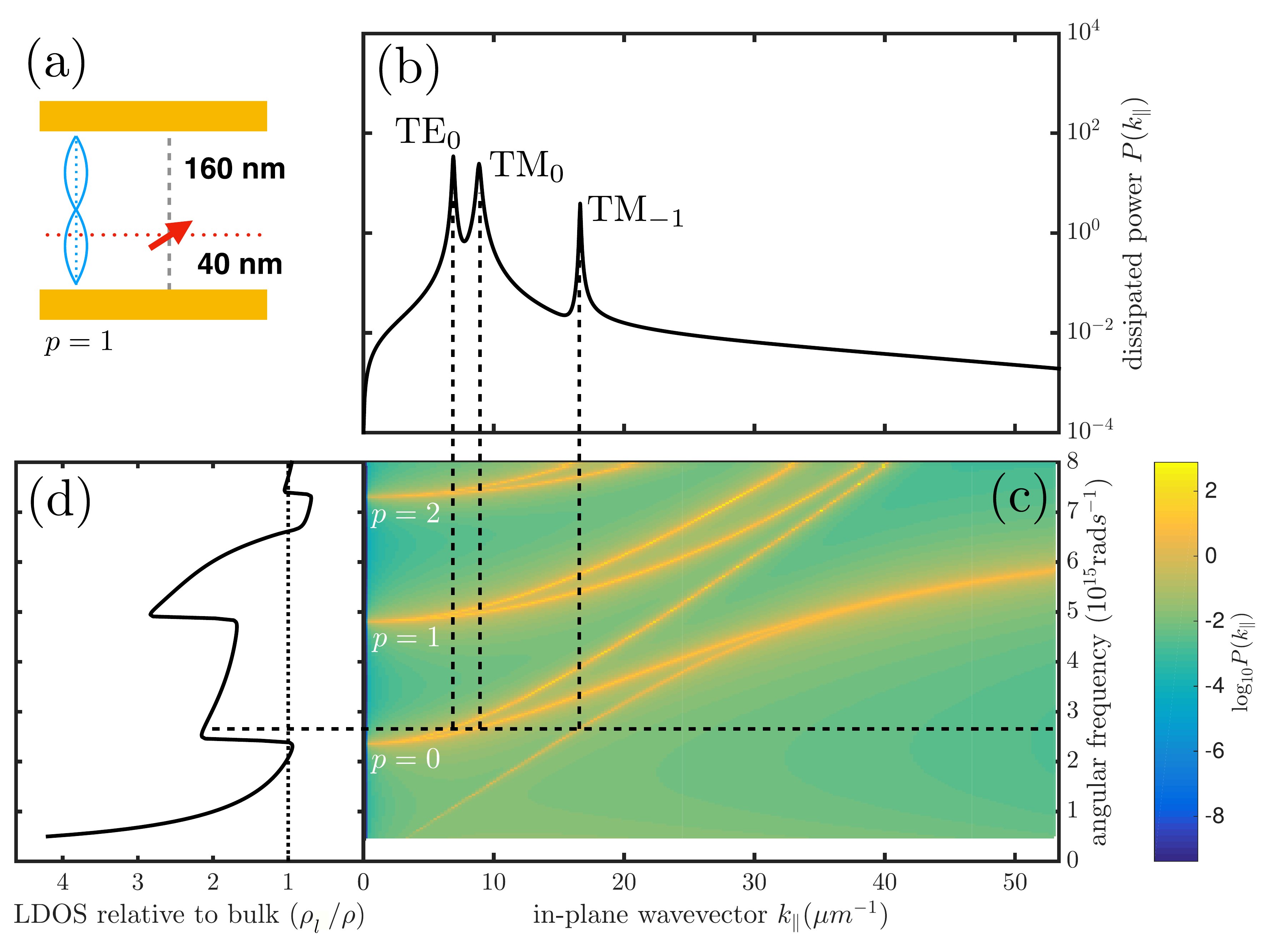}
\caption{\textbf{(a)} Two semi-infinite gold mirrors form a planar microcavity of height 200 nm, the emitter (red arrow indicates dipole moment) is located 40 nm from the lower mirror. 
The mirrors have a complex permittivity given by Eq.~(\ref{eq:mirror_permittivity}), with $\omega_p = 1.29\times10^{16}~\textrm{rad}~s^{-1}$ and $\gamma = 1.01\times10^{14}~\textrm{rad}~s^{-1}$. 
The blue lines indicate electric field amplitude of the second order $(p=2)$ cavity resonances. 
\textbf{(b)} Power dissipated by emitter at frequency $2.72\times10^{15}~\textrm{rad}~s^{-1}$ as a function of in--plane wavevector $k_{\parallel}$ (using the integrand of (\ref{eq:imperfect_cavity_ldos})), and the orientational averaged power from a quantum emitter (Eq.~\ref{eq:quantum_power_pre}). 
The dissipated power is here dominated by coupling of the emitter to three modes of the cavity, indicated by the dashed lines, where the $\textrm{TM}_{-1}$ mode corresponds to a coupled plasmon mode. 
\textbf{(c)} Power dissipation as a function of both emission frequency and in-plane wavevector. 
This represention shows the dispersion of the cavity modes. 
(d) Cavity LDOS (Eq.~\ref{eq:imperfect_cavity_ldos}) relative to that of a bulk homogeneous medium ($\epsilon = 2.49$). 
As the frequency is increased the local density of states increases sharply as each new set of modes is introduced. 
\label{fig:microcavity_figure}}
\end{figure}

We begin by finding the Green function corresponding to an emitter above a single lossy mirror and embedded in a homogeneous medium. We first decompose the Green function for a homogeneous medium (Eq.~\ref{eq:homogeneous_gf}) into a sum over plane waves as,
\begin{align}
	\overleftrightarrow{\boldsymbol{G}}_0(\boldsymbol{r},\boldsymbol{r}_0,\omega)&=\left[\frac{1}{{\rm n}^{2}k_0^{2}}\boldsymbol{\nabla}\otimes\boldsymbol{\nabla}+\boldsymbol{1}\right]\frac{{\rm e}^{{\rm i}{\rm n}k_0|\boldsymbol{r}-\boldsymbol{r}_{0}|}}{4\pi|\boldsymbol{r}-\boldsymbol{r}_{0}|},\nonumber\\
    &=\left[\frac{1}{{\rm n}^{2}k_0^{2}}\boldsymbol{\nabla}\otimes\boldsymbol{\nabla}+\boldsymbol{1}\right]\int\frac{d^{2}\boldsymbol{k}_{\parallel}}{(2\pi)^{2}}\frac{{\rm i}}{2k_z}{\rm e}^{{\rm i}k_z|z-z_0|}{\rm e}^{{\rm i}\boldsymbol{k}_{\parallel}\cdot(\boldsymbol{r}_{\parallel}-\boldsymbol{r}_{\parallel 0})},\label{eq:gf_decomposition}
\end{align}
where $k_z=\sqrt{{\rm n}^{2}k_0^{2}-k_{\parallel}^{2}}$. 
The above expression can be verified through direct substitution into Eq.~(\ref{eq:green_function_definition}).
To introduce the reflection from the surface, we note that just above the surface $(z_0 > z > 0)$, the above expression for the Green function takes the form,
\begin{equation}
	\overleftrightarrow{\boldsymbol{G}}_0(\boldsymbol{r},\boldsymbol{r}_0,\omega)=\sum_{\zeta=1,2}\int\frac{d^{2}\boldsymbol{k}_{\parallel}}{(2\pi)^{2}}\frac{{\rm i}\boldsymbol{e}_{\zeta}^{(-)}\otimes\boldsymbol{e}_{\zeta}^{(-)}}{2k_z}{\rm e}^{{\rm i}k_z(z_0-z)}{\rm e}^{{\rm i}\boldsymbol{k}_{\parallel}\cdot(\boldsymbol{r}_{\parallel}-\boldsymbol{r}_{\parallel 0})}~~(z_0 > z > 0),~~\label{eq:incident_field}
\end{equation}
where we used the identity $\boldsymbol{1} - \boldsymbol{e}_{\boldsymbol{k}}\otimes\boldsymbol{e}_{\boldsymbol{k}} = \sum_{\zeta=1,2}\boldsymbol{e}_{\zeta}\otimes\boldsymbol{e}_{\zeta}$ (which is simply $\hat{\boldsymbol{e}}_{x}\otimes\hat{\boldsymbol{e}}_{x}+\hat{\boldsymbol{e}}_{y}\otimes\hat{\boldsymbol{e}}_{y}+\hat{\boldsymbol{e}}_{z}\otimes\hat{\boldsymbol{e}}_{z}=\boldsymbol{1}$, expressed in a different basis), with the two unit vectors,
\begin{align}
	\boldsymbol{e}_{1}^{(\pm)}&=\frac{1}{k_{\parallel}}\left(k_x\boldsymbol{e}_{y}-k_y\boldsymbol{e}_{x}\right),\nonumber\\
    \boldsymbol{e}_{2}^{(\pm)}&=\frac{1}{{\rm n}k_0}\left(k_\parallel\boldsymbol{e}_{z}\mp k_z\boldsymbol{e}_{\boldsymbol{k}_{\parallel}}\right),
\end{align}
the former denoting `TE' polarization (the electric field is purely transverse to the mirror), and the latter `TM' polarization (the magnetic field is purely transverse to the mirror).
Evidently (\ref{eq:incident_field}) is a sum of waves of two polarizations incident on our mirror at $z=0$.
To include the effect of the mirror, we simply need to include the reflected part of the field, \textit{i.e.},
\begin{equation}
\overleftrightarrow{\boldsymbol{G}}(\boldsymbol{r},\boldsymbol{r}_0,\omega) = \overleftrightarrow{\boldsymbol{G}}_0(\boldsymbol{r},\boldsymbol{r}_0,\omega)+\sum_{\zeta=1,2}\int\frac{d^{2}\boldsymbol{k}_{\parallel}}{(2\pi)^{2}}\frac{{\rm i}\boldsymbol{e}_{\zeta}^{(+)}\otimes\boldsymbol{e}_{\zeta}^{(-)}}{2k_z}r_{\zeta}(\omega,k_{\parallel}){\rm e}^{{\rm i}k_z(z_0+z)}{\rm e}^{{\rm i}\boldsymbol{k}_{\parallel}\cdot(\boldsymbol{r}_{\parallel}-\boldsymbol{r}_{\parallel\,0})}.\label{eq:gf_modified}
\end{equation}
The PLDOS can now be calculated from its definition in terms of the Green function (\ref{eq:defn_pldos_green}), which gives,
\begin{equation}
	\rho_{\rm p}(\boldsymbol{e}_{d},\boldsymbol{r}_{0},\omega) = \frac{{\rm n}^{3}\omega^{2}}{3\pi^{2}c^{3}}+\frac{\omega{\rm n}^{2}}{\pi c^{2}}{\rm Im}\left[\sum_{\zeta=1,2}\int\frac{d^{2}\boldsymbol{k}_{\parallel}}{(2\pi)^{2}}\frac{{\rm i}(\boldsymbol{e}_{d}\cdot\boldsymbol{e}_{\zeta}^{(+)})(\boldsymbol{e}_{d}\cdot\boldsymbol{e}_{\zeta}^{(-)})}{k_z}r_{\zeta}(\omega,k_{\parallel}){\rm e}^{2{\rm i}k_z z_0}\right],\label{eq:pldos_imperfect_mirror}
\end{equation}
where we used the expression for the PLDOS in a homogeneous medium (\ref{eq:pldos_gf_fs}).
We can therefore see that the PLDOS can be broken up into two parts; one due to the homogeneous environment, the other due to the reflection.
In general the integral involving $r(\omega,k_\parallel)$ has to be evaluated numerically and can have quite a complicated dependence on the position and orientation of the emitter, see Fig~\ref{fig:microcavity_figure}.
It is difficult to say very much more about the general properties of (\ref{eq:pldos_imperfect_mirror}), except that the integral can be broken up into two rather different pieces 
\begin{multline}
	\rho_{\rm p}(\boldsymbol{e}_{d},\boldsymbol{r}_{0},\omega)=\frac{{\rm n}^{3}\omega^{2}}{3\pi^{2}c^{3}}\\
    +\frac{\omega{\rm n}^{2}}{\pi c^{2}}\sum_{\zeta=1,2}\int_0^{2\pi}\frac{d\theta_k}{2\pi}\int_0^{{\rm n}k_0}\frac{k_\parallel dk_\parallel}{2\pi}\frac{(\boldsymbol{e}_{d}\cdot\boldsymbol{e}_{\zeta}^{(+)})(\boldsymbol{e}_{d}\cdot\boldsymbol{e}_{\zeta}^{(-)})}{k_z}{\rm Re}\left[r_{\zeta}(\omega,k_{\parallel}){\rm e}^{2{\rm i}k_z z_0}\right]\\
    +\frac{\omega{\rm n}^{2}}{\pi c^{2}}\sum_{\zeta=1,2}\int_0^{2\pi}\frac{d\theta_k}{2\pi}\int_{{\rm n}k_0}^{\infty}\frac{k_\parallel dk_{\parallel}}{2\pi}\frac{(\boldsymbol{e}_{d}\cdot\boldsymbol{e}_{\zeta}^{(+)})(\boldsymbol{e}_{d}\cdot\boldsymbol{e}_{\zeta}^{(-)})}{\kappa_z}{\rm Im}\left[r_{\zeta}(\omega,k_{\parallel})\right]{\rm e}^{-2\kappa_z z_0},
    \label{eq:pldos_broken_integral}
\end{multline}
where $\kappa_z={\rm i}k_z=\sqrt{k_{\parallel}^{2}-{\rm n}^{2}k_0^{2}}$ is the decay constant in between the plates.  The first of these two integrals depends on the reflection of propagating waves ($k_z$ is real) that first impinge on the mirror before travelling off to infinity. 
The second integral is over evanescent (exponentially decaying) waves that are bound to the surface of the mirror: this term becomes increasingly important as the emitter is brought close to the surface. 
Interestingly, this evanescent contribution to the PLDOS is only non--zero when the reflection coefficient of the mirror has a non--zero imaginary part. 
Taking, for example an ideal mirror and a dipole oriented perpendicular to the mirror plane $\boldsymbol{e}_{d}=\boldsymbol{e}_{z}$, the emitter only couples to the TM polarization, which has a real valued reflection coefficient $r_2=+1$.
The PLDOS then has a contribution only from the propagating waves and may be reduced to 
\begin{align}
	\rho_{\rm p}(\boldsymbol{e}_{d},\boldsymbol{r}_{0},\omega)&=\frac{\omega^{2}{\rm n}^{3}}{2\pi^{2} c^{3}}\int_0^{1}dx\frac{x^{3}}{\sqrt{1-x^{2}}}\left[1+\cos\left(2\sqrt{1-x^2}{\rm n}k_0z_0\right)\right] \nonumber\\
    &=\frac{\omega^{2}{\rm n}^{3}}{\pi^{2} c^{3}}\int_0^{\pi/2}d\theta_k\sin^{3}(\theta_k)\cos^{2}\left(\cos(\theta_k)\frac{{\rm n}\omega}{c}z_0\right).\label{eq:pldos_imperfect_mirror_sc}
\end{align}
This is in agreement with equation (\ref{eq:out_of_plane_integral}) of section~\ref{sec:mirror_emission} for the case where ${\rm n}=1$.  More generally the reflection coefficients for electromagnetic waves incident onto an interface between a material with parameters $\epsilon_{1},\mu_1$, and $\epsilon_{2},\mu_2$ are:
\begin{align}
	r_{1}(\omega,k_\parallel)&=\frac{\mu_2\sqrt{\epsilon_1\mu_1 k_0^{2}-k_\parallel^{2}}-\mu_1\sqrt{\epsilon_2\mu_2 k_0^{2}-k_\parallel^{2}}}{\mu_2\sqrt{\epsilon_1\mu_1 k_0^{2}-k_\parallel^{2}}+\mu_1\sqrt{\epsilon_2\mu_2 k_0^{2}-k_\parallel^{2}}}\nonumber\\
    r_{2}(\omega,k_\parallel)&=\frac{\epsilon_2\sqrt{\epsilon_1\mu_1 k_0^{2}-k_\parallel^{2}}-\epsilon_1\sqrt{\epsilon_2\mu_2 k_0^{2}-k_\parallel^{2}}}{\epsilon_2\sqrt{\epsilon_1\mu_1 k_0^{2}-k_\parallel^{2}}+\epsilon_1\sqrt{\epsilon_2\mu_2 k_0^{2}-k_\parallel^{2}}}\label{eq:fresnel}
\end{align}
known as the \emph{Fresnel} coefficients~\cite{landau2004} (we assume that our emitter is embedded in medium $1$, where ${\rm Re}[\epsilon_{1}\mu_{1}]>0$). These reflection coefficients have a non--zero imaginary part, for example when (a) either medium $1$ or medium $2$ exhibits dissipation (complex $\epsilon_{1,2}$ and $\mu_{1,2}$); or (b) $\epsilon_{1}\mu_{1}k_0<k_\parallel<\epsilon_{2}\mu_{2}k_0$, corresponding to waves that are totally internally reflected in medium $2$, and exponentially decaying in medium $1$.

There are some situations where there is a very large contribution to the PLDOS from the second integral in (\ref{eq:pldos_broken_integral}).  When dissipation is significant, this enhancement of the amount of power leaving the dipole is not just due to the emission of radiation.  Power can be lost via both radiative and non--radiative decay (the latter typically leading to heating).  In section~\ref{sec:rate} we shall discuss how to estimate the radiative and non--radiative contributions to the local density of states when such non--radiative channels are important.  As an example, suppose we have an emitter very close to the surface and oriented so that $\boldsymbol{e}_{d}=\boldsymbol{e}_{z}$.  Due to its proximity to the surface, the second integral over $k_{\parallel}$ in (\ref{eq:pldos_broken_integral}) will include contributions from large values of $k_{\parallel}$, where the imaginary part of $r_{2}$ is approximately,
\begin{equation}
	\frac{k_{\parallel}}{{\rm n}k_0}\gg1:\;{\rm Im}[r_{2}]\sim{\rm Im}\left[\frac{\epsilon_{2}-\epsilon_{1}}{\epsilon_{2}+\epsilon_{1}}\right].
\end{equation}
Therefore when $|\epsilon_{2}+\epsilon_{1}|\ll1$, there will be a very large contribution to the PLDOS from evanescent waves, and this condition is met when medium $2$ is metallic, \textit{i.e.}, (${\textrm Re}[\epsilon_{2}]<0$). 
The condition $|\epsilon_{2}+\epsilon_{1}|=0$, is the condition for the asymptotic limit of the surface plasmon dispersion. 
There is consequently a high density of available modes for emission. 
This indicates that the power radiated by an emitter can be drastically modified in the vicinity of a surface, and this is the origin of effects such as the enhanced fluorescence reported in~\cite{tam2007}. 
\par
Note that the example of emission between parallel plates illustrates the general decomposition (\ref{eq:G0Gs}), where the Green function at a given point in a medium can be written as the sum of two parts, $\overleftrightarrow{\boldsymbol{G}}=\overleftrightarrow{\boldsymbol{G}}_0+\overleftrightarrow{\boldsymbol{G}}_S$ (equation \eqref{eq:gf_modified} is another example).
\par

\subsubsection{Planar microcavity with imperfect mirrors\label{sec:non-perfect-cavity}}
As a final example, consider a cavity made of identical imperfect mirrors separated by a distance $h$.
We can determine the Green function in this cavity geometry in an analogous fashion to the case of a single mirror, the difference here is that we need to consider multiple reflections. 
For instance, the reflection coefficient appearing in (\ref{eq:gf_modified}) must include the effect of the wave reflecting from the lower surface then again from the upper one, reflecting a second time from the lower surface and so on.  We can sum these multiple reflections as a geometric series.  For example, the waves emitted down onto the surface at $z=0$ undergo the following sequence of reflections,
\begin{multline}
	\boldsymbol{e}_{\zeta}^{(-)}{\rm e}^{-{\rm i}k_z (z-z_0)}+\boldsymbol{e}_{\zeta}^{(+)}r_{\zeta}{\rm e}^{{\rm i}k_z (z+z_0)}+\boldsymbol{e}_{\zeta}^{(-)}r_{\zeta}^{2}{\rm e}^{{\rm i}k_z (2h+z_0-z)}+\boldsymbol{e}_{\zeta}^{(+)}r_{\zeta}^{3}{\rm e}^{{\rm i}k_z (z+z_0+2h)}+\dots\\
    =\boldsymbol{e}_{\zeta}^{(-)}{\rm e}^{-{\rm i}k_z (z-z_0)}+\frac{\boldsymbol{e}^{(+)}_{\zeta}r_{\zeta}{\rm e}^{{\rm i}k_z(z+z_0)}}{1-r_{\zeta}^{2}{\rm e}^{2{\rm i}k_z h}}+\frac{\boldsymbol{e}^{(-)}_{\zeta}r_{\zeta}^{2}{\rm e}^{{\rm i}k_z(2h+z_0-z)}}{1-r_{\zeta}^{2}{\rm e}^{2{\rm i}k_z h}},\label{eq:reflections_1}
\end{multline}
and similarly the waves emitted upwards onto the surface at $z=h$ undergo the sequence,
\begin{multline}
	\boldsymbol{e}_{\zeta}^{(+)}{\rm e}^{{\rm i}k_z (z-z_0)}+\boldsymbol{e}_{\zeta}^{(-)}r_{\zeta}{\rm e}^{{\rm i}k_z (2h-z_0-z)}+\boldsymbol{e}_{\zeta}^{(+)}r_{\zeta}^{2}{\rm e}^{{\rm i}k_z (2h-z_0+z)}+\boldsymbol{e}_{\zeta}^{(-)}r_{\zeta}^{3}{\rm e}^{{\rm i}k_z (4h-z_0-z)}+\dots\\
    =\boldsymbol{e}_{\zeta}^{(+)}{\rm e}^{{\rm i}k_z (z-z_0)}+\frac{\boldsymbol{e}_{\zeta}^{(-)}r_{\zeta}{\rm e}^{{\rm i}k_z(2h-z_0-z)}}{1-r_{\zeta}^{2}{\rm e}^{2{\rm i}k_z h}}+\frac{\boldsymbol{e}_{\zeta}^{(+)}r_{\zeta}^{2}{\rm e}^{{\rm i}k_z(2h-z_0+z)}}{1-r_{\zeta}^{2}{\rm e}^{2{\rm i}k_z h}},\label{eq:reflections_2}
\end{multline}
Identifying the final two terms in equations (\ref{eq:reflections_1}--\ref{eq:reflections_2}) as the contributions to the scattered Green function $\overleftrightarrow{\boldsymbol{G}}_{S}$, the full electromagnetic Green function in a cavity composed of imperfect mirrors can be written as,
\begin{multline}
\overleftrightarrow{\boldsymbol{G}}(\boldsymbol{r},\boldsymbol{r}_0,\omega)=\overleftrightarrow{\boldsymbol{G}}_0(\boldsymbol{r},\boldsymbol{r}_0,\omega)
\\
+\sum_{\zeta=1,2}\int\frac{d^{2}\boldsymbol{k}_{\parallel}}{(2\pi)^{2}}\frac{{\rm i}{\rm e}^{{\rm i}\boldsymbol{k}_{\parallel}\cdot(\boldsymbol{r}_{\parallel}-\boldsymbol{r}_{\parallel\,0})}}{2k_z}\left[\frac{\boldsymbol{e}_{\zeta}^{(+)}\otimes\boldsymbol{e}_{\zeta}^{(-)}r_{\zeta}{\rm e}^{{\rm i}k_z(z+z_0)}}{1-r_{\zeta}^{2}{\rm e}^{2{\rm i}k_z h}}+\frac{\boldsymbol{e}_{\zeta}^{(-)}\otimes\boldsymbol{e}_{\zeta}^{(-)}r_{\zeta}^{2}{\rm e}^{{\rm i}k_z(2h+z_0-z)}}{1-r_{\zeta}^{2}{\rm e}^{2{\rm i}k_z h}}\right]\\
+\sum_{\zeta=1,2}\int\frac{d^{2}\boldsymbol{k}_{\parallel}}{(2\pi)^{2}}\frac{{\rm i}{\rm e}^{{\rm i}\boldsymbol{k}_{\parallel}\cdot(\boldsymbol{r}_{\parallel}-\boldsymbol{r}_{\parallel\,0})}}{2k_z}\left[\frac{\boldsymbol{e}_{\zeta}^{(-)}\otimes\boldsymbol{e}_{\zeta}^{(+)}r_{\zeta}{\rm e}^{{\rm i}k_z(2h-z_0-z)}}{1-r_{\zeta}^{2}{\rm e}^{2{\rm i}k_z h}}+\frac{\boldsymbol{e}_{\zeta}^{(+)}\otimes\boldsymbol{e}_{\zeta}^{(+)}r_{\zeta}^{2}{\rm e}^{{\rm i}k_z(2h-z_0+z)}}{1-r_{\zeta}^{2}{\rm e}^{2{\rm i}k_z h}}\right].\label{eq:cavity_gf}
\end{multline}
As a general point it is worth noting the possibility of zeros in the denominators $1-r_{\zeta}^{2}{\rm e}^{2{\rm i}k_z h}$ of the above expressions.  These singular contributions to the integrand come from the series of multiple reflections (\ref{eq:reflections_1}--\ref{eq:reflections_2}) between the mirrors, and correspond to modes that are bound between the two mirrors.  However, for lossy systems these singular points always occur at complex frequencies.  From the definition (\ref{eq:pldos_defn}), and its Green function for (\ref{eq:defn_pldos_green}), we can thus derive the following expression for the PLDOS 
\begin{multline}
\rho_{\rm p}(\boldsymbol{e}_{d},\boldsymbol{r}_{0},\omega)=\frac{\omega^{2}{\rm n}^{3}}{3\pi^{2} c^{3}}
\\
+\frac{\omega{\rm n}^{2}}{\pi c^{2}}{\rm Re}\left\{\sum_{\zeta=1,2}\int\frac{d^{2}\boldsymbol{k}_{\parallel}}{(2\pi)^{2}}\frac{1}{k_z}\left[\frac{(\boldsymbol{e}_{d}\cdot\boldsymbol{e}_{\zeta}^{(+)})(\boldsymbol{e}_{\zeta}^{(-)}\cdot\boldsymbol{e}_{d})r_{\zeta}{\rm e}^{2{\rm i}k_z z_0}}{1-r_{\zeta}^{2}{\rm e}^{2{\rm i}k_z h}}+\frac{(\boldsymbol{e}_{d}\cdot\boldsymbol{e}_{\zeta}^{(-)})^{2}r_{\zeta}^{2}{\rm e}^{2{\rm i}k_z h}}{1-r_{\zeta}^{2}{\rm e}^{2{\rm i}k_z h}}\right]\right\}\\
+\frac{\omega{\rm n}^{2}}{\pi c^{2}}{\rm Re}\left\{\sum_{\zeta=1,2}\int\frac{d^{2}\boldsymbol{k}_{\parallel}}{(2\pi)^{2}}\frac{1}{k_z}\left[\frac{(\boldsymbol{e}_{d}\cdot\boldsymbol{e}_{\zeta}^{(-)})(\boldsymbol{e}_{\zeta}^{(+)}\cdot\boldsymbol{e}_{d})r_{\zeta}{\rm e}^{2{\rm i}k_z(h-z_0)}}{1-r_{\zeta}^{2}{\rm e}^{2{\rm i}k_z h}}+\frac{(\boldsymbol{e}_{d}\cdot\boldsymbol{e}_{\zeta}^{(+)})^{2}r_{\zeta}^{2}{\rm e}^{2{\rm i}k_z h}}{1-r_{\zeta}^{2}{\rm e}^{2{\rm i}k_z h}}\right]\right\},
\label{eq:imperfect_cavity_pldos}
\end{multline}
which is rather complicated to evaluate.  If---as in figure~\ref{fig:microcavity_figure}---the orientation of the emitter is randomly varying then the LDOS governs the emission, and Eq.~(\ref{eq:imperfect_cavity_pldos}) simplifies to
\begin{multline}
\frac{\rho_{\rm l}(\boldsymbol{r}_{0},\omega)}{{\rm n}^{3}\rho_0}=1+\frac{1}{2}{\rm Re}\int_{0}^{\infty}\frac{u du}{\sqrt{1-u^{2}}}\bigg[\frac{r_{1}\left[{\rm e}^{2{\rm i}k_z z_0}+{\rm e}^{2{\rm i}k_z(h-z_0)}\right]+2r_{1}^{2}{\rm e}^{2{\rm i}k_z h}}{1-r_{1}^{2}{\rm e}^{2{\rm i}k_z h}}\\
+\frac{\left(2u^{2}-1\right)r_{2}\left[{\rm e}^{2{\rm i}k_z z_0}+{\rm e}^{2{\rm i}k_z(h-z_0)}\right]+2r_{2}^{2}{\rm e}^{2{\rm i}k_z h}}{1-r_{2}^{2}{\rm e}^{2{\rm i}k_z h}}\bigg],\label{eq:imperfect_cavity_ldos}
\end{multline}
where we have written $k_{\parallel}={\rm n}k_0 u$. 
After a considerable translation of notation (including a minus sign difference in the definition of $r_2$) and an application of the identity
\begin{equation}
	1={\rm Re}\int_{0}^{\infty}\frac{u d u}{\sqrt{1-u^{2}}},
\end{equation}
expression (\ref{eq:imperfect_cavity_ldos}) (for unit quantum efficiency, see Sec~\ref{sec:rate}) is in agreement with the expressions given by Chance, Prock and Silbey~\cite{Chance_AdvChemPhys_1978_37_1} based on the damping of the motion of the current in a radiating antenna.
Figure~\ref{fig:microcavity_figure} shows an evaluation of this quantity for a randomly oriented emitter in a gold microcavity where $\mu_{1,2}=1$, $\epsilon_{1}={\rm n}^{2}=2.49$ (appropriate for the organic matrix in which the Eu$^{3+}$ ions reside), and
\begin{equation}
	\epsilon_{2}(\omega)=1+\frac{\omega_p^{2}}{\omega^{2}+{\rm i}\omega\gamma}.\label{eq:mirror_permittivity}
\end{equation}
\noindent where $\omega_p$ is the plasma frequency for the gold, and $\gamma$ is the damping constant~\cite{Fox_OPS}
As this figure shows, the main contribution to the LDOS (and to the PLDOS for that matter) comes from the points in frequency and wavevector where $|1-r_{1,2}^{2}{\rm e}^{2{\rm i}k_z h}|\sim 0$, which are the equivalent of the cavity modes (\ref{eq:rho_par_cavity}--\ref{eq:rho_perp_cavity}) included in the PLDOS for a cavity made of a pair of perfect mirrors.   This is the final example we shall consider here before we treat the quantum mechanics of a small emitter.
\par
Given that the power being emitted from the antenna is proportional to the partial local density of states, which is in turn proportional to the imaginary part of the Green function, the radiated power relative to a homogeneous environment is given by,
\begin{equation}
	\text{Relative radiated power}=\frac{3\rho_{\rm p}(\boldsymbol{e}_{d},\boldsymbol{r}_{0},\omega)}{{\rm n}^{3}\rho_{0}(\omega)}=1+\frac{{\rm Im}\left[\boldsymbol{e}_{d}\cdot\overleftrightarrow{\boldsymbol{G}}_S(\boldsymbol{r}_0,\boldsymbol{r}_0,\omega)\cdot\boldsymbol{e}_{d}\right]}{{\rm Im}\left[\boldsymbol{e}_{d}\cdot\overleftrightarrow{\boldsymbol{G}}_0(\boldsymbol{r}_0,\boldsymbol{r}_0,\omega)\cdot\boldsymbol{e}_{d}\right]},\label{eq:relradpow}
\end{equation}
where we used the expression for the partial local density in a homogeneous environment (\ref{eq:pldos_gf_fs}).  The denominator in the fraction on the right of (\ref{eq:relradpow}) was already calculated as $k_0{\rm n}/6\pi$ in (\ref{eq:ImG0}), and the quantity $\boldsymbol{e}_{d}\cdot\overleftrightarrow{\boldsymbol{G}}_S(\boldsymbol{r}_0,\boldsymbol{r}_0,\omega)\cdot\boldsymbol{e}_{d}$ is proportional to the field that is scattered from the environment and acts back on the dipole.  Using our definition for the electric field in terms of the Green function (\ref{eq:E_field_Gf}), for the case of a point dipole $\tilde{\boldsymbol{j}}=-{\rm i}\omega\boldsymbol{e}_{d}\tilde{d}\delta^{(3)}(\boldsymbol{r}-\boldsymbol{r}_{0})$, this scattered field is given by,
\begin{equation}
\boldsymbol{E}_{S}=\mu_0\mu\omega^{2}\tilde{d}\,\overleftrightarrow{\boldsymbol{G}}_S(\boldsymbol{r}_0,\boldsymbol{r}_0,\omega)\cdot\boldsymbol{e}_{d},
\end{equation}
and thus the relative radiated power (\ref{eq:relradpow}) is given by
\begin{equation}
	\text{Relative radiated power}=1+\frac{6\pi\epsilon_0}{\mu \tilde{d}k_0^{3}{\rm n}}{\rm Im}\left[\boldsymbol{e}_{d}\cdot\boldsymbol{E}_{S}\right],\label{eq:relradpow2}
\end{equation}
where $\tilde{d}$ has been chosen as real valued. 
Note that in addition to the amplitudes of $\boldsymbol{e}_{d}$ and $\boldsymbol{E}_{S}$, their relative phase and orientation are important. 
In non--magnetic media $\mu=1$, and in c.g.s units $\epsilon_0=1/4\pi$, so that (\ref{eq:relradpow2}) is the same expression given by Chance, Prock and Silbey~\cite{Chance_AdvChemPhys_1978_37_1}, for the case of unit quantum efficiency, we will return to discuss quantum efficiency of the emitter in section 8.

\subsection{Calculation of densities of states}
Having discussed the intricacies of the various densities of states and how they can be obtained from experimental observations, and conversely, how the densities of states serve to interpret experimental observations (which we will explore further in Section~\ref{sec:rate}), it is useful to briefly outline some of the convenient methods to compute the densities of states for a number of common nanophotonic systems studied in experiments. 

\begin{table}[ht!]
\caption{\label{tab:Numerical_methods}
Examples of typical nanophotonic systems studied in experiments listed with suitable theoretical and numerical methods to compute the relevant ((P)L)DOS. 
Entries in the table indicate which methods are suited for which system, with the relevant equation and an exemplary reference. 
DG stands for discrete Galerkin methods, FDTD for finite difference time domain, FEM for finite-element method.\\ }
\begin{centering}
\begin{tabu} to 0.98\textwidth { | X[l] | X[l] | X[l] | }
\hline 
\textbf{System \textbackslash~~Method} & Mode counting & Green Function; Fields  \tabularnewline
\hline 
Mirror, lossless & Eqs.~\ref{eq:pldos_in_plane},~\ref{eq:pldos_out_of_plane} & Eq.~\ref{eq:defn_pldos_green}; Ref.~\cite{Chance_AdvChemPhys_1978_37_1} \tabularnewline
\hline 
Mirror, lossy & Not possible  & Eq.~\ref{eq:defn_pldos_green}; Ref.~\cite{Chance_AdvChemPhys_1978_37_1}  \tabularnewline
\hline 
Planar cavity \\ lossless mirrors & Eq.~\ref{eq:rho_par_cavity}  & Ref.~\cite{Haroche_1992_CQED} \tabularnewline
\hline 
Planar cavity \\ lossy mirrors & Not possible & Eq.~\ref{eq:imperfect_cavity_pldos}   \tabularnewline
\hline 
Sphere, lossless & ...  & Multipoles~\cite{Chew_PRA_1988_38_3410, Hafner_COMPEL_1983_2_1} \tabularnewline
\hline 
Sphere, lossy & Not possible & Green~\cite{tai1993},  DG~\cite{Busch2011LPR}, FDTD~\cite{oskooi2010}, FEM~\cite{Monk2003book} \tabularnewline
\hline 
Infinite photonic crystal 2D, 3D & Eq.~\ref{eq:pldos_defn}, plane wave expansion~\cite{Johnson_2001_OE}  & DG~\cite{Busch2011LPR}, FEM~\cite{Monk2003book}  \tabularnewline
\hline 
Finite photonic crystal 2D, 3D & Ref.~\cite{Hasan2018PRL} & DG~\cite{Busch2011LPR}, FDTD~\cite{oskooi2010}, FEM~\cite{Monk2003book, Devashish2017PRB} \tabularnewline
\hline
Plasmonic surface lattice & Not possible  & FDTD~\cite{Guo_Optica_2016_3_289}
\tabularnewline
\hline
\end{tabu}
\par\end{centering}
\end{table}
The main classes of methods listed in Table~\ref{tab:Numerical_methods} are distinguished as on one hand mode counting methods, and on the other hand methods based on Green functions or direct expressions of fields. 
While there are subtle and not-so subtle differences between Green function and direct field methods, we decide to lump them into one category, since Eq.~(\ref{eq:field_from_G}) illustrates nicely that there is a direct relation between the electric field $\boldsymbol{E}$ and the Green function $\overleftrightarrow{\boldsymbol{G}}$, whereas the integration over all space entails interesting complications whenever methods are put into practical numerical code. 
Table~\ref{tab:Numerical_methods} lists the nanophotonic systems that we consider, which we will distinguish as "lossless" to indicate a system with a purely real dielectric function, and "lossy" is meant to indicate a system with a complex dielectric function. 
In case of a mirror (both lossless and lossy) and a planar microcavity (both lossless and lossy), the densities of states and emission rates have been discussed extensively in the previous sections, hence Table~\ref{tab:Numerical_methods} lists the relevant equations. 
For the lossy mirror and lossy planar cavity, we have also seen that no mode counting pertains, therefore these entries are left empty, as is also the case for a lossy sphere, and a plasmonic surface lattice~\cite{Wang_MatToday_2018_21_303}. 
For a dielectric sphere, mode counting should in principle be feasible but it is likely rather tedious, therefore the entry is left open. 
For the cases that we do not discuss explicitly, such as a sphere, a photonic crystals and a plasmonic surface lattice, we provide a few references where interested readers can find entry points into a relevant numerical method. 

\section{The quantum dipole emitter and the PLDOS}\label{sec:quantum emitter}
\subsection{The quantum viewpoint}\label{quantum viewpoint}
\par
An asymmetric distribution of charge around the nucleus of an atom leads to a dipole moment (or a higher order multipole).  An oscillation of this dipole moment will give rise to the emission of radiation, just as in the case of the classical antenna described above.  However, atomic systems are subject to quantum mechanical laws not classical ones, and atoms do not continuously radiate.  The electrons within the quantum mechanical atom can only occupy discrete energy levels.  The existence of a lowest energy level is what stops matter from radiating away to nothing.
Yet despite the gulf between classical and quantum theories, the process of radiation from an atom undergoing a transition between atomic levels remains extraordinarily similar to that governing the radiation from a classical antenna (as is evident in formulae (\ref{eq:intro_classical_power}) and (\ref{eq:intro_quantum_power})).  In this section we illustrate the role of the density of states in the emission of electromagnetic radiation from an atom.  In a later section (Sec~\ref{sec:the_difference}) we then elaborate the subtle differences between the classical and quantum mechanical descriptions of the process.
\par
In quantum mechanics we have two kinds of objects: states $|\psi\rangle$ (which describe how the system is configured) and operators e.g. $\hat{H}$ (which represent the results of measurements on the states).  There is a curious ambiguity about the time dependence of these objects.  As undergraduates we are usually taught that the Schr\"odinger equation governs the time dependence of a quantum system,
\begin{equation}
	\hat{H}|\psi(t)\rangle={\rm i}\hbar\frac{\partial |\psi(t)\rangle}{\partial t}.\qquad\text{[S]}\label{eq:schrodinger}
\end{equation}
The time dependence of the state is thus dictated by the form of the Hamiltonian operator $\hat{H}$.
This is called the \emph{Schr\"odinger picture} (when necessary here indicated by [S]).  In this picture operators are typically time independent objects. But there is no physical reason why this has to be so: we never directly measure either the states or the operators, all measured quantities are matrix elements, which are combinations of the two.  For example the average dipole moment is given by $\langle\psi(t)|\hat{d}|\psi(t)\rangle$.  The time derivative of such a matrix element can be calculated in the Schro\"dinger picture using the Schr\"odinger equation given above (\ref{eq:schrodinger}) as follows
\begin{align}
    \frac{d}{dt}\langle\psi(t)|\hat{d}|\psi(t)\rangle&=\langle\dot{\psi}(t)|\hat{d}|\psi(t)\rangle+\langle\psi(t)|\hat{d}|\dot{\psi}(t)\rangle\nonumber\\
    &=\frac{{\rm i}}{\hbar}\langle\psi(t)|\hat{H}\hat{d}-\hat{d}\hat{H}|\psi(t)\rangle.\qquad\text{[S]}\label{eq:schrodinger_matrix_element}
\end{align}
Now, because neither state nor operators are ever directly measured we can take an alternative point of view about the time dependence of a quantum system, and instead take the state to be time independent, writing (\ref{eq:schrodinger_matrix_element}) as
\begin{equation}
    \frac{d}{dt}\langle\psi|\hat{d}(t)|\psi\rangle=\frac{{\rm i}}{\hbar}\langle\psi|\hat{H}\hat{d}(t)-\hat{d}(t)\hat{H}|\psi\rangle.\qquad\text{[H]}\label{eq:heisenberg_matrix_element}
\end{equation}
In this case the dipole moment operator $\hat{d}$ for our atomic system is the time dependent object and obeys the equation of motion
\begin{equation}
	\frac{d}{dt}~\hat{d}(t)=\frac{\rm i}{\hbar}\left[\hat{H},\hat{d}(t)\right].\qquad\text{[H]}\label{eq:heisenberg}
\end{equation}
This version of the time evolution is called the \emph{Heisenberg picture} (when necessary here indicated by [H]). Provided an operator is time independent in the Schr\"odinger picture, equation (\ref{eq:heisenberg}) governs its time dependence in the Heisenberg picture. 
It is purely a matter of taste or convenience which picture one chooses to use.  The advantage of using the Heisenberg picture in the present work is that it is more reminiscent of classical physics, and so easier to connect with the discussion of section~\ref{sec:classical-dipole}.
\par
To illustrate this connection with classical physics take just one of the simple harmonic oscillators $c_n$ into which we decomposed the classical electromagnetic field (\ref{eq:driven_oscillator}).  In the absence of anything else the Hamiltonian operator for this oscillator will be given by the usual sum of kinetic and potential energy terms
\begin{equation}
    \hat{H}=\frac{1}{2\epsilon_{0}}\left[\hat{\pi}_{n}^{2}+\epsilon_{0}^{2}\omega_{n}^{2}\hat{c}_{n}^{2}\right].\label{eq:lone_oscillator}
\end{equation}
The operator $\hat{\pi}_{n}$ is the momentum of the oscillator, and as in our classical expression (\ref{eq:driven_oscillator}), the permittivity of free space $\epsilon_0$ plays the role of the oscillator mass.  Using the Heisenberg equation of motion (\ref{eq:heisenberg}) and the Hamiltonian (\ref{eq:lone_oscillator}) to calculate the time derivatives of the oscillator momentum and displacement operators one finds that the form of the operator equations of motion are exactly the same as in classical physics
\begin{align}
    \frac{d\hat{c}_{n}}{dt}&=\frac{{\rm i}}{\hbar}[\hat{H},\hat{c}_{n}]=\frac{\hat{\pi}_{n}}{\epsilon_0}\nonumber\\
    \frac{d\hat{\pi}_{n}}{dt}&=\frac{{\rm i}}{\hbar}[\hat{H},\hat{\pi}_{n}]=-\epsilon_0\omega_{n}^{2}\hat{c}_{n}\label{eq:eqm_example}
\end{align}
where we used the usual commutation relation between position and momentum
\begin{equation}
	\left[\hat{c}_{n},\hat{\pi}_{m}\right]={\rm i}\hbar\delta_{nm}.\label{eq:oscillator_commutation}
\end{equation}
Combining the two equations of motion (\ref{eq:eqm_example}) into a single equation for the oscillator amplitude $\hat{c}_{n}$, we then have the familiar Newtonian form for simple harmonic motion
\begin{equation}
    \frac{d^{2}\hat{c}_{n}}{d t^{2}}+\omega_{n}^{2}\hat{c}_{n}=0.\label{eq:eqm_example2}
\end{equation}
Equations (\ref{eq:eqm_example}--\ref{eq:eqm_example2}) show that the Heisenberg picture is, at least for simple harmonic oscillators, equivalent to `classical physics with hats on'.  Using this picture we shall thus show that the quantum mechanical calculation of the radiation from an atom is very similar to our classical calculation of the radiation from a small antenna.
%
\subsection{The Hamiltonian}
\par
The Hamiltonian of our system consists of three pieces: an atomic part, an electromagnetic field part, and a part describing the interaction between the atom and the field
\begin{equation}
	\hat{H}=\hat{H}_{\rm atom}+\hat{H}_{\rm field}+\hat{H}_{\rm int}.\label{eq:full_hamiltonian}
\end{equation}
There is no need to worry about the distinction between the \emph{total} Hamiltonian in the Schr\"odinger and Heisenberg pictures; since it commutes with itself, its time derivative (\ref{eq:heisenberg}) is zero and it is thus the same time independent object in both pictures.
\par
In our classical calculation we assumed that the current flowing through the antenna was somehow fixed, ignoring the influence of the emitted field on the current.  Although in our quantum mechanical calculation we shall also assume the field does not significantly perturb the atom, we shall now calculate the dynamics of both the atom and the field.  The simplest model of an atom (sketched in Fig~\ref{fig:cartoon-emission}), assumes there are only two possible energy levels $E_{\rm at}=\pm\hbar\omega/2$, corresponding to a ground state $|0\rangle_{\rm at}$ and an excited state $|1\rangle_{\rm at}$.  We write the Hamiltonian of such a model atom as
\begin{equation}
	\hat{H}_{\rm atom}=\frac{\hbar\omega}{2}\hat{s}_{z}\qquad\text{[H]}.\label{eq:H_atom}
\end{equation}
where $\hat{s}_{z}$ is a time dependent operator with eigenvalues, $\pm1$.  In the Schr\"odinger picture $\hat{s}_{z}$ becomes time independent and can be represented by the Pauli matrix $\sigma_{z}={\rm diag}[1,-1]$.  Given that any Hermitian operator acting on a two level system can be represented as a combination of the three Pauli matrices $\sigma_{x}$, $\sigma_{y}$ and $\sigma_{z}$, and the identity $\boldsymbol{1}_{2}$, we introduce two more operators, $\hat{s}_{x}$ and $\hat{s}_{y}$, which satisfy the commutation relations
\begin{equation}
	[\hat{s}_{i},\hat{s}_{j}]=2{\rm i}\hat{s}_{k},\label{eq:pauli_commutation}
\end{equation}
where $(i,j,k)$ are $(x,y,z)$, $(y,z,x)$, or $(z,x,y)$.  As with $\hat{s}_{z}$, the two operators $\hat{s}_{x}$ and $\hat{s}_{y}$ are respectively given by the Pauli matrices $\sigma_{x}$ and $\sigma_{y}$ in the Schr\"odinger picture. 

To specify the interaction Hamiltonian $\hat{H}_{\rm int}$ we need to introduce the dipole moment operator $\hat{d}$, which is the means by which the atom couples to the electromagnetic field.  As we just discussed, in general this operator can be expressed as a combination of $\hat{s}_{x}$, $\hat{s}_{y}$, $\hat{s}_{z}$ and $\boldsymbol{1}_{2}$.  However, both $\hat{s}_{z}$ and $\boldsymbol{1}_{2}$ commute with the atomic part of the Hamiltonian (\ref{eq:H_atom}) and thus these parts of $\hat{d}$ cannot lead to any atomic transitions.  Because we are interested in the emission of radiation from the atom, we therefore neglect these contributions, which is physically equivalent to saying that there is no permanent dipole moment associated with either energy level.  We thus write the dipole moment operator in the following form
\begin{equation}
	\hat{d}={\rm Re}[\mathcal{P}]\hat{s}_x+{\rm Im}[\mathcal{P}]\hat{s}_y,\label{eq:dipole_operator2}\qquad\text{[H]}
\end{equation}
As the calculation will make clear, the complex number $\mathcal{P}$ (known as the \emph{transition dipole moment}) plays the same role as the complex classical dipole moment $\tilde{d}$ introduced in Eq. (\ref{eq:monochromatic_dipole}), \textit{i.e.}, it's amplitude and phase represent the amplitude and phase of the oscillating atomic dipole moment.  Strictly speaking it is $\mathcal{P}/\sqrt{2}$ that plays the role of $\tilde{d}$.  We shall return to this factor of $\sqrt{2}$ in Sec.~\ref{sec:the_difference}.
\par
As we saw in our discussion of the classical antenna (\ref{eq:driven_oscillator}), the electromagnetic field behaves as a collection of simple harmonic oscillators, one for every mode of the system.  We can write the electric field in exactly the same form as (\ref{eq:potentials}) and (\ref{eq:field_expansion})
\begin{align}
	\hat{\boldsymbol{A}}(\boldsymbol{r},t)&=\sum_{n}\hat{c}_{n}(t)\boldsymbol{\mathcal{E}}_{n}(\boldsymbol{r})\nonumber\\
    \hat{\boldsymbol{E}}(\boldsymbol{r},t)&=-\sum_{n}\frac{d\hat{c}_{n}}{dt}\boldsymbol{\mathcal{E}}_{n}(\boldsymbol{r})-\boldsymbol{\nabla}\phi(\boldsymbol{r},t).\label{eq:quantum_field_expansion}
\end{align}
where, for the sake of simplicity, we have taken the field modes $\boldsymbol{\mathcal{E}}_{n}$ as real functions of position.  The expansion coefficients $c_{n}$ for the classical antenna have now become operators $\hat{c}_{n}$, which we again interpret as the amplitudes of a set of simple harmonic oscillators.  In the absence of any interaction with matter, each of these simple harmonic oscillators has the Hamiltonian (\ref{eq:lone_oscillator}) and the operators $\hat{c}_{n}$ and $\hat{\pi}_{n}$ obey the equations of motion (\ref{eq:eqm_example}).
\par
In order to deduce the Hamiltonian operator for the field plus the atom--field interaction, we first deduce the corresponding classical Hamiltonian.  The collection of oscillators representing the electromagnetic field is driven by the time variation of the dipole moment according to the classical equation of motion Eq. (\ref{eq:driven_oscillator}).  This classical equation of motion can also be written in the form of Hamilton's equations of motion
\begin{align}
    \frac{d c_{n}}{d t}&=\frac{1}{\epsilon_{0}}\left(\pi_{n}+d(t)\boldsymbol{e}_{d}\cdot\boldsymbol{\mathcal{E}}_{n}\right)=\frac{\partial H}{\partial\pi_{n}}\nonumber\\
    \frac{d\pi_{n}}{d t}&=-\epsilon_0\omega_{n}^{2}c_{n}=-\frac{\partial H}{\partial c_{n}}\label{eq:classical_Hamiltonian}.
\end{align}
where $\pi_{n}$ is the momentum variable conjugate to the oscillator amplitude $c_{n}$.  After a little thought one can see that a suitable classical Hamiltonian is given by
\begin{equation}
    H=\frac{1}{2\epsilon_0}\sum_{n}\left[\left(\pi_{n}+d(t)\boldsymbol{e}_{d}\cdot\boldsymbol{\mathcal{E}}_{n}(\boldsymbol{r}_{0})\right)^{2}+\epsilon_0^{2}\omega_{n}^{2}c_{n}^{2}\right].
\end{equation}
Therefore the corresponding quantum mechanical Hamiltonian describing the field plus interaction energy is
\begin{align}
	\hat{H}_{\rm field}+\hat{H}_{\rm int}=\frac{1}{2}\sum_{n}\left[\epsilon_0^{-1}\left(\hat{\pi}_{n}+\hat{d}\boldsymbol{e}_{d}\cdot\boldsymbol{\mathcal{E}}_{n}(\boldsymbol{r}_{0})\right)^{2}+\epsilon_0\omega_{n}^{2}\hat{c}_{n}^{2}\right]\label{eq:field_interaction_hamiltonian}
\end{align}
which in the limit $\mathcal{P}\to0$ reduces to a sum over the Hamiltonians (\ref{eq:lone_oscillator}) for the free electromagnetic field.  Using the Heisenberg equations of motion (\ref{eq:heisenberg}), we find again that the operators obey the classical equations of motion
\begin{align}
	\frac{d\hat{c}_{n}}{dt}&=\epsilon_{0}^{-1}\left(\hat{\pi}_{n}+\hat{d}\boldsymbol{e}_{d}\cdot\boldsymbol{\mathcal{E}}_{n}\right)\nonumber\\
    \frac{d\hat{\pi}_{n}}{dt}&=-\epsilon_{0}\omega_{n}^{2}\hat{c}_{n}\label{eq:quantum_first_order}
\end{align}
which when combined together yield a single second-order equation for the oscillator amplitude $\hat{c}_{n}$
\begin{equation}
	\frac{d^{2}\hat{c}_{n}}{d t^{2}}+\omega_{n}^{2}\hat{c}_{n}=\frac{1}{\epsilon_{0}}\frac{d\hat{d}}{d t}\boldsymbol{e}_{d}\cdot\boldsymbol{\mathcal{E}}_{n}(\boldsymbol{r}_{0}).\label{eq:quantum_eqn_motion}
\end{equation}
As emphasized in our discussion surrounding equations (\ref{eq:lone_oscillator}--\ref{eq:eqm_example2}), the operator equation of motion (\ref{eq:quantum_eqn_motion}) is formally identical to the classical equation (\ref{eq:driven_oscillator}).  We take the sum of (\ref{eq:H_atom}) and (\ref{eq:field_interaction_hamiltonian}) as the full Hamiltonian of our system.  
Before moving on we remind the reader that, inherent in this Hamiltonian is the same dipole approximation we made for the classical antenna: we are assuming that, relative to the wavelength of the emitted radiation, the electron only moves a small amount within the atom.
%
%
\subsection{An approximate solution to the operator equations of motion}
\par
Having specified all the terms within the Hamiltonian of our system (\ref{eq:full_hamiltonian}), we now look to solve the equations of motion for the field operators $\hat{c}_{n}$ and the dipole moment $\hat{d}$. 
It is difficult to do this exactly, so we shall make an approximation.
We assume that the coupling between the atom and the field is weak, expanding the solution in powers of the dipole strength $\mathcal{P}$, and dropping any terms higher than first-order. 
The physical meaning of this approximation is that the oscillation of the atomic dipole moment is not significantly altered by the electromagnetic field.
\par
In the Heisenberg picture the field amplitude operators $\hat{c}_{n}$ and $\hat{\pi}_{n}$ are functions of time, as are the atomic operators $\hat{s}_{x}$, $\hat{s}_{y}$, $\hat{s}_{z}$ and $\hat{d}$.
We already have the equations of motion for the field operators $\hat{c}_{n}$ and $\hat{\pi}_{n}$ in equations (\ref{eq:quantum_first_order}), and the time evolution of $\hat{d}$ is fixed by that of $\hat{s}_{x}$ and $\hat{s}_{y}$ via (\ref{eq:dipole_operator2}).
The equations of motion for these atomic operators follows from the Heisenberg equation of motion (\ref{eq:heisenberg}), combined with the commutation relations (\ref{eq:pauli_commutation}),
\begin{align}
	\frac{d\hat{s}_x}{dt}&=\frac{{\rm i}}{\hbar}\left[\hat{H},\hat{s}_{x}\right]=-\omega\hat{s}_{y}+\frac{2{\rm Im}[\mathcal{P}]}{\hbar\epsilon_0}\sum_n\hat{\pi}_n\boldsymbol{e}_{d}\cdot\boldsymbol{\mathcal{E}}_n(\boldsymbol{r}_0)\hat{s}_{z},\nonumber\\
    \frac{d\hat{s}_y}{dt}&=\frac{{\rm i}}{\hbar}\left[\hat{H},\hat{s}_{y}\right]=\omega\hat{s}_{x}-\frac{2{\rm Re}[\mathcal{P}]}{\hbar\epsilon_0}\sum_n\hat{\pi}_n\boldsymbol{e}_{d}\cdot\boldsymbol{\mathcal{E}}_n(\boldsymbol{r}_0)\hat{s}_{z},\nonumber\\
    \frac{d\hat{s}_z}{dt}&=\frac{{\rm i}}{\hbar}\left[\hat{H},\hat{s}_{z}\right]=\frac{2}{\hbar\epsilon_0}\sum_n\hat{\pi}_n\boldsymbol{e}_{d}\cdot\boldsymbol{\mathcal{E}}_n(\boldsymbol{r}_0)\left({\rm Re}[\mathcal{P}]\hat{s}_{y}-{\rm Im}[\mathcal{P}]\hat{s}_{x}\right).\label{eq:spin_eqm}
\end{align}
We can solve these equations of motion to successive orders in $\mathcal{P}$, in this case stopping at the first-order.  To zeroth-order $\hat{s}_{z}$ is constant and can thus be represented as the Pauli spin matrix $\sigma_{z}$
\begin{equation}
    \hat{s}_{z}^{(0)}=\sigma_{z}.
\end{equation}
Meanwhile to this order $\hat{s}_{x}$ and $\hat{s}_{y}$ obey the coupled first-order equations
\begin{align}
    \frac{d\hat{s}_{x}^{(0)}}{dt}&=-\omega\hat{s}_{y}^{(0)}\nonumber\\
    \frac{d\hat{s}_{y}^{(0)}}{dt}&=\omega\hat{s}_{x}^{(0)}
\end{align}
which---assuming that at $t=0$, $\hat{s}_{x,y}=\sigma_{x,y}$---have the solution
\begin{align}
    \hat{s}_{x}^{(0)}(t)&=\sigma_{x}\cos(\omega t)-\sigma_{y}\sin(\omega t)\nonumber\\
    \hat{s}_{y}^{(0)}(t)&=\sigma_{y}\cos(\omega t)+\sigma_{x}\sin(\omega t)
\end{align}
Substituting these expressions into our definition of the dipole moment operator (\ref{eq:dipole_operator2}), we have to zeroth-order electric dipole operator
\begin{equation}
	\hat{d}^{(0)}=\left(\begin{matrix}0&\mathcal{P}^{\star}{\rm e}^{{\rm i}\omega t}\\\mathcal{P}{\rm e}^{-{\rm i}\omega t}&0\end{matrix}\right)\label{eq:dipole0}.
\end{equation}
This illustrates that in the absence of interaction with the electromagnetic field, the dipole moment of the atom is oscillating with frequency $\omega$.  This is the analogue of the classical current (\ref{eq:monochromatic_dipole}), and will lead to radiation at the next order, when we include the coupling to the field.  To zeroth-order in the interaction, the electromagnetic field is represented by a set of undriven simple harmonic oscillators with amplitudes $\hat{c}_{n}^{(0)}$ and momenta $\hat{\pi}_{n}^{(0)}$.  To this order the operator equations of motion (\ref{eq:quantum_first_order}) reduce to the usual equations of motion for a simple harmonic oscillator, Eq. (\ref{eq:eqm_example}).  The solutions are sums of complex exponentials
\begin{align}
	\hat{c}^{(0)}_{n}&=\sqrt{\frac{\hbar}{2\omega_{n}\epsilon_0}}\left[\hat{a}_{n}{\rm e}^{-{\rm i}\omega_{n} t}+\hat{a}_{n}^{\dagger}{\rm e}^{{\rm i}\omega_{n} t}\right]\nonumber\\
	\hat{\pi}^{(0)}_{n}&=-{\rm i}\sqrt{\frac{\hbar\omega_{n}\epsilon_0}{2}}\left[\hat{a}_{n}{\rm e}^{-{\rm i}\omega_{n} t}-\hat{a}_{n}^{\dagger}{\rm e}^{{\rm i}\omega_{n} t}\right]
	\label{eq:field0},
\end{align}
where the prefactors of e.g. $\sqrt{\hbar/2\omega_n\epsilon_0}$ are chosen so that the operators $\hat{a}_{n}$ and $\hat{a}_{n}^{\dagger}$ obey the usual commutation relations for the raising and lowering operators of a simple harmonic oscillator,
\begin{equation}
	[\hat{a}_{n},\hat{a}_{m}^{\dagger}]=\delta_{nm}.
\end{equation}
We have thus solved the system to zeroth-order.  Now for the inclusion of the interaction.  Differentiating the first two of equations (\ref{eq:spin_eqm}) with respect to time and substituting for $d\hat{s}_{x,y}/dt$ we find that to first-order the dipole moment operator $\hat{d}$ (\ref{eq:dipole_operator2}) obeys the equation of motion for a driven harmonic oscillator,
\begin{align}
	\frac{d^{2}\hat{d}^{(1)}}{d t^{2}}+\omega^{2}\hat{d}^{(1)}&=\frac{2\omega|\mathcal{P}|^{2}}{\hbar\epsilon_0}\sum_{n}\hat{\pi}_{n}^{(0)}\boldsymbol{e}_{d}\cdot\boldsymbol{\mathcal{E}}_{n}(\boldsymbol{r}_{0})\hat{s}_{z}^{(0)}\nonumber\\
	&=\frac{2\omega|\mathcal{P}|^{2}}{\hbar\epsilon_0}\sum_{n}\hat{\pi}_{n}^{(0)}\boldsymbol{e}_{d}\cdot\boldsymbol{\mathcal{E}}_{n}(\boldsymbol{r}_{0})\sigma_{z}\label{eq:quantum_dipole_evolution}
\end{align}
To this order the oscillators comprising the electromagnetic field similarly obey
\begin{equation}
	\frac{d^{2}\hat{c}_{n}^{(1)}}{d t^{2}}+\omega_{n}^{2}\hat{c}_{n}^{(1)}=\frac{1}{\epsilon_{0}}\frac{d\hat{d}^{(0)}}{d t}\boldsymbol{e}_{d}\cdot\boldsymbol{\mathcal{E}}_{n}(\boldsymbol{r}_{0}).\label{eq:field_oscillators_first_order}
\end{equation}
Both the field modes $\hat{c}_n$ and the dipole moment therefore satisfy Eqns. (\ref{eq:quantum_dipole_evolution}) and (\ref{eq:field_oscillators_first_order}) which are differential equations of exactly the same form.  Moreover, we've already solved exactly the same equation for the classical antenna!  Using our earlier solution (\ref{eq:solution_part1}) we find that the general form of the field operators $\hat{c}_{n}$ and the dipole operator are,
\begin{align}
	\hat{c}_{n}(t)=\hat{c}_{n}^{(0)}(t)+\frac{1}{\omega_{n}\epsilon_0}\boldsymbol{e}_{d}\cdot\boldsymbol{\mathcal{E}}_{n}(\boldsymbol{r}_0)\int_{0}^{t}\sin(\omega_n(t-t'))\frac{d\hat{d}^{(0)}(t')}{dt'}dt',\label{eq:sol1}
\end{align}
and
\begin{align}
	\hat{d}(t)=\hat{d}^{(0)}(t)+\frac{2 |\mathcal{P}|^{2}}{\hbar\epsilon_0}\sum_{n}\boldsymbol{e}_{d}\cdot\boldsymbol{\mathcal{E}}_{n}(\boldsymbol{r}_0)\int_{0}^{t}\sin(\omega(t-t'))\hat{\pi}_{n}^{(0)}(t')\sigma_{z}dt'.\label{eq:sol2}
\end{align}
There are several notable differences between (\ref{eq:sol1}) and (\ref{eq:sol2}), and the classical solution (\ref{eq:solution_part1}). 
Firstly, we haven't made the assumption---as we did for the classical antenna---that the coupling between the atom and the field is suddenly `switched on' at $t=0$.  This is just to avoid the complicated atomic dynamics due to the sudden change in coupling with the field. 
A more fundamental difference is that we have been forced to introduce the zeroth-order part of the field operator, $\hat{c}_{n}^{(0)}(t)$, which is the solution to the equation of motion (\ref{eq:quantum_eqn_motion}) with zero on the right hand side. 
In the classical calculation we explicitly neglected such terms, setting the electromagnetic field to be exactly zero before the antenna was switched on. 
However, in quantum mechanics it is not possible to set the operator $\hat{c}_{n}$ to zero (or the operators $\hat{s}_{x}$, $\hat{s}_{y}$ and $\hat{s}_{z}$ for that matter) because these operators must fulfill the commutation relation (\ref{eq:oscillator_commutation}) at all times. 
Therefore the terms $\hat{c}^{(0)}_{n}$ must always be included, whether a source of radiation is present or not. 
This is an expression of the uncertainty principle: we cannot simultaneously know both the `positions' $c_{n}$ and `momenta' $\pi_{n}$, of the oscillators making up the electromagnetic field. 
There is thus no way that the field can be zero at any time. 
We can think of these terms $\hat{c}^{(0)}_{n}$ and $\hat{d}^{(0)}$ as a kind of ever--present quantum noise, where the value of the atomic and field variables are never at rest, instead always jiggling around.
\par
Using expressions (\ref{eq:dipole0}--\ref{eq:field0}) along with the definitions of the various operators we can find the solution to the integrals in our expressions for the field and dipole operators (\ref{eq:sol1}--\ref{eq:sol2}),
\begin{equation}
	\hat{c}_{n}(t)=\sqrt{\frac{\hbar}{2\omega_{n}\epsilon_0}}\left[\hat{a}_{n}{\rm e}^{-{\rm i}\omega_{n} t}+\hat{a}_{n}^{\dagger}{\rm e}^{{\rm i}\omega_{n} t}\right]
    +\frac{1}{\omega_{n}\epsilon_0}\boldsymbol{e}_{d}\cdot\boldsymbol{\mathcal{E}}_{n}(\boldsymbol{r}_0)\left(\begin{matrix}0&\alpha_{n}(t)\mathcal{P}^{\star}\\\alpha^{\star}_{n}(t) \mathcal{P}&0\end{matrix}\right)\label{eq:field_operator_sol},
\end{equation}
and
\begin{multline}
	\hat{d}(t)=\left(\begin{matrix}0&\mathcal{P}^{\star}{\rm e}^{{\rm i}\omega t}\\\mathcal{P}{\rm e}^{-{\rm i}\omega t}&0\end{matrix}\right)
    \\
    +\frac{2 |\mathcal{P}|^{2}}{\hbar}\left(\begin{matrix}1&0\\0&-1\end{matrix}\right)\sum_{n}\sqrt{\frac{\hbar}{2\omega_{n}\epsilon_0}}\boldsymbol{e}_{d}\cdot\boldsymbol{\mathcal{E}}_{n}(\boldsymbol{r}_0)\left(\hat{a}_{n}\beta_{n}^{\star}(t)+\hat{a}_{n}^{\dagger}\beta_{n}(t)\right),\label{eq:dipole_operator_sol}
\end{multline}
where we defined the coefficients 
\begin{align}
	\alpha_{n}(t)&={\rm i}\omega\int_{0}^{t}\sin(\omega_{n}(t-t')){\rm e}^{{\rm i}\omega t'}dt',\nonumber\\
    \beta_{n}(t)&={\rm i}\omega_{n}\int_{0}^{t}\sin(\omega(t-t')){\rm e}^{{\rm i}\omega_{n} t'}dt'.\label{eq:alpha_beta}
\end{align}
This completes our first-order determination of the atom and field operators.
%
%
\subsection{The power radiated by the atom, and the spontaneous emission rate}
\par
Now consider the process shown in the cartoon of figure~\ref{fig:cartoon-emission}.  
Our atom is initially (at time $t=0$) in the excited state $|1\rangle_{\rm at}$ and the field is initially in the ground state $|0\rangle_{\rm f}$ (\textit{i.e.}, all the simple harmonic oscillators with amplitude operators $\hat{c}_{n}$ are in their ground state). 
We calculate the average rate of power leaving the atom just as we did in the classical case (\ref{eq:antenna_power}). 
By analogy with our classical expression (\ref{eq:emitted_power}) we define a (Hermitian) power operator
\begin{equation}
	\hat{P}=-\int d^{3}\boldsymbol{r}\frac{1}{2}\left[\hat{\boldsymbol{j}}(\boldsymbol{r},t)\cdot\hat{\boldsymbol{E}}(\boldsymbol{r},t)+\hat{\boldsymbol{E}}(\boldsymbol{r},t)\cdot\hat{\boldsymbol{j}}(\boldsymbol{r},t)\right].\label{eq:power_operator}
\end{equation}
The current and field operators can be expressed in terms of the expansion of the electromagnetic field (\ref{eq:quantum_field_expansion}) and the relationship between the current and the dipole moment (\ref{eq:defn_current})).  Averaging over a time interval $[0,T]$, the average power leaving the atom is given by
\begin{equation}
\langle P^{\rm Q}\rangle=\sum_{n}\boldsymbol{e}_{d}\cdot\boldsymbol{\mathcal{E}}_{n}(\boldsymbol{r}_{0})\frac{1}{2T}\int_{0}^{T}dt\langle 0|_{\rm f}\langle 1|_{\rm at}\left[\frac{d\hat{d}(t)}{dt}\frac{d \hat{c}_{n}(t)}{dt}+\frac{d \hat{c}_{n}(t)}{dt}\frac{d\hat{d}(t)}{dt}\right]|0\rangle_{\rm f}|1\rangle_{\rm at}.
\end{equation}
where the superscript $\rm Q$ indicates that this quantity was calculated using quantum mechanics, as opposed to the classical power flow (\ref{eq:average_emitted_power}).  Inserting our first-order solution for the evolution of the field amplitude and dipole operators (\ref{eq:field_operator_sol}--\ref{eq:dipole_operator_sol}), and using the fact that the lowering operator $\hat{a}_{n}$ reduces the ground state to zero, the average power leaving the atom is given by
\begin{equation}
	\langle P^{\rm Q}\rangle=\frac{|\mathcal{P}|^{2}}{\epsilon_0}\sum_{n}\left[\boldsymbol{e}_{d}\cdot\boldsymbol{\mathcal{E}}_{n}(\boldsymbol{r}_{0})\right]^{2}\frac{1}{T}{\rm Im}\int_{0}^{T}dt\left[\frac{\omega}{\omega_{n}}\dot{\alpha}_{n}(t){\rm e}^{-{\rm i}\omega t}+\dot{\beta}_{n}(t){\rm e}^{-{\rm i}\omega_n t}\right].\label{eq:power_alpha_beta}
\end{equation}
As a final step we use the definitions of the time dependent quantities $\alpha_{n}(t)$ and $\beta_{n}(t)$ given in (\ref{eq:alpha_beta}), to evaluate the integral over time, giving
\begin{multline}
	\frac{1}{T}\int_{0}^{T}dt\left[\frac{\omega}{\omega_{n}}\dot{\alpha}_{n}(t){\rm e}^{-{\rm i}\omega t}+\dot{\beta}_{n}(t){\rm e}^{-{\rm i}\omega_n t}\right]=\\(\omega\omega_{n}+\omega^{2})\left[\frac{\sin^{2}((\omega-\omega_n)T/2)}{(\omega-\omega_n)^{2}T}+\frac{\sin^{2}((\omega+\omega_n)T/2)}{(\omega+\omega_n)^{2}T}\right].
\end{multline}
We can now take the limit of a long averaging time $T\to\infty$, and use the same formula for the delta function as we did for the classical antenna (\ref{eq:delta_formula}),
\begin{align}
	\langle P^{\rm Q}\rangle=\frac{\pi\omega^{2}|\mathcal{P}|^{2}}{\epsilon_0}\sum_{n}\left[\boldsymbol{e}_{d}\cdot\boldsymbol{\mathcal{E}}_{n}(\boldsymbol{r}_{0})\right]^{2}\delta(\omega-\omega_n)&=\frac{\pi\omega^{2}|d_{0}|^{2}}{\epsilon_0 }\rho_{\rm p}(\boldsymbol{e}_{d},\boldsymbol{r}_0,\omega)\label{eq:quantum_power_pre}\\
    &=2\mu_{0}\omega^{3}|\mathcal{P}|^{2}\boldsymbol{e}_{d}\cdot{\rm Im}\left[\overleftrightarrow{\boldsymbol{G}}(\boldsymbol{r}_{0},\boldsymbol{r}_{0},\omega)\right]\cdot\boldsymbol{e}_{d}.\label{eq:quantum_power}
\end{align}
This is Fermi's Golden Rule~\cite{Visser_AmJPhys_2009_77_487,Fermi_NM,Dirac_1930}.   Expression (\ref{eq:quantum_power}) shows that the power leaving the atom as it makes its transition from the excited to the ground state is---just like the classical antenna---proportional to the local density of states. 
The average rate at which the atom makes this transition $\Gamma^{\rm Q}$ is simply the average power divided by the photon energy $\hbar\omega$,
\begin{equation}
	\Gamma^{\rm Q}=\frac{\langle P^{\rm Q}\rangle}{\hbar\omega}=\frac{\pi\omega|\mathcal{P}|^{2}}{\hbar\epsilon_0}\rho_{\rm p}(\boldsymbol{e}_{d},\boldsymbol{r}_0,\omega).\label{eq:quantum_rate_emission}
\end{equation}
As anticipated, there is very little difference between this expression and the classical power flow in Eq.~(\ref{eq:intro_classical_power}). 
Given the dependence on the PLDOS, all of the results from section~\ref{sec:examples} carry over to the case of an atom spontaneously emitting a photon. 
Assuming the quantum transition dipole moment $\mathcal{P}$ is analogous to the classical dipole amplitude $d_0$, the only obvious difference between the rate of power emitted from the atom (\ref{eq:quantum_power_pre}) and that from a classical antenna (\ref{eq:classical_power}) is a factor of $4$. 
This factor is curious, because it suggests that a quantum mechanical system is four times more effective at emitting electromagnetic power than a classical one. 
Can this really be true?  
We shall return to this point in Sec~\ref{sec:the_difference}.

\section{Approximations and limitations} \label{failings}
\par
At this point it is worth taking a moment to reflect on what we have achieved. 
We have derived the steady state power emitted from both a classical and quantum mechanical dipole antenna, and emphasized its dependence on the environment. 
It should be remembered that both expressions Eq.~(\ref{eq:classical_power}) and Eq.~(\ref{eq:quantum_power_pre}) are only approximate, and we now give a list of the most important approximations we made:
%
%
\begin{enumerate}[i]

\item \textbf{Dipole approximation}: The assumption made in the dipole approximation is that only the electric dipole moment associated with an emitter's excited state is important. This means the emitter is small compared to the scale over which the field varies.  The typical scale of variation is the wavelength, meaning that neglecting the higher order multipole moments is typically a good approximation for sub--wavelength sized emitters.  However, in many nanophotonic structures of recent interest the optical fields are compressed into very small volumes; this is particularly the case for plasmonic resonators, see \textit{e.g.}, Ref.~\cite{Chikkaraddy_Nature_2016_535_127}. Associated with the tight field confinement is an increase in the field gradient, so much so that the electric field can no longer be considered to be uniform across the emitter; the dipole approximation fails in such circumstances~\cite{Andersen_NatPhys_2011_7_215}. This 'failure' is in fact an attractive one since it provides a way to access emitter transitions that are forbidden within the dipole approximation. 
This fascinating prospect has been investigated by a number of authors, for example \cite{Kern_PRA_2012_85_022501,Rivera_Science_2016_353_263,Cuartero-Gonzalez_ACSPhot_2018_5_3415,Neuman_NL_2018_18_2358}, including selective coupling to modes of different orbital angular momentum \cite{Rivera_Science_2016_353_263,Machado_ACSPhot_2018_5_3064}.

\item \textbf{Weak coupling}: In our classical calculation this approximation was made when we assumed that the current flowing through the antenna was not influenced by the emission of radiation. 
Quantum mechanically we made this assumption when we applied first-order perturbation theory, which is an approximation that is valid only for weak interactions. 
This assumes that the internal forces within the antenna or the atom are much larger than those due to the field. 
Were we to work in the opposite limit, where the coupling energy is comparable to or larger than the internal energy we would be in the \emph{strong coupling} regime, described for example in~\cite{Haroche_1992_CQED, Torma_RepProgPhys_2015_78_013901}. 

\item \textbf{Time averaging}: Although not an approximation \textit{per se}, we must remember that both our classical and quantum mechanical rates of emission are averages over a long time period. 
In many quantum mechanical calculations, this time averaging is evident when one neglects rapidly oscillating terms in the Hamiltonian, which is known as the `rotating--wave' approximation.
We therefore should not expect to be able to accurately predict, \textit{e.g.}, transient behaviour after an atom has been initially excited. 
This transient behaviour can be subtle; for instance the way in which the atom decays from the excited state depends on the extent to which its environment exhibits memory of the past state of the atom. 
If such a memory is negligible then the excited state decays exponentially. 
If the memory effect is not negligible, then the decay has a different behaviour, as discussed for example in~\cite{lewenstein1988, lagendijk1993lucca, Rothe_PRL_2006_96_163601}. 

\item \textbf{Effective medium approximation}: when we calculated the PLDOS Eq.~(\ref{eq:free_space_continuum_ldos}) in a homogeneous material of permittivity $\epsilon$ and permeability $\mu$, we implicitly assumed that the microscopic details of the material were not important for the rate of emission. 
We should be careful about doing this, because---for example---our dipole could be an atom placed within a material, which is nothing more than a collection of other atoms. 
How can we justify only properly calculating the dynamics of one atom?
There are some important cases where we cannot get away with this. 
For example, if the medium exhibits significant absorption then the local density of states becomes undefined in the effective medium approximation. 
We see this immediately from the expression for the partial local density of states (\ref{eq:pldos_green_function}) in terms of the Green function for a homogeneous medium (\ref{eq:homogeneous_gf}),
\begin{align}
	\rho_{p}(\boldsymbol{e}_{d},\boldsymbol{r}_{0},\omega)&=\frac{2\omega}{\pi c^{2}}\boldsymbol{e}_{d}\cdot{\rm Im}\left[\overleftrightarrow{\boldsymbol{G}}(\boldsymbol{r}_{0},\boldsymbol{r}_{0},\omega)\right]\cdot\boldsymbol{e}_{d},\nonumber\\
	&=\frac{2\omega}{\pi c^{2}}\left\{\frac{k_0{\rm n}}{4\pi}\left[1-{\rm Re}\left[\frac{1}{{\rm n}^{2}}\right]\frac{{\rm n}^{2}}{3}\right]+\frac{1}{k_0^{2}}{\rm Im}\left[\frac{1}{{\rm n}^{2}}\right]\lim_{r\to 0}\frac{\partial^{2}}{\partial z^{2}}\frac{\cos\left(k_0{\rm n}r\right)}{4\pi r}\right\},\nonumber\\
    &\to\infty.
\end{align}
The factor of $1/r$ inside the double derivative diverges in the limit $r\to0$, indicating that the effective medium approximation does not lead to a finite local density of states in the presence of any absorption. 
For the case of a classical dipole antenna, this prediction of an infinite rate of emission (and the limitations of the effective medium approximation) has been discussed by Tai~\cite{tai2000}. 
On the quantum mechanical side, the problem of predicting spontaneous emission in an absorbing medium, and the associated breakdown of the effective medium approximation is discussed by Barnett \textit{et al.} ~\cite{Barnett_JPhysB_1996_29_3763}, the same problem has also been addressed from a classical perspective~\cite{Tomas_PRA_1997_56_4197,scheel1999a}. 
Closely related to this breakdown of macroscopic electromagnetism is the need to include local-field corrections~\cite{schuurmans1998prl,scheel1999b}. 
These corrections estimate the microscopic field at the position of the emitter (the emitter couples to the exact electric field at its location, not the spatially averaged one we have used here).
\end{enumerate}

\section{A difference between classical and quantum emitters\label{sec:the_difference}}
As we have seen, the quantum mechanical behaviour of an atom coupled to the electromagnetic field has much in common with its classical counterpart, and the local density of states governs the rate of emission in both cases. 
This is because the local density of states quantifies the allowed electromagnetic modes at a particular point in space, and quantum and classical electromagnetism don't disagree about what modes are allowed, only how they are occupied. 
So after all that, is it just a matter of language? 
Is the physical process of emission from a classical antenna actually the same as that from a quantum system? 
The answer is no, as indicated by the difference of a factor of $4$ between (\ref{eq:classical_power}) and (\ref{eq:quantum_power_pre}). 
In this section we tease apart the difference in the physics of the emission of electromagnetic waves from classical and quantum dipoles.  
%
%
\begin{figure}[ht!]
\centering
\includegraphics[width=\textwidth]{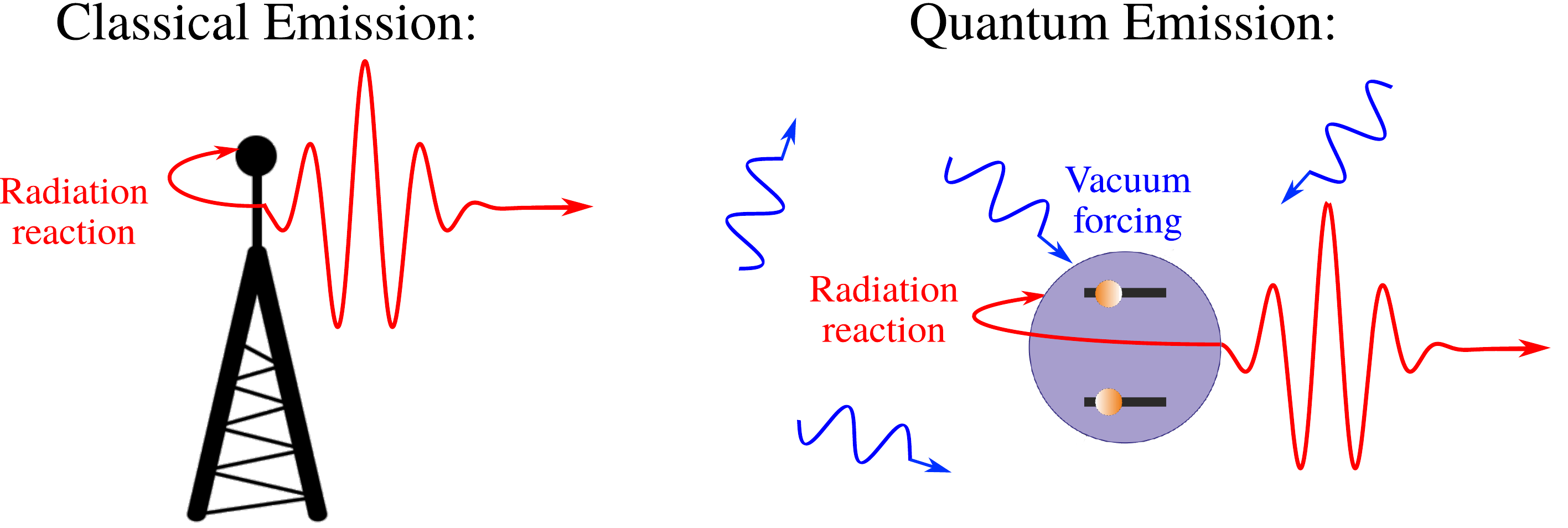}
\caption{Many textbooks attribute the process of spontaneous emission to 'vacuum fluctuations'. 
These exotic sounding fluctuations are actually only part of the story, and the remainder of the tale has a lot in common with a classical antenna. 
Emission from a classical antenna can be understood in terms of the radiation reaction, where the emitted field acts back on the antenna, pulling energy out of the current flowing up and down the wire.
The same process also happens during spontaneous emission from an atom, with the emitted radiation extracting energy from the jiggling charge. 
In addition there is an erratic background (vacuum) field that contributes to the emission process. 
In the case of a two level atom the radiation reaction effect and vacuum effect contribute equally to the emission. 
\label{fig:classical_vs_quantum_dipole}}
\end{figure}
The factor of $4$ difference between quantum and classical predictions for spontaneous emission arises from two factors of two that come from quite different places. 
One of these factors is trivial and comes from a lack of consistency in notation, the second comes from interesting physics. 
\par
To see where the first of these factors arises, consider the time average of the square of the classical dipole moment. 
From its definition (\ref{eq:monochromatic_dipole}) we see that this average equals,
\begin{equation}
\langle d(t)^{2}\rangle^{\rm C} = \frac{1}{T}\int_{0}^{T}\frac{1}{4}[d_0^{2}{\rm e}^{-2{\rm i}\omega t} + {d_0^{\star}}^{2}{\rm e}^{2{\rm i}\omega t} + 2|d_0|^{2}] = \frac{1}{2}|d_0|^{2}.\label{eq:classical_dipole_average}
\end{equation}
Meanwhile if we take the equivalent time average of the expectation value of the quantum dipole moment operator squared (defined as in (\ref{eq:dipole_operator2})), for an atom initially in the excited state, we obtain,
\begin{equation}
	\langle d(t)^{2}\rangle^{\rm Q}=\frac{1}{T}\int_{0}^{T}\langle 1| \hat{d}^{2}(t) |1\rangle=\frac{1}{T}\int_{0}^{T}\langle 1|\left[{\rm Re}[\mathcal{P}]^{2}\hat{s}_x^{2}+{\rm Im}[\mathcal{P}]^{2}\hat{s}_y^{2}\right]|1\rangle=|\mathcal{P}|^{2},\label{eq:quantum_dipole_average}
\end{equation}
where we used the fact that $\hat{s}_{x}^{2}=\hat{s}_{y}^{2} = \boldsymbol{1}_{2}$ and $\hat{s}_{x}\hat{s}_{y}+\hat{s}_{y}\hat{s}_{x} = 0$. 
Comparing the two expressions (\ref{eq:classical_dipole_average}) and (\ref{eq:quantum_dipole_average}) we see that assuming that $\mathcal{P}$ is equivalent to the classical dipole moment of the atom $d_0$ (\ref{eq:dipole_operator2}) is not quite right. 
The dipole moment $d_0$ is actually a factor of $\sqrt{2}$ smaller than we thought, and therefore expression (\ref{eq:quantum_power}) artificially looks twice as big. 
This explains one of the factors of $2$, and shows that for a proper comparison we should really reduce $\mathcal{P}$ by a factor of $\sqrt{2}$.
\par
The other factor of $2$ is much more interesting. 
we see where it comes from by comparing the general form of the classical and quantum expressions for the power emitted from the atom.  Our quantum mechanical solution for the evolution of the field amplitudes and the dipole operator (\ref{eq:sol1}--\ref{eq:sol2}) has the form,
\begin{align}
	\hat{c}_{n}(t)&=\hat{c}_{n}^{(0)}(t)+\hat{c}_{n}^{(1)}(t),\nonumber\\
    \hat{d}(t)&=\hat{d}^{(0)}(t)+\hat{d}^{(1)}(t),\label{eq:quantum_scheme}
\end{align}
where the quantities with a superscript zero indicate the motion in the absence of any atom--field coupling, and the superscript of unity indicates the effect of the interaction. 
For example $c^{(0)}_{n}$ represents the quantum mechanical `vacuum' field, which is present irrespective of the atom, while $\hat{d}^{(1)}$ represents the atomic motion induced by the `vacuum' field $\hat{c}^{(0)}$. 
The analogous classical solution for the evolution of the field and the dipole (\ref{eq:solution_part1}) only contains some of the terms present in the quantum mechanical solution, \textit{i.e.},
\begin{align}
	c_{n}(t)&=c_{n}^{(1)}(t),\nonumber\\
    d(t)&=d^{(0)}(t).\label{eq:classical_scheme}
\end{align}
The electromagnetic field is zero without the antenna (so $c_{n}^{(0)}=0$), and there is (to first-order in $d_0$) no effect of the electromagnetic field on the antenna (so $d^{(1)}$ is zero).
\par
This difference between the terms in (\ref{eq:quantum_scheme}) and (\ref{eq:classical_scheme}) is an expression of the effect of the `vacuum' electromagnetic field.  In quantum physics the field commutation relations (\ref{eq:oscillator_commutation}) imply that there is always some non--zero electromagnetic field.  This electromagnetic field causes current to flow within the atom, and this changes the way the atom emits, relative to a classical antenna.
A picture of this difference is sketched in figure~\ref{fig:classical_vs_quantum_dipole}.
In both classical and quantum emission there is a `radiation reaction' where the emitted field acts back on the current, `pulling' energy out of its motion.  However, in quantum physics there is also an ever--present vacuum field that acts on the atom, changing the rate at which the energy is being pulled out.
\par
We can identify the radiation reaction and vacuum field terms in the equation for the power leaving the atom with the $\dot{\alpha}_{n}(t)$ and $\dot{\beta}_{n}(t)$ terms in (\ref{eq:power_alpha_beta}).
From our solution for the motion of the dipole (\ref{eq:sol2}) we know that terms involving $\beta_{n}$ arise from the effect of the vacuum on the motion of the atom.
Therefore, neglecting such terms, the power leaving an atom due to `radiation reaction' is equal to,
\begin{align}
	\langle P_{\rm rr}\rangle&=\frac{|\mathcal{P}|^{2}}{\epsilon_0}\sum_{n}\left[\boldsymbol{e}_{d}\cdot\boldsymbol{\mathcal{E}}_{n}(\boldsymbol{r}_{0})\right]^{2}\frac{1}{T}{\rm Im}\int_{0}^{T}dt\left[\frac{\omega}{\omega_{n}}\dot{\alpha}_{n}(t){\rm e}^{-{\rm i}\omega t}\right],\nonumber\\
    &\stackrel{\text{long time}}{\to}\frac{\pi\omega^{2}|\mathcal{P}|^{2}}{2\epsilon_0}\sum_{n}\left[\boldsymbol{e}_{d}\cdot\boldsymbol{\mathcal{E}}_{n}(\boldsymbol{r}_{0})\right]^{2},
\end{align}
which---after taking into account the $\sqrt{2}$ difference in the definition of the dipole moment---is exactly equal to the rate of power leaving a classical antenna.
Meanwhile the contribution from the vacuum field is given by those terms involving $\beta_n$ and not $\alpha_n$,
\begin{align}
	\langle P_{\rm vac}\rangle&=\frac{|\mathcal{P}|^{2}}{\epsilon_0}\sum_{n}\left[\boldsymbol{e}_{d}\cdot\boldsymbol{\mathcal{E}}_{n}(\boldsymbol{r}_{0})\right]^{2}\frac{1}{T}{\rm Im}\int_{0}^{T}dt\left[\dot{\beta}_{n}(t){\rm e}^{-{\rm i}\omega_n t}\right],\nonumber\\
    &\stackrel{\text{long time}}{\to}\frac{\pi\omega^{2}|\mathcal{P}|^{2}}{2\epsilon_0}\sum_{n}\left[\boldsymbol{e}_{d}\cdot\boldsymbol{\mathcal{E}}_{n}(\boldsymbol{r}_{0})\right]^{2}.
\end{align}
This last result shows the curious result that the emission from a two level system is indeed twice as effective as an equivalent classical emitter, with equal contributions coming from the radiation reaction force and the vacuum field.  In short, we can understand the emission from an atom using nothing but classical physics, but we must assume the presence of a `noisy' background electromagnetic field, that serves to increase the radiation from a two level system.  This splitting of effects is well described by Milonni in his textbook, \textit{The quantum vacuum}~\cite{Milonni}.  However, the the choice of a two-level system is not the only one, other less well explored systems are also possible and will be the subject of a future report.
%
%
\begin{figure}[ht!]
\centering
\includegraphics[width=10cm]{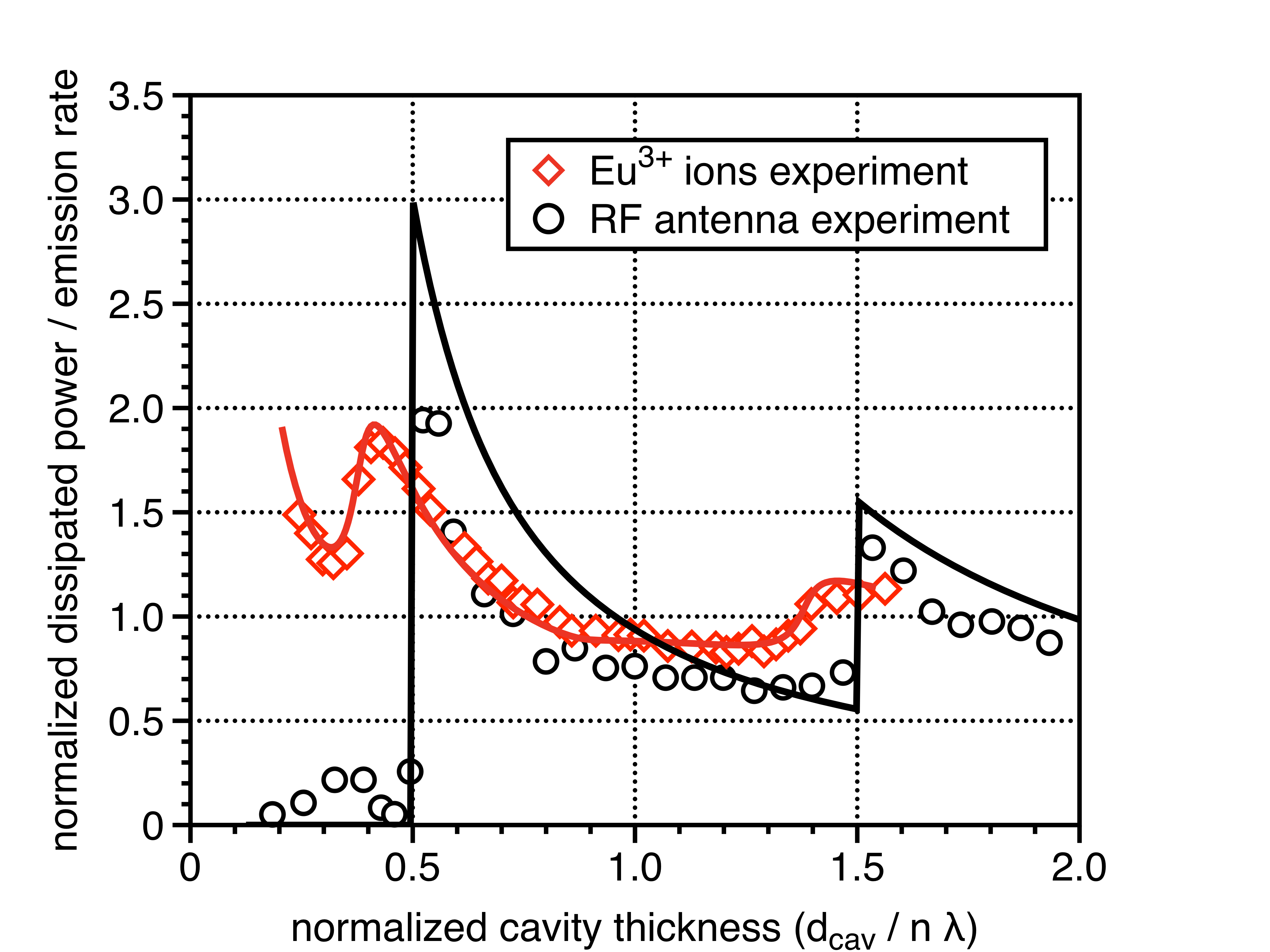}
\caption{Comparison of classical and quantum. 
Experimental data (open symbols) are shown from two experiments, one classical and one quantum. 
The classical experiment involved a radio-frequency (RF) antenna ($\lambda_0=$ 29.7 cm) located inside a planar cavity made of two large conducting panels \cite{Seeley_AmJPhys_1993_61_545}; the antenna was parallel to the cavity plane. 
The quantum experiment involved $Eu^{3+}$ ions as emitters with free-space emission wavelength $\lambda = 614$ nm, in a monolithic planar cavity made by a combination of different thin-film deposition techniques~\cite{Worthing_JAP_2001_89_615}. 
For both experiments the antenna/emitters were located in the centre of the cavity. 
In each case the data are accompanied by a line showing what is expected based on equations discussed elsewhere in the present article. 
\label{fig:seeley-worthing}}
\end{figure}

Before ending this section it is interesting to compare data from experiments based on classical to those based on quantum emitters, as is done in Fig~\ref{fig:seeley-worthing} above. 
In this figure, data have been compiled from experiments reported in the literature in two very different regimes. 
The first is classical, the experiment involving a radio frequency antenna suspended between two conducing planes that form a planar cavity. 
The second is quantum, the experiment involving the emission of light by excited $Eu^{3+}$ ions.
For the RF experiment the normalized power dissipated by the antenna is plotted as a function of the normalized cavity thickness (black circles); the power is normalized w.r.t. the power dissipated in the absence of a cavity, and the cavity thickness is normalized to the emission wavelength of the antenna.
For the experiment with $Eu^{3+}$ ions the measured emission rate (inverse fluorescence lifetime) is plotted as a function of normalized cavity thickness (red diamonds). 
Regarding the calculated data shown in Fig \ref {fig:seeley-worthing}: for the RF case (black line) the expected dependence was calculated using Eq. (\ref{eq:imperfect_cavity_pldos}) above (for the dipole moment/antenna parallel to the plane of the cavity); for the optical case (red line) the expected dependence was calculated using Eq. (\ref{eq:imperfect_cavity_ldos}).
In both cases the theoretical expectations provide a reasonable match with experiment. We should comment on the differences.
\par
The RF and optical expectations are different for two reasons. First, in the RF case the mirrors are perfect reflectors, whilst in the optical case they are thin metal films with that exhibit loss, hence the rounding of the features. 
Second, in the RF case the antenna is in the parallel orientation, whilst the $Eu^{3+}$ ions are free to tumble in space on a time scale that is faster than their emission lifetime, see Eq (\ref{eq:imperfect_cavity_ldos}). 
The loss associated with metals at optical frequencies is also responsible for the steep rise in the emission rate for normalized cavity thicknesses $d_{cav} / (n \lambda)< 0.3$, since, at these frequencies metallic cavities support coupled surface plasmon modes that have no cut-off~\cite{Gramotnev_NatPhot_2013_8_13,Smith_JMO_2008_55_2929}. 
Note that for both systems there is very little evidence of the second-order cavity mode that has a cut-off at a normalized cavity thickness of 1. 
This is because this second-order mode has a node at the centre of the cavity, the very place where the antenna/emitters are located.
Despite the differences in the details, the data from both realms in Fig \ref{fig:seeley-worthing} show much the same behaviour, reflecting the fact that the densities of states are the same in classical and quantum realms, as we have been keen to point out throughout this article.
In addition, the length scales that pertain to these systems differ by more than 5 orders of magnitude, nicely demonstrating the universal character of the effect of the local density of states on emission processes.

\section{Extracting the LDOS from the measured rate\label{sec:rate}}
%
\setcounter{footnote}{0}
%
In this section, we shall put our newly acquired know-how to use. 
We shall look at how to obtain the LDOS from experiments of spontaneous emission, which requires a careful treatment of the measurement of spontaneous emission decay, including dealing with the possibility of non-exponential decay, and considering the effect of the quantum efficiency of the emission process. 

\subsection{Measurement of emission rates}
\par
So far we have been concerned with the concept and formalism of the local and partial densities of states. 
In this section we discuss how the local and partial densities of states may be determined from experiment. These densities of states are not quantities that we can measure directly, instead we need to determine them indirectly via the effect they have on light-matter processes.
The best known of these is the spontaneous emission of light by a dipole emitter, but as already discussed, others include black-body radiation and the Casimir effect.  
We have already studied in detail the emission by a classical dipole antenna, and by a quantum emitter (Sections~\ref{sec:classical-dipole} and~\ref{sec:quantum emitter}). 
As we have seen, in both cases the rate of emission has the same dependence on the local and partial densities of states (see equations \eqref{eq:intro_classical_power} and \eqref{eq:intro_quantum_power}).
We will find that several assumptions are necessary to determine the local or partial density of states from an experiment.
\begin{figure}[ht!]
\centering
\includegraphics[width=0.67\columnwidth]{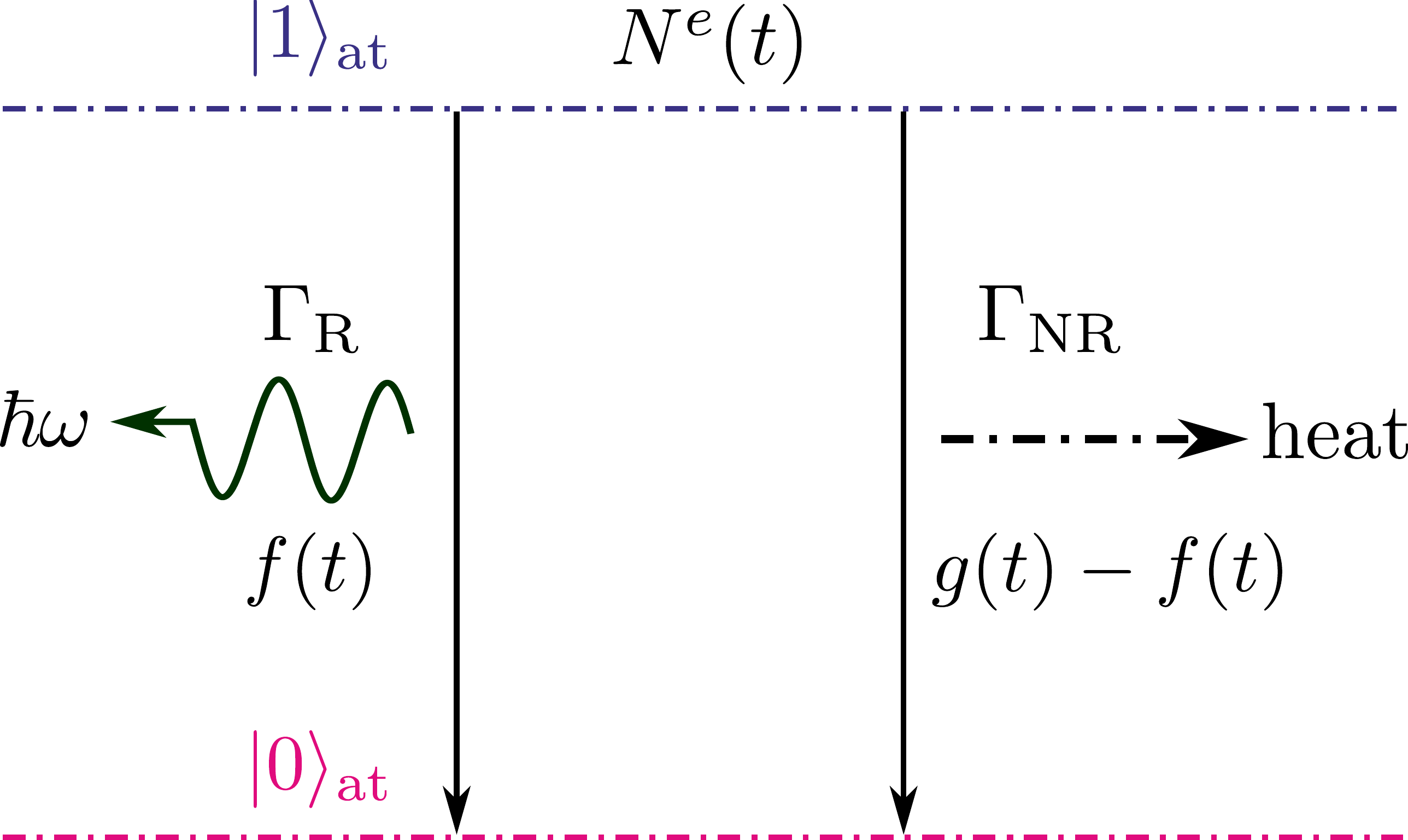}
\caption{Schematic of the decay of an excited state $\ket{1}_{\rm at}$ of $N^e$ emitters to their ground state $\ket{0}_{\rm at}$, and the associated experimentally observable parameters. 
Photons with energy $\hbar \omega$ are emitted, with $f(t)$ the rate of photon emission as a function of time. 
The non-radiative decay (such as heat) occurs at a rate $h(t) = g(t) - f(t)$. 
\label{fig:two-levels_rates_intensities}}
\end{figure}
\par
Experiments typically determine the spontaneous emission decay rate through recording the intensity of the luminescence following pulsed excitation. 
This is usually done repeatedly, and the arrival times of photons are recorded using time--correlated single photon counting over many cycles of pulsed excitation~\cite{Becker2005book}. 
Note that determining the local or partial density of states via emission rates has the benefit that the rate is independent of both the direction in which the emitted light is collected \cite{Mony_JPCC_2018_122_24917}, and the collection efficiency of the detection optics. 
The problem is that the emission process is never perfect: not all excited emitters produce a photon. 
Instead some lose their energy in a non--radiative way. 
Non--radiative decay refers to the dissipation of the emitter's energy into non-optical modes, primarily into heat (phonons). 
The total decay of emitters from the excited state is thus a combination of radiative and non-radiative decay, as sketched in Fig.~\ref{fig:two-levels_rates_intensities}.  Given that the local or partial density of states only affects the radiative decays, one needs to be able to eliminate these non--radiative contributions.
\par
As an additional complication, changing the local optical environment may alter both radiative and non-radiative rates~\cite{Pelton_NatPhot_2015_9_427}. 
In the presence of an interface as considered in sections~\ref{sec:mirror_emission} and~\ref{sec:non-perfect-mirror}, the radiative rate will be different because the allowed electromagnetic modes will be altered.
If the interface is between the host medium and a metal of finite conductivity then surface plasmon modes may be present, offering another decay route for energy from the excited molecule. 
However, we still regard this as a modification to radiative decay, even though the surface plasmon mode might be highly damped; this is because surface plasmon modes can be recovered to produce light \cite{Worthing_APL_2001_79_3035}. 
But introducing the same (metal) surface may also alter the non-radiative rate since the energy of the excited molecule may now be lost to heat in the metal through Joule heating, Landau damping etc.~\cite{Ford_PhysRep_1984_113_195}. 
The length scale at which significant modifications to the non--radiative rate occurs---\textit{i.e.}, the range of molecule-surface separations for which such interactions are significant---is small, typically a few nanometres~\cite{Smith_AdvFuncMat_2005_15_1839}.
In many situations these short range interactions are negligible, and the non-radiative decay rate can be taken to be independent of any structuring of the local environment.  We make this assumption throughout this section.
\par
Let us start by considering the the time dependence of an ensemble of $N^e$ nominally identical emitters, \textit{e.g.}, quantum dots, in their excited (emissive) state at time $t=0$.  Here we concentrate on quantum emission.  We write the probability that an emitter decays in the time interval $t\longrightarrow t+dt$, as $\Gamma(t)dt$.
If at time $t$ there are $N^e(t)$ emitters in the excited state then the number that decay between $t\longrightarrow t+dt$ is $N^e(t) \Gamma(t) d t$, and this must be equal to minus the rate of change of emitters in the excited state, multiplied by the time interval $d t$, or equivalently
\begin{equation}
\label{eq:rate_a}
\frac{d}{dt}[N^e(t)]=-N^e(t) \Gamma(t)
\end{equation}
\noindent
If the rate of decay, $\Gamma$, is independent of time, the number of emitters remaining in the excited state at time $t$ is given by,
\begin{equation}
\label{eq:rate_c}
N^e(t) = N^e_0~\exp(-\Gamma t),
\end{equation}
\noindent
which is the solution to (\ref{eq:rate_a}), where $N^e_0$ is the number of emitters in the excited state at time $t=0$.  The number of emitters in the excited state thus decays in a single exponential fashion.
As already discussed, in the presence of non-radiative decay, we need to consider both the radiative $(\Gamma_{\rm R})$ and non-radiative $(\Gamma_{\rm NR})$ contributions to the total decay rate $(\Gamma)$, \textit{i.e.},
\begin{equation}
\label{eq:rate_d}
\Gamma = \Gamma_{\rm R} + \Gamma_{\rm NR}.
\end{equation}
\noindent the degree of non--radiative decay is quantified via the quantum efficiency $\eta$, which is defined as
\begin{equation}
    \eta=\frac{\Gamma_{\rm R}}{\Gamma_{\rm R}+\Gamma_{\rm NR}}.\label{eq:qe_defn}
\end{equation}
If the quantum efficiency equals unity, the emitter can only decay through releasing a photon, whereas a quantum efficiency of zero indicates that the decay is entirely non--radiative.  Having separated the decay rate into these two parts, the rate of decay from the excited state can also be written as a sum of two terms
\begin{equation}
\frac{d}{dt}[N^e(t)]=-\Gamma_{\rm R}N_0^e~\exp(-\Gamma t)-\Gamma_{\rm NR}N_0^e~\exp(-\Gamma t)
\end{equation}
which we write as
\begin{equation}
    g(t)=f(t)+h(t)
\end{equation}
\noindent \textit{i.e.}, the total rate of decay $g(t)$ is equal to the sum of the number of photons produced per unit time, $f(t)$ and the number of non-radiative decays per unit time, $h(t)$. 
Note that in practice we will not capture all of the photons that are produced, our measured photon rate will thus be less than $f(t)$ by some factor. 
Figure~\ref{fig:two-levels_rates_intensities} shows how the observable $f(t)$ is related to the decay of a quantum emitter from its excited state $|1\rangle_{\rm at}$ to its ground state $|0\rangle_{\rm at}$. 
The number of excited emitters at time $t$ is $N^e(t)$, and this number can be probed by transient absorption spectroscopy~\cite{Foggi1995JPC,Klimov1998PRL,Neuwahl2000CPL}. 
Meanwhile, the non-radiative decay events $h(t)$ can be probed using photothermal techniques~\cite{Rosencwaig1980book,Grinberg2003PRB}. 
\par
It is instructive to calculate the average arrival time of an emitted photon, $\langle t\rangle$, this being one of the quantities that can be determined in the aforementioned experiments. 
Given that the instantaneous probability of photon emission is proportional to $f(t)dt$, this is given by

\begin{equation}
\label{rate_h}
\langle t\rangle = \tau_{\rm ave} = \frac{\int^{\infty}_0f(t)~t ~dt}{\int^{\infty}_0f(t) dt},
\end{equation}

\noindent where we note that $\int^{\infty}_0f(t) dt$ is the total number of photons produced over all time.  Carrying out the integration we find $\int^{\infty}_0f(t) dt$ to be,

\begin{equation}
\label{rate_i}
\int^{\infty}_0f(t) dt =
\Gamma_{\rm R}N^e_0\int^{\infty}_0 \exp(-\Gamma t) dt = N^e_0\frac{\Gamma_{\rm R}}{\Gamma},
\end{equation}

\noindent whilst evaluating $\int^{\infty}_0f(t)tdt$ gives,

\begin{equation}
\label{rate_j}
\int^{\infty}_0f(t) t dt = N^e_0\frac{\Gamma_R}{\Gamma^2},
\end{equation}

\noindent so that the average arrival time $\langle t\rangle$ is, from (\ref{rate_h})

\begin{equation}
\label{rate_k}
\langle t\rangle = \tau_{\rm ave} = \frac{1}{\Gamma},
\end{equation}

\noindent The average arrival time is thus governed by the total decay rate $\Gamma$, rather than the radiative decay rate $\Gamma_{\rm R}$.  However, the local density of states only affects the radiative decay rate.  To find the local or partial density of states one must separate out the radiative and non--radiative contributions to this decay rate. 
 



\begin{figure}[ht!]
\centering
\includegraphics[width=1.0\columnwidth]{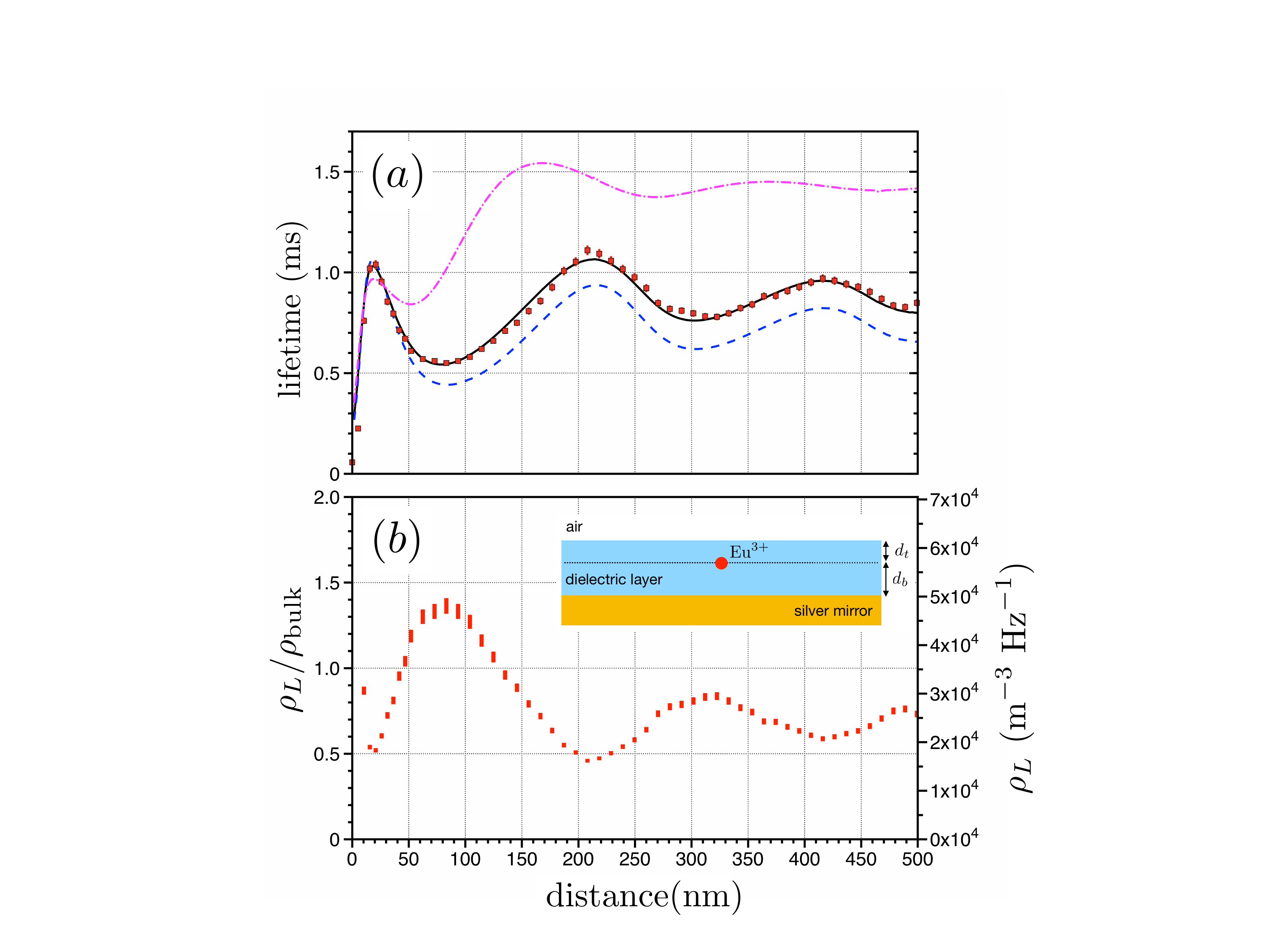}
\caption{
Analysis of a Drexhage-type experiment involving $\rm{Eu^{3+}}$ ions (emission wavelength 615 nm) in a dielectric film above a silver mirror, see inset in panel (b). 
As the distance between the $\rm{Eu^{3+}}$ ions and the silver mirror $d_b$ is varied, the lifetime of the $\rm{Eu^{3+}}$ excited state changes. 
(a) Results from Ref.~\cite{Amos_PRB_1997_55_7249} are re-plotted as red data points. 
Calculations based on equation (\ref{eq:imperfect_cavity_ldos}) are shown for different emitter orientations: 
for perpendicular (dash-dot magenta line), parallel (dashed blue line), and an isotropic combination (black solid line); the last gives the best match to the experimental data. 
Two further parameters are obtained: the quantum efficiency (found to be $\eta = 0.70\pm0.03$), and $\Gamma_{\rm{bulk}}$, \textit{i.e.}, the spontaneous emission rate of the $\rm{Eu^{3+}}$ ions in the bulk ($d_b \rightarrow \infty$) dielectric medium (found to be $\Gamma_{\rm{bulk}} = 1450\pm40~\rm{s}^{-1}$). 
\label{fig:Drexhage}}
\end{figure}

To determine the partial or local density of states experimentally we perform a relative measure, comparing the spontaneous emission rate in two different systems.  We take these two systems to be (i) an unstructured `bulk' environment; and (ii) a structured environment, where we want to know the partial or local density of states.  We write the bulk decay rate as $\Gamma_{\rm b}$, so that Eq. (\ref{eq:rate_d}) becomes, 
\begin{equation}
\label{eq:bulk}
\Gamma_{\rm b} = \Gamma_{\rm R} + \Gamma_{\rm NR}.
\end{equation}
We assume that the quantum efficiency $\eta_{\rm b}=\Gamma_{\rm R}/\Gamma_{\rm b}$ (defined in eq.~(\ref{eq:qe_defn})) of this decay process is known.
If we now consider structuring the local optical environment then the decay rate $\Gamma$ is modified,
\begin{equation}
\label{eq:struc}
\Gamma = \Gamma^\prime_{\rm R} + \Gamma^\prime_{\rm NR} = \frac{\rho_{\rm l,p}(\boldsymbol{e}_{d},\boldsymbol{r}_{0},\omega)}{\rho_{\rm b}(\omega)}\Gamma_{\rm R} + \Gamma_{\rm NR},
\end{equation}
where $\rho_{\rm l,p}(\boldsymbol{e}_{d},\boldsymbol{r}_{0},\omega)/\rho_{\rm b}(\omega)$ is the extent to which the structuring the local environment alters the local/partial (as appropriate) density of states compared to the bulk, and thus the radiative decay rate.
As discussed above we have assumed that structuring the environment does not alter the non-radiative decay rate, we have thus replaced $\Gamma^\prime_{\rm NR}$ by $\Gamma_{\rm NR}$ in \eqref{eq:struc}. 
Combining equations \eqref{eq:bulk}, \eqref{eq:qe_defn} and \eqref{eq:struc} we can thus find an expression for the local or partial density of states in terms of the relative emission rate and the quantum efficiency,
\begin{equation}
\label{eq:beta_1}
\frac{\rho_{\rm l,p}(\boldsymbol{e}_{d},\boldsymbol{r}_{0},\omega)}{\rho_{\rm b}(\omega)} = 
\frac{\Gamma/\Gamma_{\rm b} - (1 - \eta_{\rm b})}{\eta_{\rm b}}.
\end{equation}
Thus, provided we can determine the quantum efficiency $\eta_{\rm b}$ and the normalized decay rate $\Gamma/\Gamma_{\rm b}$, then we can find the structured environment contribution to the local/partial density of states. 
We can also relate our expression to the vacuum density of states and the refractive index of our 'bulk' medium, $\rm n$ through \eqref{eq:pldos_gf_fs} as,
\begin{equation}
\label{eq:beta_2}
\frac{\rho_{\rm l,p}(\boldsymbol{e}_{d},\boldsymbol{r}_{0},\omega)}{{\rm n}^3\rho_0(\omega)} = 
\frac{\Gamma/\Gamma_{\rm b} - (1 - \eta)}{\eta}.
\end{equation}
\noindent where $\rm n$ is the refractive index of the bulk medium.
\par
An example of how measured lifetimes (rates) may be used to determine the partial/local density of states is shown in Fig~\ref{fig:Drexhage}. 
These data were taken from \cite{Amos_PRB_1997_55_7249} and replicate the classic experiment reported by Drexhage in the 1970s \cite{Drexhage_ProgOpt_1974_12_163}. 
In this experiment a layer of ${\rm Eu}^{3+}$ ions is embedded in a dielectric layer above a silver mirror. 
The fluorescence lifetime (rate) is measured as a function of the distance between the ${\rm Eu}^{3+}$ ions and the silver mirror. 
The figure reveals that the lifetime oscillates as a function of this distance, the oscillations becoming smaller in amplitude as the distance increases.
From our discussions in Sec~\ref{sec:examples} we can understand this result by thinking of the ${\rm Eu}^{3+}$ ions as antennas that are driven by the reflected field, \textit{i.e.}, the scattering view point of Fig~\ref{fig:emission_pictures}.
In figure~\ref{fig:Drexhage}a, the experimental data are plotted together with three calculated data sets, using an approach based on equations~\eqref{eq:imperfect_cavity_pldos} and~\eqref{eq:imperfect_cavity_ldos} given above~\cite{Chance_AdvChemPhys_1978_37_1}.
Fitting theory to experiment is done by simulating the experimental data using~\eqref{eq:imperfect_cavity_pldos} and~\eqref{eq:imperfect_cavity_ldos}, varying three adjustable parameters: the quantum efficiency, the bulk decay rate (\textit{i.e.}, the rate in the absence of a mirror), and the dipole orientation  
The quantum efficiency is determined largely by the amplitude of the oscillations.
Notice also that the data are best fit by a model in which the emitters sample all directions in space on a timescale that is faster than the decay rate, see Eq.~(\ref{eq:imperfect_cavity_ldos}) above. Fig.~\ref{fig:Drexhage}(b) makes use of Eq.~(\ref{eq:beta_2}) to extract the local density of states. 
The sharp rise in the density of states at small emitter-mirror distances is due to the surface plasmon mode supported by the silver/dielectric interface. 
Such a conclusion is supported by an analysis similar to that shown in Fig~\ref{fig:microcavity_figure} (not shown, but available in \cite{Amos_PRB_1997_55_7249}). 
A similar approach was adopted recently by van Dam \textit{et al.} to investigate the internal emission efficiency of silicon nanoparticles \cite{vanDam_ACSPhot_2018_5_2129}. 

The change in rate of spontaneous emission for emitters in front of a mirror originally pioneered by Drexhage \cite{Drexhage_ProgOpt_1974_12_163} has become a powerful system in which to explore (P)LDOS variations in a number of arenas including acoustics \cite{Langguth_PRL_2016_116_224301} and circuit QED \cite{Hoi_NatPhys_2015_11_1045}, there is also now a nice 2D version of the Drexhage experiment \cite{Brechbuhler_PRL_2018_121_113601}.

\subsection{Exponential versus non--exponential decay}
\par
From the above analysis it seems relatively straightforward to extract the (P)LDOS from the relative decay rate and the quantum efficiency, via eq.~(\ref{eq:beta_1}). 
However, this is only the case if the decay is well represented by a single exponential as in eq.~(\ref{eq:rate_c}). 
In an experiment the measured decay curve $f(t)$ is a histogram of the arrival times of single photons after many excitation--detection cycles. 
This histogram should be fit with an appropriate statistical function, the decay rate---or more generally the distribution of decay rates---of the process may then be deduced by fitting the modelled function to the experimental data. 
In the simplest case, when the system is characterized by a single decay rate $\Gamma$, the decay curve is described by a single-exponential function (see Figure~\ref{fig:pigments} and the discussion above). 
In many cases, however, the decay is much more complex and may be very different from a single-exponential decay~\cite{Foggi1995JPC,James1986CPL,Siemiarczuk1990JPC,Brochon1990CPL,Crooker2002PRL,Wlodarczyk2003BJ,Wuister2004AnChem,Lodahl2004Nature,Rothe_PRL_2006_96_163601}. 
Non--exponential decay usually means that the decay is characterized by a distribution of rates (either radiative, non-radiative, or both) instead of a single rate\footnote{In the case of strong coupling in cavity quantum electrodynamics, the decay of even a single emitter is not single-exponential~\cite{Vasa_NatPhot_2013_7_128}.}~\cite{vanDriel_2007_PRB_75_035329}. 
For example, ensembles of quantum dots in photonic crystals experience the spatial and orientational variations of the PLDOS leading to strongly non-single-exponential character of the decay~\cite{Nikolaev2007PRB}. 
The problem of describing relaxation processes that do not obey a simple single-exponential decay is a very general one, and to help deal with it the decay function $f(t)$ is best described using a distribution of rates, of which the log-normal distribution has been found particularly useful, see for details Ref.~\cite{vanDriel_2007_PRB_75_035329}. 

\par
We note that in many reports too little attention is paid in selecting the right function $f(t)$ to fit to the data, as this is essential to adequately interpret non-exponential decay. 
The widely adopted use of stretched exponentials to interpret non-exponential decay (see \textit{e.g.}, ~\cite{Reil_JOpt_2007_9_S437,vanDam_ACSPhot_2018_5_2129}) has a number of problems, notably regarding proper normalization, proper definition, and physical interpretation, see~\cite{vanDriel_2007_PRB_75_035329}.

\subsection{Role of the quantum efficiency}\label{sec:quantum_efficiency}
\par
The quantum efficiency associated with an emissive state determines the effectiveness with which the density of states may change the emission rate. This can easily be seen by re-arranging equation (\ref{eq:beta_1}) to give

\begin{equation}
\label{eq:quant_eff_1}
\frac{\Gamma}{\Gamma_{\rm b}}=(1-\eta)+\eta\frac{\rho_{\rm l,p}(\boldsymbol{e}_{d},\boldsymbol{r}_{0},\omega)}{\rho_{\rm b}(\omega)}.
\end{equation}
The first term $(1-\eta)$ on the right hand side of eq.~\eqref{eq:quant_eff_1} represents the part of the overall decay probability that is non--radiative in character and can therefore not be changed by altering the density of optical states; the second term represents radiative decay and this component is subject to change through the density of states.
\par

\begin{figure}[ht!]
\centering
\includegraphics[width=0.7\columnwidth]{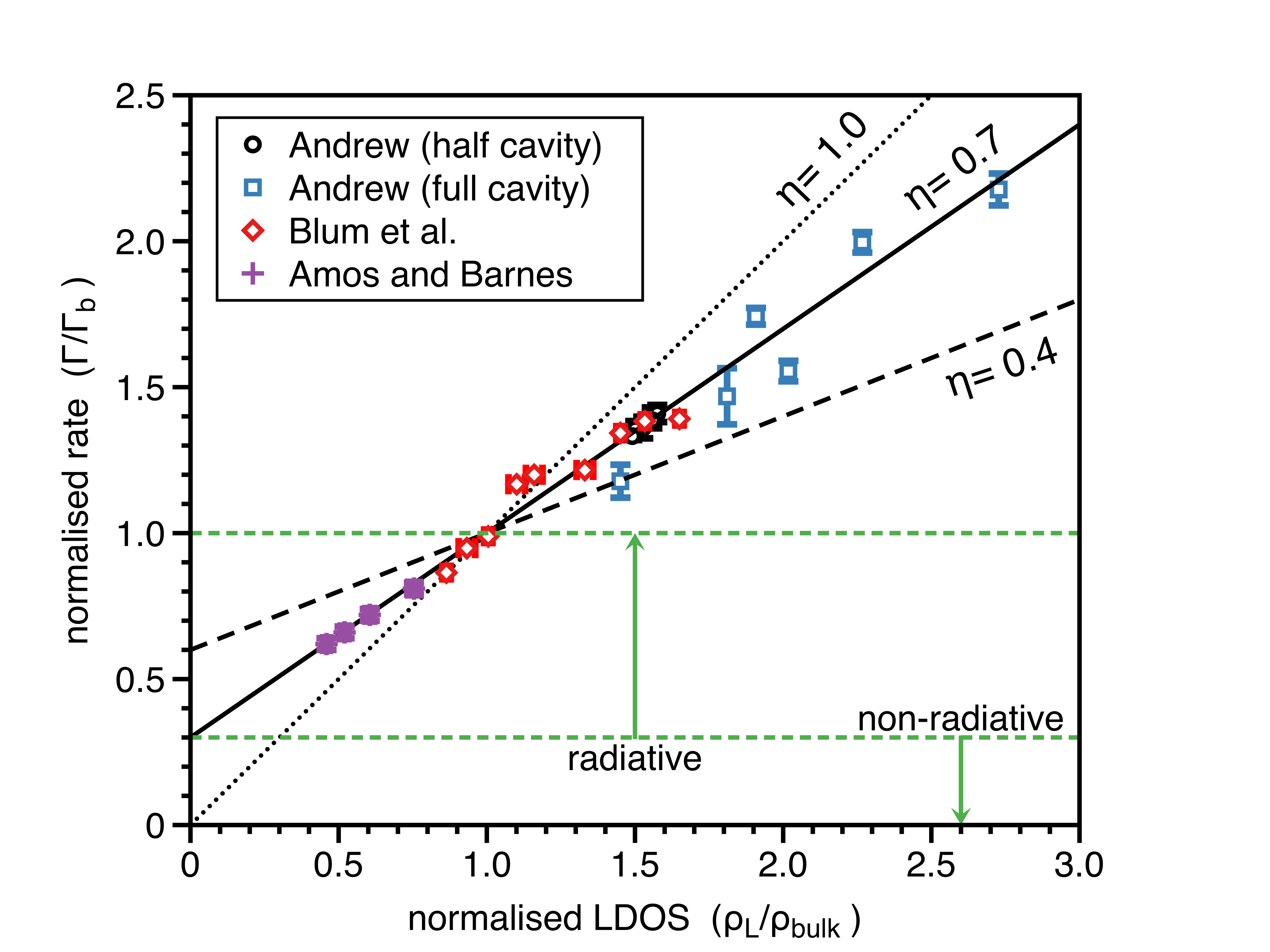}
\caption{
Relationship between decay rate and LDOS.
The normalized decay rate is plotted against the normalized local density of states for two different emitters: ${\rm Eu}^{3+}$ ions and the dye Alexa488. 
Both emitters are randomly oriented, and thus the LDOS rather than the PLDOS is appropriate. 
The full (black circles) and half--cavity (blue squares) data are unpublished results for ${\rm Eu}^{3+}$ ions (kindly provided by Piers Andrew).
The data from Blum \textit{et al.}~\cite{Blum_2012_PRL_109_203601} are for the dye Alexa488. 
Four data points from figure~\ref{fig:Drexhage} have also been added to increase the range of normalized LDOS covered.
These two emitters have the same quantum efficiency ($\sim 0.7$), whereas the decay rates differ by nearly 5 orders of magnitude: ${\rm Eu}^{3+}: \Gamma \sim10^3~{\rm s}^{-1}$; Alexa488: $\Gamma \sim10^8~{\rm s}^{-1}$.
\label{fig:rates-vs-LDOS}}
\end{figure}

One way to visualize the influence of non-radiative decay and the role of the quantum efficiency on the emission process is to make use of Eq.~(\ref{eq:quant_eff_1}) and plot the normalized decay rate, $\Gamma/\Gamma_{\rm b}$ against the normalized partial/local density of states, $\rho_{\rm l,p}(\boldsymbol{e}_{d},\boldsymbol{r}_{0},\omega)/\rho_{\rm b}(\omega)$. 
This is done in Fig~\ref{fig:rates-vs-LDOS} where data are compiled primarily from two sources. 
Data from samples that made use of ${\rm Eu}^{3+}$ ions as the emitters are shown from previously unpublished data kindly provided by Piers Andrew. 
Data presented by Blum \textit{et al.} \cite{Blum_2012_PRL_109_203601} are also shown, for the emissive dye molecules (Alexa488).
These two data sets were chosen because the ${\rm Eu}^{3+}$ ions and the Alexa488 dye have the same quantum efficiency $(\sim 0.7)$, whereas their decay rates differ by nearly 5 orders of magnitude (${\rm Eu}^{3+}~10^3\sim s^{-1}$, Alexa488 $~10^8\sim s^{-1}$). 
The data in Fig.~\ref{fig:rates-vs-LDOS} thus show that the effect of the LDOS on spontaneous emission is universal in character, it is not important what the absolute value of the decay rate is.
Below, we examine the consequences of low and high values of the quantum efficiency for two different kinds of measurement~\cite{Koenderink_PSSA_2003_197_648, El-Dardiry_PRA_2011_83_031801}.

\subsubsection{Continuous excitation}\label{sec:quantum_efficiency_CW}
\begin{figure}[ht!]
\centering
\includegraphics[width=0.5\columnwidth]{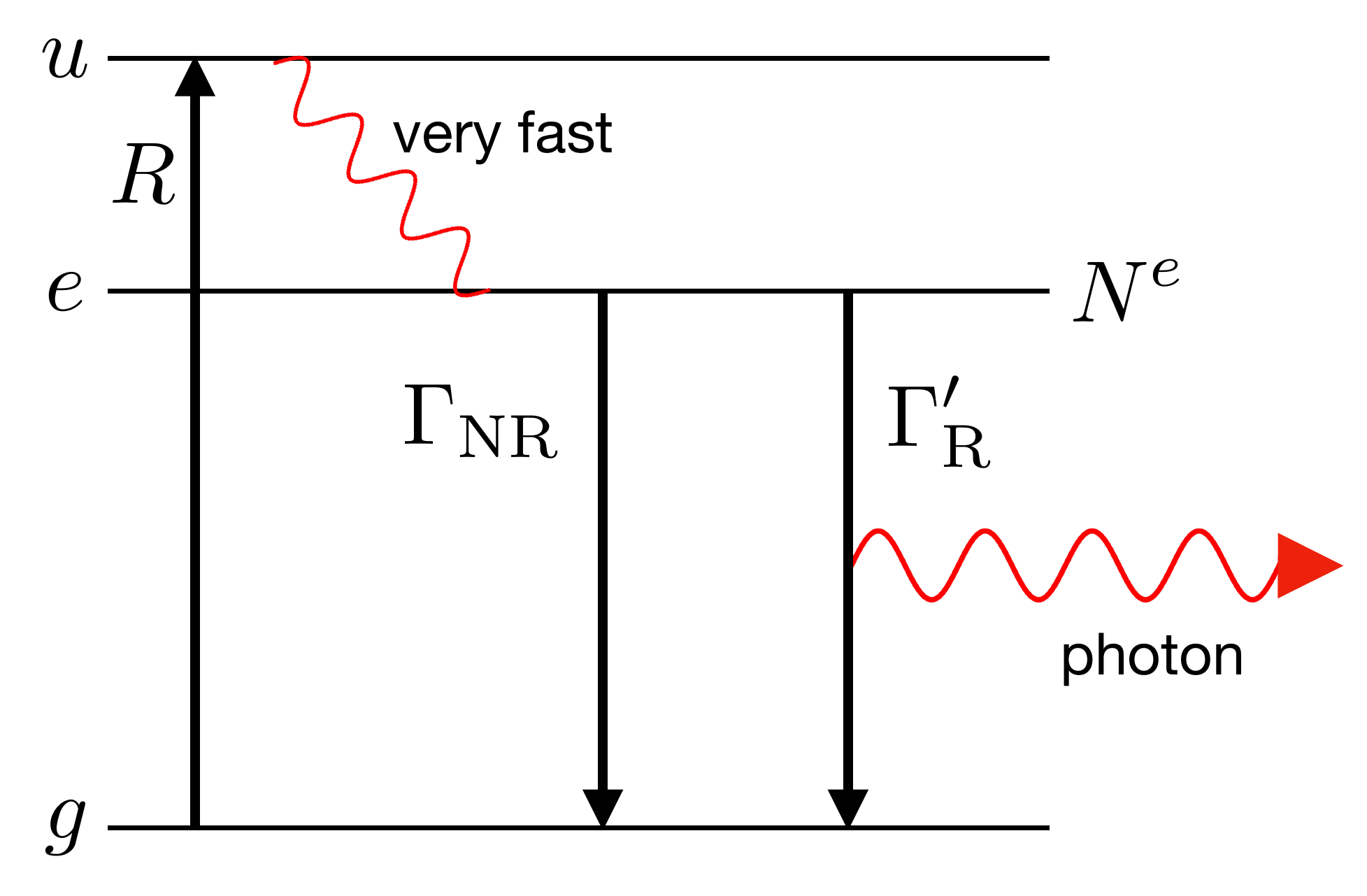}
\caption{
The energy 3-level scheme considered in the analysis of the effect of the (P)LDOS on emitters that are excited continuously, \textit{e.g.}, by a CW laser incident on an ensemble of emitters.
The incident laser leads to some of the emitters in the ground state (g) being excited to the upper state (u), from where they relax very quickly to the excited state (e).
Emitters in the excited state decay back to the ground state either non-radiatively, or radiatively.

\label{fig:3-level}}
\end{figure}

Here we need to be explicit about what the excitation mechanism is not! We are \textbf{not} considering resonant excitation, \textit{i.e.}, exciting emitters by illuminating them with an electromagnetic (optical) field at the same frequency at which they emit, such a situation is known as resonance fluorescence and is not an appropriate way to interrogate emitters if one wishes to learn something about the LDOS they experience. That apart, the mechanism of excitation is not really important, but to be concrete we consider a 3-level emitter-scheme as shown in figure \ref{fig:3-level}.

We consider an ensemble of emitters~\cite{Koenderink_PSSA_2003_197_648}. 
At any moment the number of emitters in the excited state (e) is $N^e$. Emitters in the ground state (g) are pumped at a rate $R$ to an upper state (u) from where they relax very quickly to the excited state.
We consider the non-saturation regime, \textit{i.e.}, where pump rates are such that at any given time nearly all of the emitters are in the ground state.
The pump rate $R$ is thus independent of the population $N^e$.
The rate of change of the number of emitters in the excited state has two components: the state is populated by pumping, and depleted by decay. 
We thus have 
\begin{equation}
\label{eq:rate-eqn}
\frac{dN^e}{dt} = R - \Gamma N^e,
\end{equation}
where $\Gamma$ is the decay rate in our structured local environment \footnote{There is scope for some confusion here, since both $R$ and $\Gamma$ are referred to as rates, and yet $\Gamma$ in eq.~\eqref{eq:rate-eqn} is multiplied by $N^e$.
Both $R$ and $\Gamma$ are indeed rates, but $\Gamma$ is the rate for one emitter, the rate for $N^e$ emitters is thus $\Gamma N^e$.}, see~eq.~\eqref{eq:struc}. 
In steady state ${dN^e}/{dt}=0$ so that eq.~\eqref{eq:rate-eqn} becomes
\begin{equation}
\label{eq:eqn-steadystate-R}
R = \Gamma N^e.
\end{equation}
It is, however, more informative to write this equation as
\begin{equation}
\label{eq:rate-eqn-steadystate}
N^e = R/\Gamma,
\end{equation}
because the meaning of this equation is that for a given pump rate $R$ the number of emitters in the excited state, under steady-state conditions, is $R/\Gamma$.
Now, the intensity $I$ of the emission produced by the emitters is equal to 
\begin{equation}
\label{eq:intensity-1}
I = \Gamma'_R N^e,
\end{equation}
where as before, $\Gamma'_R$ is the radiative rate in our modified environment. 
It is useful to make use of the quantum efficiency $\eta$, given by \eqref{eq:qe_defn} as $\eta = \Gamma'_R/(\Gamma'_R+\Gamma_{NR})$.
We can thus write the intensity as
\begin{equation}
\label{eq:intensity-2}
I = \eta \Gamma N^e,
\end{equation}
which in turn, through the use of \eqref{eq:rate-eqn-steadystate} can be written as 
\begin{equation}
\label{eq:intensity-3}
I = \eta \Gamma N^e.= \eta \Gamma \frac{R}{\Gamma}.
\end{equation}
This equation tells us how the emitted intensity under CW excitation depends on the pump rate.
It is instructive to look at two regimes, $\eta \sim 1$ (high efficiency), and $\eta \ll 1$ (low efficiency).\\[10pt] 

\noindent (1) High efficiency $(\eta = 1)$. 
In this case eq.~\eqref{eq:intensity-3} becomes $I=R$, \textit{i.e.}, the emitted intensity is determined solely by the pump rate $R$, whereas the (P)LDOS plays no role.
In the radio antenna world this is known as a constant power source (CPS), the power radiated by the antenna is independent of the load it is subjected to~\cite{El-Dardiry_PRA_2011_83_031801}.
This lack of LDOS dependence may at first sight seem strange since a cursory examination of eq.~\eqref{eq:eqn-steadystate-R} might lead one to think that the LDOS \textit{does} play a role, since a change in LDOS leads to a change in $\Gamma$. 
This highlights the need to be careful to consider which are the dependent variables in an equation such as eq.~\eqref{eq:eqn-steadystate-R}; in this case it is \textit{not} the pump rate that is determined by the number of emitters in the excited state $N^e$, rather it is the number of emitters in the excited state that is determined by the pump rate, \textit{i.e.}, it is \eqref{eq:rate-eqn-steadystate} that should be used to guide our thinking here.
Summarizing, for $\eta \sim 1$, \textit{i.e.}, for efficient emitters, the emission intensity depends only on the pump rate and no information can be gained about the LDOS from CW measurements.\\[10pt]

\noindent (2) Low quantum efficiency $(\eta \ll 1)$. 
In this case eq.~\eqref{eq:intensity-3} becomes 
\begin{equation}
\label{eq:intensity-4}
I = \frac{\Gamma'_R}{\Gamma_{NR}}R,
\end{equation}
since for $\eta \ll 1$, ${\Gamma'_R}/{(\Gamma'_R+\Gamma_{NR})}\approx {\Gamma'_R}/{\Gamma_{NR}}$.
By using eq.~\eqref{eq:quant_eff_1} we find that
\begin{equation}
\label{eq:intensity-5}
I = \Gamma_R \frac{\rho_{\rm l,p}}{\rho_{\rm b}} R,
\end{equation}
so that in this poor emitter limit the intensity \textit{is} a good probe of the LDOS.
In the radio-antenna world this is known as a constant amplitude source (CAS).
In practice things may not be so simple, \textit{i.e.}, a low quantum efficiency emitter interrogated using CW excitation may not in fact be such a good probe of the LDOS since: (i) the signal strength from poor emitters will be low and, (ii) a change in LDOS is likely to lead to a change in the radiation pattern, something that is not easy to take into account, see e.g., Ref.~\cite{Koenderink_PRL_2002}. 

\subsubsection{Pulsed excitation}\label{sec:quantum_efficiency_pulsed}
We now consider the case of time-resolved measurements, where the rate or lifetime is measured following pulsed (rather than CW) excitation, and where the excitation pulse is considered to be much shorter than the decay rate $\Gamma$. 
Looking at Eq.~\eqref{eq:quant_eff_1} we see that if the quantum efficiency is high, the LDOS significantly alters the decay rate, thus time-resolved emission experiments provide a convenient probe of the LDOS.
If the quantum efficiency is low, however, then Eq.~\eqref{eq:quant_eff_1} shows that the LDOS has little effect on the overall rate; the time resolved emission is dominated by non-radiative decay. 
Thus in this case, difficult as it may be due to possible problems with radiation patterns and so on, CW measurements may be preferred.\\[10pt]

\subsection{Bandwidths, time-scales, and the LDOS}
In experiments and in real devices we should note that it is important to consider not just the central frequency (see equation~\ref{eq:SPE__text_book}) and the radiative emission rate (see equation~\ref{eq:SPE_rate_PLDOS}), we need also to consider the spectral bandwidth of the spontaneous emission process.. 
Commonly, the spectral bandwidth has two contributions, known as homogeneous and inhomogeneous broadening, where the latter results from studying many emitters (an ensemble), whilst the former is the broadening associated with a single emitter. 
Inhomogeneous effects may for instance be caused by strain in a solid sample varying from one emitter site to another. 
The homogeneous bandwidth $\Delta \omega_{hom}$ is typically interpreted as a combination of the finite excited state lifetime of any emitter, and the dephasing time. 
The latter is the time scale over which an oscillation maintains its phase, before being disturbed by external fluctuations such as Brownian motion~\cite{Loudon_TQTL}. 

As a final thought about time-scales, we note that there is very way to view the LDOS by considering it to be a correlation time of the so-called 'bath' (the environment) that surrounds the oscillating emitter~\cite{lagendijk1993lucca}. 
One can show that this correlation time equals the product of the LDOS and the wavelength cubed, see also~\cite{Nikolaev_PhD_2006}. As an example, for an LDOS of 10$^4$ $s~m^{-3}$ and a wavelength of 1 $\mu$m, the correlation time is $\sim$ 10$^{-14}~s$, this compares with typical emission lifetimes encountered in nanophotonics of 10$^{-12}$ -- 10$^{-3}~s$.
When the correlation time is much faster than a characteristic time of the emitter (like the lifetime), as in the case of this simple example, the bath responds faster than the emitter influences its environment, and in this case Markovian dynamics pertain.

\section{Summary and conclusions}\label{sec:conclusions}
In this didactic article we have reviewed the concept of the local density of optical states, showing that there are in fact three quantities, namely the PLDOS, the LDOS, and the DOS that contain decreasing levels of information about the available modes in a system. 
Having shown that the partial local density of states (PLDOS) governs the radiation from both classical and quantum emitters, we illustrated several examples in detail, where the PLDOS, LDOS and DOS can be calculated exactly; from the simplest case of a homogeneous medium, to the relatively complicated case of a planar cavity that has lossy mirrors. 
We illustrated the very close relationship between the process of quantum and classical emission of radiation, showing that quantum emitters are more efficient than expected classically due to the presence of the quantum vacuum. 
We have also presented data to show that the effect of the LDOS on spontaneous emission is universal in character by comparing data that spans five orders of magnitude in terms of emission rate, and in terms of emission wavelength.

\section{Acknowledgments}
We would like to acknowledge useful discussions with Malcolm Longair, Keith Burnett, Piers Andrew, Tom Philbin, Ad Lagendijk, Allard Mosk, Christian Blum, Femius Koenderink, Ivan Nikolaev, Jean-Michel G\'erard, Martijn Wubs, Merel Leistikow, Mischa Megens, Pepijn Pinkse, Peter Lodahl, Rudolf Sprik, and many, many others over the course of many years.
SARH acknowledges financial support from a Royal Society TATA University Research Fellowship (RPG-2016-186), WLV thanks FOM, NWO, STW, MESA+ for support. WLB is indebted to both the Leverhulme Trust and the Royal Society for their support. 
WLB would like to thank the Royal Society for access to a first edition of Dirac, \textit{Principles of Quantum Mechanics}, and is indebted to Piers Andrew for successfully digging out data from more 20 years ago.
Finally, we are grateful to both Femius Koenderink and Dean Patient for providing valuable and detailed feedback on early versions of this manuscript.\\

\section{References}

\bibliography{LDOS.bib}
%
%
\section*{Glossary of symbols}
Symbols are arranged as follows: physical constants, geometric quantities, rates and numbers, fields, material parameters, quantum operators and states, general notation. 
\begin{longtable}{ll}
$\hbar$& Planck's constant, units J s \\
$\epsilon_{0}$& Dielectric permittivity of free space, units m$^{-3}$ kg$^{-1}$ s$^4$ A$^2$ \\
$\mu_0$& Magnetic permeability of free space, units m kg s$^{-2}$ A$^{-2}$ \\
$c$&Speed of light in vacuum, units m s$^{-1}$ \\
$\boldsymbol{r}=(x,y,z)$& Position of observer, or point of calculation, units m\\
$\boldsymbol{r}_{0}=(x_0,y_0,z_0)$& Position of emitter, units m\\
$\boldsymbol{e}_{i}$& Unit vector in the direction $i$\\
$\perp$& Out of plane orientation\\
$\parallel$& in-plane orientation\\
$\boldsymbol{k}=(k_x,k_y,k_z)$& Wave vector, units rad m$^{-1}$\\
$k$&Magnitude of $\boldsymbol{k}$, units rad m$^{-1}$ \\
$\kappa$&Decay constant (imaginary part of wave number), units rad m${^{-1}}$\\
$\lambda$& Wavelength, units m\\
$L$& Spatial period (with periodic boundary conditions), units m\\
$h$& Planar cavity width, units m\\
$\ell$& Length of wire, units m\\
$V$& Volume (occupied by field modes), units m$^3$\\
$n_{x},n_{y},n_{z}$&Mode numbering\\
$\zeta$& Polarization of mode\\
$f$& Frequency, units s$^{-1}$\\
$\omega$& Angular frequency, units rad s$^{-1}$\\
$\omega_{p}$& Plasma frequency, units rad s$^{-1}$\\
$\gamma$& Damping constant, units rad s$^{-1}$\\
$k_0$& Free space wave number, units rad m$^{-1}$\\
$p$& Probability of decay from excited state\\
$n(t)$& Number of excited emitters\\
$g(t)$& Total decay intensity, units s$^{-1}$\\
$f(t)$& Radiative decay intensity, units s$^{-1}$\\
$\Gamma$& Rate of decay from excited state, units s$^{-1}$\\
$\Gamma_{0}$& Free space decay rate, units s$^{-1}$\\
$\Gamma_{\rm b}$& Bulk decay rate, units s$^{-1}$\\
$\Gamma_{\rm R}$, $\Gamma_{\rm R}'$& Radiative decay rate (prime indicates structured environment), units s$^{-1}$\\
$\Gamma_{\rm NR}$, $\Gamma_{\rm NR}'$& Non--radiative decay rate (prime indicates structured environment), units s$^{-1}$\\
$N^e$& Number of emitters in the excited state\\
$R$& Pump rate, units s$^{-1}$\\ 
$\eta$& Quantum efficiency\\
$\eta_b$& Quantum efficiency in bulk\\
$\tau$& Lifetime of excited state, units $s$\\
$t_{1/2}$& Half-life of excited state, units $s$\\
$T$& Time interval (for time averaging), units $s$\\
$\rho_{\rm p}$&Partial local density of states, units m$^{-3}$ s\\
$\rho_{\rm b}$&Partial local density of states in bulk medium, units m$^{-3}$ s\\
$\rho_{\rm l}$&Local density of states, units m$^{-3}$ s\\
$\rho$& Number density of states, units m$^{-3}$ s\\
$\rho_0$& Number density of states in free space, units m$^{-3}$ s\\
$N$&Number of modes\\
$N^e$&Number of emitters in the excited state\\
$N^e_0$&Number of emitters in the excited state at $t=0$\\
$\boldsymbol{d}=(d_x,d_y,d_z)$&Dipole moment vector, units C m\\
$\boldsymbol{e}_{d}$& Unit vector in direction of $\boldsymbol{d}$\\
$d$& Amplitude of $\boldsymbol{d}$, units C m\\
$\boldsymbol{E}=(E_{x},E_{y},E_{z})$& Electric field vector, units N C$^{-1}$ \\
$\boldsymbol{B}=(B_{x},B_{y},B_{z})$& Magnetic field vector, units kg s$^{-1}$ C$^{-1}$\\
$\boldsymbol{A}=(A_{x},A_{y},A_{z})$& Magnetic vector potential, units kg m s$^{-1}$ C$^{-1}$\\
$\phi$& Electric scalar potential, units J C$^{-1}$\\
$\sigma$&Electric charge density, units C m$^{-3}$\\
$\boldsymbol{j}=(j_x,j_y,j_z)$&Electric current density, units C m$^{-2}$ s$^{-1}$ \\
$I$&Electrical current, units C s$^{-1}$\\
$\boldsymbol{v}=(v_x,v_y,v_z)$&Velocity, units m s$^{-1}$\\
$\overleftrightarrow{\boldsymbol{G}}$&Dyadic Green function, units m$^{-1}$\\
$\overleftrightarrow{\boldsymbol{G}}_0$&Dyadic Green function in a homogeneous medium, units m$^{-1}$\\
$\sigma_{s}$&Total scattering cross section, units m$^{2}$\\
$f(\theta,\phi)$&Scattering amplitude, units m\\
$\boldsymbol{\mathcal{E}}_{n}$&Eigenmode of vector Helmholtz equation, units m$^{-3/2}$\\
$f_{n}$& Basis set of functions\\
$v_n$& Expansion coefficients of voltage, units V m$^{-1}$\\
$i_n$& Expansion coefficients of current, units C s$^{-1}$\\
$\omega_n$& Frequency of above eigenmode, units s$^{-1}$\\
$c_{n}$& Coefficient in mode expansion, units C$^{-1}$ m$^{5/2}$ s$^{-1}$\\
$P$& Emitted power, units J s$^{-1}$\\
$P_{0}$& Emitted power in free space, units J s$^{-1}$\\
$\alpha$& Polarizability, units C$^{2}$ m N$^{-1}$\\
$\epsilon$& Relative permittivity\\
$\chi$&Electric susceptibility\\
$\mu$& Relative permeability\\
$\rm n$ &Refractive index\\
$Z$& Electrical impedance, units $\Omega$\\
$Z_{nm}$& Impedance matrix, units $\Omega$\\ 
$r_{i}$& Reflection coefficient for the $i^{\rm th}$ polarization\\
$\hat{H}$&Hamiltonian operator\\
$|\psi\rangle$&General state vector\\
$\hat{s}_{x,y,z}$&Atomic operators\\
$|0\rangle_{\rm at},|1\rangle_{\rm at}$&Atomic states\\
$|0\rangle_{\rm f}$&Electromagnetic vacuum (ground) state\\
$\hat{\pi}_{n}$&Momentum operator, conjugate to $\hat{c}_{n}$\\
$\mathcal{P}$&Complex transition dipole moment, units C m$^{-1}$ \\
$d_{1},d_{2}$&Diagonal elements of dipole operator\\
$\alpha_{n},\beta_{n}$&Auxiliary quantities determining evolution of dipole operator\\
$q$&Quantum efficiency\\
$\tilde{\dots}$&Complex amplitude at fixed frequency, \textit{e.g.}, $A(t)={\rm Re}[\tilde{A}{\rm e}^{-{\rm i}\omega t}]$\\
$\hat{\dots}$&Operator corresponding to classical quantity beneath\\
$\dots^{\rm Q}$&Calculated using quantum mechanics\\
$\dots^{\rm C}$&Calculated using classical electromagnetism\\
$\dots^{(0)}$&zeroth-order in perturbation theory\\
$\dots^{(1)}$&first-order in perturbation theory\\
$\text{[S]}$&Equation in Schr\"odinger picture\\
$\text{[H]}$&Equation in Heisenberg picture\\
\end{longtable}

\end{document}